\documentclass[a4paper,11pt]{article}
\pdfoutput=1

\usepackage{jheppub}
\usepackage{graphicx}
\usepackage{amsmath}
\usepackage{xspace}
\usepackage{color}
\usepackage[utf8]{inputenc} 
\usepackage{placeins}

\newcommand{\eq}[1]{eq.~\eqref{eq:#1}}

\renewcommand{\sec}[1]{section~\ref{sec:#1}}

\newcommand{\app}[1]{appendix~\ref{app:#1}}
\newcommand{\fig}[1]{figure~\ref{fig:#1}}

\newcommand{\Tab}[1]{table~\ref{tab:#1}}
\newcommand{\mycites}[1]{refs.~\cite{#1}}
\newcommand{\mycite}[1]{ref.~\cite{#1}}

\newcommand{\CP}{$\mathcal{CP}$}

\newcommand{\Zsym}{$\mathbb{Z}_2$ }
\newcommand{\real}[1]{\mathrm{Re}\left(#1\right)}
\newcommand{\imag}[1]{\mathrm{Im}\left(#1\right)}
\newcommand{\df}{\mathrm{d}}

\newcommand{\nn}{\nonumber}

\newcommand{\braket}[1]{\left\langle #1\right\rangle}
\newcommand{\code}[1]{\texttt{#1}}

\usepackage{feynmp-auto}

\allowdisplaybreaks[3]

\title{2-loop RG evolution of \CP-violating 2HDM}

\author{Joel Oredsson,}
\author{Johan Rathsman}

\emailAdd{joel.oredsson@thep.lu.se}
\emailAdd{johan.rathsman@thep.lu.se}

\affiliation{Department of Astronomy and Theoretical Physics, Lund University,
S\"{o}lvegatan 14A 223 62 Lund, Sweden}

\abstract{
    We use the recently developed code \code{2HDME} to perform a 2-loop
    renormalization group analyzis of the \CP~violating Two-Higgs-doublet model
    (2HDM).  Using parameter scans of several scenarios of \Zsym symmetry
    breaking, we investigate the properties of 2HDMs under renormalization group
    evolution.  Collider data constraints are implemented with \code{HiggsBounds}
    and \code{HiggsSignals} and we include all the important Barr-Zee diagram
    contributions to the electron's electric dipole moment to put limits on the
    \CP~violation that is allowed both in the scalar and the Yukawa sector. As a
    result, we see that the \CP~violation spreads easily across the sectors
    during renormalization group evolution when one breaks the \Zsym symmetry in
    either sector, putting additional constraints on the \CP~violating
    parameters.
}
\keywords{Two-Higgs-Doublet Model, CP violation, electric dipole moment,
renormalization group equation, 2-loop}

\begin{document}

\preprint{
\begin{flushright}
LU TP 19-43\\
\today
\end{flushright}
}

\maketitle

%%%%%%%%%%%%%%%%%%%%%%%%%%%%%%%%%%%%%%%%%%%%%%%%%%%%%%%%%%%%%%%%%%%%%%%%%%%%%%%%
\section{Introduction}\label{sec:intro}
%%%%%%%%%%%%%%%%%%%%%%%%%%%%%%%%%%%%%%%%%%%%%%%%%%%%%%%%%%%%%%%%%%%%%%%%%%%%%%%%

The Two-Higgs-Doublet Model (2HDM) is one of the most studied minimalistic
extensions of the Standard Model (SM) of particle physics and serves as an
effective theory for many Beyond the SM (BSM) models. It offers a rich scalar
sector with its three neutral plus a charged pair of Higgs bosons. One of the
neutral ones should make up the 125 GeV scalar particle that has been discovered
at the Large Hadron Collider (LHC) by the ATLAS \cite{Aad:2012tfa} and CMS
\cite{Chatrchyan:2012xdj} collaborations; that, so far, resembles the SM Higgs
boson \cite{Khachatryan:2016vau}.  According to the LHC data, it is ruled out
that the discovered particle is a pure \CP~odd scalar
\cite{Khachatryan:2014kca}; however, there is still the possibility that it is a
mixture of \CP~even and \CP~odd states. 

While the 2HDM is well studied in the literature, it is most often the
\CP~conserving one even though the 2HDM exhibits the possibility of
\CP~violation. It is well known that, to fulfill the Sakharov's criteria for
baryogenesis \cite{Sakharov:1967dj}, new sources of \CP~violation are needed to
sufficiently explain the excess of matter over anti-matter in the universe.  The
possibility for \CP~violation in the 2HDM is therefore an intriguing feature, which
was the original motivation for studying the 2HDM in the first place
\cite{Lee:1973iz}.  

There are several experiments that limit the amount of \CP~violation one can
have in the 2HDM. One of the most troublesome observables is that one
easily generates a large Electric Dipole Moments (EDMs) for particles; which
are severely constrained by experiments.  The electron's EDM (eEDM) has recently
received an upper limit from the ACMEII collaboration \cite{Andreev:2018ayy},
\begin{align}\label{eq:ACMEII}
  |d_e| < 1.1 \times 10^{-29}~\text{e cm}.
\end{align}

Another well known problem of the general 2HDM, not exclusive to the
\CP~violating case, is the presence of Flavor-Changing-Neutral currents (FCNCs)
and one popular solution is to impose a \Zsym symmetry on the model
\cite{Glashow:1976nt,Paschos:1976ay}; although, often one allows for a soft
\Zsym symmetry breaking term.

In recent years, the softly broken \Zsym symmetric 2HDM has been confronted with
data from the LHC as well as the limits coming from EDMs \cite{Shu:2013uua,
Jung:2013hka, Inoue:2014nva, Cheung:2014oaa, Ipek:2013iba, Bian:2014zka,
Chen:2015gaa, Fontes:2017zfn, Egana-Ugrinovic:2018fpy}.  In this work, we
investigate the complex, \CP~violating, 2HDM by probing its behavior under
Renormalization Group (RG) evolution, while also revisiting the constraints from
collider data and the eEDM. We focus on RG effects such as how different choices
of \Zsym~symmetry can affect the energy range of validity and look for any
symmetry breaking, or \CP~violation, that spreads across the Yukawa and scalar
sectors.

There are many studies that analyze the 2HDM using 1-loop RG Equations (RGEs),
\textit{e.g.} \mycites{Dev:2014yca, Penuelas:2017ikk, Botella:2018gzy,
Gori:2017qwg, Basler:2017nzu,Ferreira:2015rha,Bijnens:2011gd}, but also some
that use the 2-loop ones \cite{Chowdhury:2015yja, Krauss:2018thf,
Oredsson:2018yho}.  We use the code \code{2HDME} \cite{Oredsson:2018vio} to
perform the RG evolution at 2-loop order. It is essential to go to 2-loop order
if one is interested in studying how all sectors affect each other during RG
evolution; since the quartic couplings enter the Yukawa couplings' RG Equations
(RGEs) first at this order.

To investigate the parameter space of the 2HDM, we set up numerical parameter
scans of different physical scenarios; each with their own level of \Zsym
symmetry in each sector.  In addition to looking for Landau poles in the RG
evolution, we perform tree-level checks of unitarity and stability.  We also
compute the oblique parameters $S$, $T$ and $U$; as well as the branching ratios
for all Higgs decays with a modified version of \code{2HDMC}
\cite{Eriksson:2009ws}.  To check whether a parameter point is excluded by
collider data, we use the codes \code{HiggsBounds} \cite{Bechtle:2008jh,
Bechtle:2011sb, Bechtle:2013wla} and \code{HiggsSignals} \cite{Bechtle:2013xfa}. 
Finally, we use the eEDM as an additional constraint on the amount of
\CP~violation.  We calculate this observable  by summing up all the relevant
Barr-Zee diagrams \cite{Barr:1990vd} that contribute. This calculation is also
implemented in a recent update of \code{2HDME}. 

This paper is structured as follows: we begin in \sec{2HDM} by giving a brief
review of the 2HDM and present the notation that we use. We describe the
different scenarios of parameter scans in \sec{scenarios}. The constraints that
we implement are listed in \sec{constraints}. To illustrate the characteristic change
in behavior of the 2HDM when allowing for \CP~violation, we vary the amount of
\CP~violation in an example point in \sec{examplePoint} and look at various
observables. The main results of the parameter scans are presented in
\sec{results} and we subsequently summarize our conclusions in
\sec{conclusions}. Some plots of the generic basis of 2HDM in the first scenario
are collected in \app{GenBase}. In \app{BarrZee}, we list all the formulas
for every Barr-Zee diagram that we use to calculate the eEDM.

%%%%%%%%%%%%%%%%%%%%%%%%%%%%%%%%%%%%%%%%%%%%%%%%%%%%%%%%%%%%%%%%%%%%%%%%%%%%%%%%
\section{The 2HDM}\label{sec:2HDM}
%%%%%%%%%%%%%%%%%%%%%%%%%%%%%%%%%%%%%%%%%%%%%%%%%%%%%%%%%%%%%%%%%%%%%%%%%%%%%%%%

Since 2HDM is one of the most studied BSM theories, we will only briefly
describe it here and for a full review we refer to \mycite{Branco:2011iw}.
Throughout this work, we use the notation employed in the basis independent
treatment of the 2HDM in 
\mycites{Davidson:2005cw,Haber:2006ue,Haber:2010bw}.

The most general gauge invariant renormalizable scalar potential for two
hypercharge $+1/2$ Higgs doublets, $\Phi_{1,2}$, can be written
\begin{align}\label{eq:GenericPotential}
        -\mathcal{L}_V=&m_{11}^2\Phi_1^\dagger\Phi_{1} +
        m_{22}^2\Phi_2^\dagger\Phi_{2} -
        (m_{12}^2\Phi_1^\dagger\Phi_{2}+\text{h.c.}) +
        \frac{1}{2}\lambda_1\left(\Phi_1^\dagger\Phi_{1}\right)^2
        +\frac{1}{2}\lambda_2\left(\Phi_2^\dagger\Phi_{2}\right)^2\nonumber\\
        &+\lambda_3\left(\Phi_1^\dagger\Phi_{1}\right)\left(\Phi_2^\dagger\Phi_2\right)
        +\lambda_4\left(\Phi_1^\dagger\Phi_{2}\right)\left(\Phi_2^\dagger\Phi_1\right)
		\nonumber\\
        &+\left[\frac{1}{2}\lambda_5\left(\Phi_1^\dagger\Phi_2\right)^2
        +\lambda_6\left(\Phi_1^\dagger\Phi_1\right)\left(\Phi_1^\dagger\Phi_2\right)
        +\lambda_7\left(\Phi_2^\dagger\Phi_2\right)\left(\Phi_1^\dagger\Phi_2\right)+\text{h.c.}\right],
\end{align}
where $m_{12}^2$ and $\lambda_{5,6,7}$ are potentially complex
while all the other parameters are real; resulting in a total of 14 degrees of
freedom. 

After electroweak symmetry breaking, $SU(2)\times U(1)_Y\rightarrow
U(1)_{\text{em}}$, both the scalar fields acquire a Vacuum Expectation Value
(VEV).  Using global $SU(2)_L$
and $U(1)$ rotations, the fields' VEV take the forms
\begin{align}
\braket{\Phi_1} = \frac{1}{\sqrt{2}}\left(\begin{array}{c} 0 \\ v_1 \end{array}\right)
  &&
  \text{and}
  &&
\braket{\Phi_2} = \frac{1}{\sqrt{2}}\left(\begin{array}{c} 0 \\ v_2e^{i\xi} \end{array}\right),
\end{align}
where $v=\sqrt{v_1^2+v_2^2}\approx 246$ GeV and we define $\tan\beta \equiv
t_\beta \equiv v_2/v_1$.  By convention, we take $0\leq \beta \leq \pi/2$ and
$0\leq \xi\leq 2\pi$.  Note that, if the two Higgs fields are identical by
having equal quantum numbers, one is free to perform a Higgs flavor basis
transformation and $\tan\beta$ is an unphysical parameter
\cite{Haber:2006ue}.

Minimizing the potential results in the tadpole equations
\begin{align}
  m_{11}^2 =~&m_{12}^2e^{i\xi}t_\beta - \frac{1}{2}v^2\left[\lambda_1 c_\beta^2 
  +(\lambda_3+\lambda_4+\lambda_5e^{2i\xi})s_\beta^2 \right.\nn\\
  &\hspace{3cm}\left.+(2\lambda_6e^{i\xi}+\lambda_6^*e^{-i\xi})s_\beta c_\beta+ \lambda_7s_\beta^2 t_\beta e^{i\xi}\right],\\
  m_{22}^2 =~&m_{12}^2e^{i\xi}t_\beta^{-1} - \frac{1}{2}v^2\left[\lambda_2 s_\beta^2 + (\lambda_3+\lambda_4+\lambda_5^*e^{-2i\xi})c_\beta^2 \right.\nn\\
  &\hspace{3cm}\left.+(\lambda_7e^{i\xi}+2\lambda_7^*e^{-i\xi})s_\beta c_\beta+ \lambda_6^*c_\beta^2 t_\beta^{-1} e^{-i\xi}\right],\\
  \text{Im}(m_{12}^2 e^{i\xi}) =~& \frac{1}{2}v^2\left[\text{Im}(\lambda_5e^{2i\xi})s_\beta c_\beta+\text{Im}(\lambda_6e^{2i\xi})c_\beta^2+ \text{Im}(\lambda_7e^{i\xi}) s_\beta^2\right].
\end{align}
These are used to fix $m_{11}^2$, $m_{22}^2$ and $\xi$.

%------------------------------------------------------------------------------
\subsection{The Higgs basis}\label{sec:HiggsBasis}
%------------------------------------------------------------------------------

In \eq{GenericPotential} the general scalar potential for the 2HDM is written in
the generic basis.  Another basis is the Higgs basis
\cite{Branco:1999fs,Davidson:2005cw}, where only one Higgs field gets a VEV.
The Higgs basis fields in terms of the previously defined generic basis fields
are\footnote{The bar notation keep tracks of complex conjugation. That is, replacing a barred index to an unbarred corresponds to complex conjugation \cite{Davidson:2005cw,Haber:2006ue, Haber:2010bw}.}
\begin{align}
	H_1 \equiv \hat{v}_{\bar{a}}^* \Phi_a, && H_2 \equiv \hat{w}_{\bar{a}}^* \Phi_a,
\end{align}
where $\hat{w}_b \equiv \hat{v}_{\bar{a}}^*\epsilon_{ab}$ (
$\epsilon_{12}=-\epsilon_{21}=1$) and
\begin{align}
    \hat{v}_a \equiv \left( \begin{array}{c} c_\beta
    \\ s_\beta e^{i\xi} \end{array}\right).
\end{align}
These fields acquire the VEVs
\begin{align}
	\braket{H_1^0} = v/\sqrt{2} , && \braket{H_2^0} = 0.
\end{align}
The scalar potential in the Higgs basis takes a similar form as in the generic
basis,
\begin{align}\label{eq:HiggsPotential}
    -\mathcal{L}_V =~& Y_1 H_1^\dagger H_1 + Y_2 H_2^\dagger H_2 +
    \left(Y_3H_1^\dagger H_2 + \text{h.c.}\right) + \frac{1}{2}Z_1(H_1^\dagger H_1)^2 +
    \frac{1}{2}Z_2(H_2^\dagger H_2)^2\nn \\ &+ \frac{1}{2}Z_3(H_1^\dagger
    H_1)(H_2^\dagger H_2)+ \frac{1}{2}Z_4(H_1^\dagger H_2)(H_2^\dagger H_1)\nn
    \\ &+\left\{\frac{1}{2}Z_5(H_1^\dagger H_2)^2 + \left[Z_6(H_1^\dagger H_1) +
    Z_7(H_2^\dagger H_2)\right]H_1^\dagger H_2 + \text{h.c.}\right\}, 
\end{align}
where $Y_3$ and $Z_{5,6,7}$ are potentially complex.
The tree-level tadpole equations are given by
\begin{align}
	Y_1 = -\frac{1}{2}Z_1 v^2, && Y_3 = -\frac{1}{2}Z_6 v^2.
\end{align}
The Higgs basis is unique up to a rephasing of $H_2$.  During a Higgs flavor
transformation of the generic basis, $\Phi_a \rightarrow U_{a\bar{b}}\Phi_b$,
the Higgs fields transform as \cite{Haber:2006ue}
\begin{align}
	H_1 \rightarrow H_1, && H_2 \rightarrow (\det U) H_2.
\end{align} 
Thus from inspection of the Higgs potential in \eq{HiggsPotential}, it follows
that $Y_{1,2}, Z_{1-4}$ are invariant, while
\begin{align}
	\{Y_3, Z_{6,7}\} \rightarrow (\det U)^{-1} \{Y_3, Z_{6,7}\}, && Z_5\rightarrow (\det U)^{-2} Z_5
\end{align} 
are pseudo-invariants under the Higgs flavor transformation.

The Higgs doublets are expanded around the VEV and parameterized as
\begin{align}
    H_1 =\left(\begin{array}{c} G^+ \\  \frac{1}{\sqrt{2}}(v+\phi_1^0 + i G^0) \end{array}\right)
  &&
  \text{and}
  &&
    H_2 =\left(\begin{array}{c} G^+ \\  \frac{1}{\sqrt{2}}(\phi_2^0 + i a^0)
    \end{array}\right),
\end{align}
where $G^{0,+}$ are Goldstone bosons that will be eaten by $Z$ and $W^\pm$.
The physical scalar degrees of freedom, after electroweak symmetry breaking,
correspond to three neutral ones that we will order according to their mass and
denote as $h_{1,2,3}$; and one $U(1)_{em}$ charged pair of Higgs bosons that we
will denote as $H^{\pm}$.  In the \CP-conserving case, the neutral mass
eigenstates have definite \CP~properties; while all the neutral Higgs bosons mix
and have indefinite \CP~properties in the \CP~violating case. The neutral mass
matrix in the $\phi_1^0 - \phi_2^0 - a^0$ basis is
\small
\begin{align}
    \mathcal{M} = v^2 \left( \begin{array}{ccc} Z_1 & \real{Z_6} & -\imag{Z_6}\\
        \real{Z_6} & \frac{1}{2}\left[Z_3+Z_4+\real{Z_5}\right]+Y_2/v^2 &
        -\frac{1}{2} \imag{Z_5}\\
        -\imag{Z_6} & -\frac{1}{2}\imag{Z_5} &
    \frac{1}{2}\left[Z_3+Z_4-\real{Z_5}\right]+Y_2/v^2 \end{array}\right),
\end{align}
\normalsize
which can be diagonalized with the rotation matrix
\begin{align}
    R = \left( \begin{array}{ccc} c_{12}c_{13} &
        -s_{12}c_{23}-c_{12}s_{13}s_{23} & -c_{12}c_{23}s_{13}+s_{12}s_{23} \\
        s_{12}c_{13} & c_{12}c_{23}-s_{12}s_{13}s_{23} &
        -s_{12}c_{23}s_{13}-c_{12}s_{23} \\
    s_{13} & c_{13}s_{23} & c_{13}c_{23} \end{array}\right),
\end{align}
where $s_{ij}(c_{ij})$ denotes $\sin\theta_{ij}(\cos\theta_{ij})$. From these
angles, one can
construct the Higgs flavor independent quantities 
$q_{kl}$ \cite{Davidson:2005cw}:
\begin{center}
\begin{tabular}{c|cc}
$k$	 & $q_{k1}$	& $q_{k2}$\\
\hline 
1 & $c_{12}c_{13}$ & $-s_{12}-ic_{12}s_{13}$\\
2 & $s_{12}c_{13}$ & $c_{12}-is_{12}s_{13}$\\ 
3 & $s_{13}$ & $ic_{13}$\\
4 & $i$ & $0$\\ 
\end{tabular},
\end{center}
which we will use to parametrize various couplings.
The angle $\theta_{23}$ is however not invariant under a  $U(2)$ Higgs flavor
transformation, but instead obeys
\begin{align}
    e^{i\theta_{23}} \rightarrow (\det U)^{-1}e^{i\theta_{23}}.
\end{align}

%------------------------------------------------------------------------------
\subsection{The Yukawa sector}\label{sec:YukawaSector}
%------------------------------------------------------------------------------

In this work, we do not include any mechanism to provide masses for the
neutrinos.  The Yukawa sector that couples the Higgs fields to the fermion
fields is in the generic basis
\begin{align}
    -\mathcal{L}_Y=&\bar{Q}_L^0\cdot\tilde{\Phi}_{\bar{a}}\eta_a^{U,0}U_R^0+\bar{Q}_L^0\cdot\Phi_a\eta_{\bar{a}}^{D,0\dagger}D_R^0
    + \bar{L}_L^0\cdot\Phi_a\eta_{\bar{a}}^{L,0\dagger}E_R^0 + \text{h.c.}~,
\end{align}
where the left-handed fermion fields in the weak eigenbasis are
\begin{align}
    Q_L^0 \equiv \left( \begin{array}{c} U_L^0 \\ D_L^0 \end{array} \right), &&
L_L^0 \equiv \left( \begin{array}{c} \nu_L^0 \\ E_L^0 \end{array} \right)
\end{align}
and $\tilde{\Phi}\equiv i\sigma_2 \Phi^*$.

In the Higgs basis, the Yukawa sector takes the form
\begin{align}
	-\mathcal{L}_Y =~& \bar{Q}_L \tilde{H}_1 \kappa^U U_R + \bar{Q}_L H_1 \kappa^{D\dagger} D_R + \bar{L}_L H_1 \kappa^{L\dagger} E_R\nn \\
	&+ \bar{Q}_L \tilde{H}_2 \rho^U U_R + \bar{Q}_L H_2 \rho^{D\dagger} D_R + \bar{L}_L H_2 \rho^{L\dagger} E_R + \text{h.c.},
\end{align}
where we have performed a biunitary transformation to go to the fermion mass
eigenbasis such that the $\kappa^F=V_L^F\kappa^{F,0}V_R^{F\dagger}$ matrices are diagonal. 
In the end, the $\kappa^F$ matrices are related to $\eta^F$ by
\begin{align}
	\kappa^U =~& \hat{v}_{\bar{a}}^*\eta_a^U=\frac{\sqrt{2}}{v}\text{diag}(m_u,m_c,m_t),\nn \\
	\kappa^D =~& \hat{v}_{\bar{a}}^*\eta_a^D=\frac{\sqrt{2}}{v}\text{diag}(m_d,m_s,m_b),\nn \\
	\kappa^L =~& \hat{v}_{\bar{a}}^*\eta_a^L=\frac{\sqrt{2}}{v}\text{diag}(m_e,m_\mu,m_\tau),
\end{align}
and $\rho^F=\hat{w}_{\bar{a}}^*\eta_a^F$, where
\begin{align}
	\eta_a^F \equiv& V_L^F\eta_a^{F,0}V_R^{F\dagger}.
\end{align}
The unitarity transformation matrices are defined by
\begin{align}
	F_L \equiv V_L^F F_L^0, && F_R \equiv V_R^F F_R^0,
\end{align}
where $F\in \{U,D,E\}$ denotes each fermion species.  The CKM matrix is composed
out of the left-handed transformation matrices, $V_{CKM} \equiv
V_L^UV_L^{D\dagger}$.

The $\kappa^F$ matrices are, of course, invariant under Higgs flavor
transformations, while $\rho^F$ transforms as
\begin{align}
    \rho^F \rightarrow (\det U)\rho^F.
\end{align}
In general, each $\rho^F$ is left as an arbitrary 3-by-3 complex matrix.

\subsection*{Couplings to mass eigenstates}

We parameterize the couplings of neutral Higgs bosons, $k=1,2,3$, to fermions as
\begin{align}
    -\mathcal{L} = \bar{F} \left(c_k^F + \tilde{c}_{k}^Fi\gamma_5\right)F h_k,
\end{align}
where $F$ corresponds to $U$, $D$ and $L$, which are the Dirac fermions as
vectors in generation space.  These couplings can be expressed in a
basis-independent way as
\begin{align}
    c_k^U =~& \frac{1}{\sqrt{2}}\left[\kappa^U q_{k1} +
    \frac{1}{2}( q_{k2}^*e^{i\theta_{23}}\rho^U 
+ q_{k2}e^{-i\theta_{23}}\rho^{U\dagger}) \right],\\
\tilde{c}_k^U =~& \frac{i}{2\sqrt{2}}\left( q_{k2}e^{-i\theta_{23}}\rho^{U\dagger}
 - q_{k2}^* e^{i\theta_{23}} \rho^U \right),\\
 c_k^D =~& \frac{1}{\sqrt{2}}\left[\kappa^D q_{k1} + \frac{1}{2}( 
 q_{k2}^*e^{i\theta_{23}}\rho^D +q_{k2}e^{-i\theta_{23}}\rho^{D\dagger})\right],\\
 \tilde{c}_k^D =~&
 \frac{i}{2\sqrt{2}}\left(q_{k2}^*e^{i\theta_{23}}\rho^{D} -
 q_{k2}e^{-i\theta_{23}}\rho^{D\dagger}\right),\\
 c_k^L =~& \frac{1}{\sqrt{2}}\left[\kappa^L q_{k1} + \frac{1}{2}( 
 q_{k2}^*e^{i\theta_{23}}\rho^L +q_{k2}e^{-i\theta_{23}}\rho^{L\dagger})\right],\\
 \tilde{c}_k^L =~&
 \frac{i}{2\sqrt{2}}\left(q_{k2}^*e^{i\theta_{23}}\rho^{L} -
 q_{k2}e^{-i\theta_{23}}\rho^{L\dagger}\right),
\end{align}

The couplings of charged Higgs to fermions is of the form
\begin{align}
    -\mathcal{L} = \bar{U}\left(c_{H^+}^Q + \tilde{c}_{H^+}^Q i \gamma_5\right)D H^+
                    + \bar{\nu}\left(c_{H^+}^L + \tilde{c}_{H^+}^L i \gamma_5 \right) E H^+ + \text{h.c.},
\end{align}
where
\begin{align}
	c_{H^+}^Q =~& \frac{1}{2}\left(V_{CKM}\rho^{D\dagger} - \rho^{U\dagger}V_{CKM}\right),\\
 	\tilde{c}_{H^+}^Q =~& -\frac{i}{2} \left(V_{CKM}\rho^{D\dagger} + \rho^{U\dagger} V_{CKM}\right),\\
	c_{H^+}^L =~& \frac{1}{2}\rho^{L\dagger},\\
 	\tilde{c}_{H^+}^L =~& -\frac{i}{2}\rho^{L\dagger}.
\end{align}

The three scalar coupling of neutral to charged Higgs, is parameterized as
\begin{align}
	\mathcal{L} = - \lambda_{kH^\pm} v h_k H^+H^-,
\end{align}
with
\begin{align}
	\lambda_{kH^\pm} = q_{k1}Z_3 + \text{Re}\left(q_{k2}e^{-i\theta_{23}}Z_7\right).
\end{align}

Finally, we write the coupling of neutral Higgs to vector bosons as
\begin{align}
	\mathcal{L} = g_{kVV} h_k\left(\frac{2m_W^2}{v}W_\mu W^\mu 
					+	\frac{m_Z^2}{v}Z_\mu Z^\mu\right),
\end{align}
where $g_{kVV} = q_{k1}$.

\subsection*{Flavor-changing-neutral currents and \Zsym~symmetry}

Since the $\rho^F$ matrices are in general completely arbitrary, the 2HDM
suffers from FCNCs at tree-level.  The most popular solution is to impose a
\Zsym symmetry on the 2HDM \cite{Glashow:1976nt, Paschos:1976ay}.  By making one
Higgs odd and the other even under the \Zsym symmetry, there are four different
choices of charge assignments of the fermions as listed in \Tab{Z2symmetries}.
With such a symmetry, the $\rho^F$ matrices become proportional to the diagonal
$\kappa^F$ matrices; hence solving the problem of having tree-level FCNCs.  

\begin{table}[h!]
		\centering
    		\begin{tabular}{|c|cccccc|}\hline
		Type	 & $U_R$	& $D_R$	& $L_R$ & $a^U$ & $a^D$ & $a^L$\\
		\hline 
		I & + & + & + &$\cot\beta$ & $\cot\beta$ & $\cot\beta$\\
		II & + & $-$ & $-$ &$\cot\beta$ & $-\tan\beta$ & $-\tan\beta$\\
		Y & + & $-$ & + &$\cot\beta$ & $-\tan\beta$ & $\cot\beta$\\
		X & + & + & $-$ &$\cot\beta$ & $\cot\beta$ & $-\tan\beta$\\ \hline
		\end{tabular}
        \caption{Different \Zsym symmetries that can be imposed on the 2HDM.
        $\Phi_1$ is odd($-1$) and $\Phi_2$ is even($+1$). For every type of
    \Zsym symmetry, the $\rho^F$ matrices become proportional to the diagonal
mass matrices, $\rho^F = a_F \kappa^F$. }
		\label{tab:Z2symmetries}
\end{table}

One can also make the ansatz of having an aligned Yukawa sector by itself
\cite{Pich:2009sp}. Then one has
\begin{align}\label{eq:alignedYuk}
    \rho^F = a_F \kappa^F,
\end{align}
with $a_F$ being completely arbitrary complex coefficients. It is well
known that this alignment ansatz is not stable during RG evolution
\cite{Jung:2010ik,Ferreira:2010xe,Botella:2015yfa,Oredsson:2018yho}, but at one
particular energy scale it results in diagonal Yukawa couplings.
Though, if one allows for complex $a_F$ coefficients, one runs into the trouble
of inducing a large eEDM \cite{BowserChao:1997bb}; as we will show in more
detail later. 

Another solution is the Cheng-Sher ansatz \cite{PhysRevD.35.3484}, where one
parameterizes the $\rho^F$ Yukawa couplings as
\begin{align}
    \rho^F \equiv \lambda_{ij}^F \frac{\sqrt{2m_im_j}}{v}.
\end{align}
This allows for mass suppressed FCNCs when the $\lambda_{ij}^F$ are of the same
magnitude. Neutral meson oscillations sets a rough upper limit of
$\lambda_{i\neq j}^F \lesssim 0.1$ \cite{Bijnens:2011gd}. We will use this
parameterization when looking at the sizes of non-diagonal Yukawa couplings,
since it gives a clear estimate of how large they are.

%%%%%%%%%%%%%%%%%%%%%%%%%%%%%%%%%%%%%%%%%%%%%%%%%%%%%%%%%%%%%%%%%%%%%%%%%%%%%%%%
\section{\CP~violation scenarios}\label{sec:scenarios}
%%%%%%%%%%%%%%%%%%%%%%%%%%%%%%%%%%%%%%%%%%%%%%%%%%%%%%%%%%%%%%%%%%%%%%%%%%%%%%%%

The scalar potential and vacuum are \CP-conserving if and only if
\cite{Davidson:2005cw,Gunion:2005ja,Lavoura:1994fv,Botella:1994cs} 
\begin{align}\label{eq:CPVparams}
	\imag{Z_5^* Z_6^2} = \imag{Z_5^* Z_7^2} = \imag{Z_6^*Z_7} = 0.
\end{align}
These quantities are of course base invariant. Furthermore, we will use these as
a measure of the amount of \CP~violation in the scalar sector.

To get a quantitative estimate on the amount of \CP~violation that is allowed in
the 2HDM and see how it affects the RG evolution, we set up parameter scans for
a number of physical scenarios with different levels of \Zsym symmetry.  In this
work, we only investigate bottom-up RG running; in that we impose different
starting conditions at the EW scale and then run up. One could also consider
scenarios where the starting conditions are fixed at some high energy UV scale
and one instead run down to the EW scale. In a way, that would seem more
natural, since a more symmetric model at the UV scale might be a more realistic
scenario. However, we assume the RG effects to be symmetrical, \textit{e.g.} the
symmetry breaking parameters spread in equal amounts in bottom-up and top-down
running. This has also been checked in some of the cases below. To perform
top-down running is computationally more expensive; since one has to fit the
evolved parameters to physical observables.  This is the reason why we limit
ourselves to bottom-up scenarios to investigate the RG effects.

Imposing an exact \Zsym symmetry fixes the Yukawa structure and forbids the
$m_{12}^2$ and $\lambda_{6,7}$ parameters in the scalar potential.  With these
being forbidden, the only potentially complex parameter is $\lambda_5$; which
can be rendered real by a Higgs flavor transformation.  Hence, the strict \Zsym
symmetric 2HDM does not allow for any explicit \CP~violation.  One can, however,
allow for a softly \Zsym breaking non-zero $m_{12}^2$ term. Then, one cannot
rotate away all the complex phases and hence we will only investigate scenarios
with at least a softly broken \Zsym symmetry.  We will also require aligned
VEVs, without loss of generality, by setting $\xi=0$ and we use one of the
tadpole equations to fix the phase of $m_{12}^2$.

When performing RG evolution of a complex 2HDM, an interesting question is how
the phases spread during the evolution.  By inspection of the 2-loop
RGEs\footnote{These can be found in \code{C++} form in the source code of
\code{2HDME} \cite{Oredsson:2018vio}. Since they are very lengthy, we do not
show them in this article.}, one finds that there is no parameter that depends
on the phase of $\lambda_5$ in the softly broken \Zsym symmetry case.  One needs
a hard \Zsym breaking in either the Yukawa or scalar sector to allow for
parameters being rendered complex during the RG running.  

We construct the following scenarios for investigation:
%------------------------------------------------------------------------------
\subsection*{Scenario I: softly broken \Zsym symmetry}
%------------------------------------------------------------------------------

The simplest scenario is the softly broken \Zsym symmetric 2HDM.  This is also
the most studied 2HDM.  Here, we scan over the free parameters in the scalar
potential. This is done in the generic basis with a flat random distribution.
The Yukawa sector is fixed to type I or type II.

We will also restrict ourselves to scenarios where all $|\lambda_i|\lesssim 2$.
The opposite case with large scalar couplings is often problematic in that it
exhibits large radiative corrections and many tree-level calculations cannot be
trusted \cite{Braathen:2017jvs,Oredsson:2018yho,Kainulainen:2019kyp}.  In the
\CP~conserving case, this is related to the decoupling limit
\cite{Haber:1989xc,Gunion:2002zf}, where the lightest Higgs boson resembles the
125 GeV SM one and the others are heavier.

The parameter ranges are:
\begin{align}
    \lambda_i \in [-2,2], && |m_{12}^2| \in [10^2, 2\times 10^5], && \beta \in
    [\text{atan}(0.5), \text{atan}(50)],
\end{align}
where $\lambda_5$ has a random phase and the phase of $m_{12}^2$ is fixed from
one of the tadpole equations.

%------------------------------------------------------------------------------
\subsection*{Scenario II: hard \Zsym symmetry breaking in the scalar potential}
%------------------------------------------------------------------------------

This is the same as scenario I, but with an addition of small hard \Zsym symmetry
breaking complex parameters, $\lambda_{6,7}$, in the scalar potential at the
electroweak scale. We restrict these to be in the range $|\lambda_{6,7}|< 0.5$,
with random phases. 

%------------------------------------------------------------------------------
\subsection*{Scenario III: hard \Zsym symmetry breaking in the Yukawa sector}
%------------------------------------------------------------------------------

Here, we have a \CP~conserved scalar potential with a softly broken \Zsym
symmetry. The parameters of the potential are distributed as in scenario I, but
all are real. The Yukawa sector will be aligned as in \eq{alignedYuk}
at the EW scale. The $a_F$ parameters are equal to the \Zsym symmetric values in
magnitude; however, we let them be complex with independent phases. We will
investigate type I, II and X as listed in \Tab{Z2symmetries}.

This can be seen as a hard \Zsym symmetry breaking in the Yukawa sector and
consequently $\lambda_{6,7}$ and non-diagonal Yukawa couplings will be generated
in the RG running. 

%%%%%%%%%%%%%%%%%%%%%%%%%%%%%%%%%%%%%%%%%%%%%%%%%%%%%%%%%%%%%%%%%%%%%%%%%%%%%%%%
\section{Constraints}\label{sec:constraints}
%%%%%%%%%%%%%%%%%%%%%%%%%%%%%%%%%%%%%%%%%%%%%%%%%%%%%%%%%%%%%%%%%%%%%%%%%%%%%%%%

There is a considerable amount of freedom when choosing the parameters of the
2HDM.  To constrain the parameter space we will use a number of theoretical and
experimental constraints.

It is well known that the mass of $m_{H^\pm}$ gets a lower bound from 
weak radiative $B$-meson decays \cite{Deschamps:2009rh, Mahmoudi:2009zx,
Hermann:2012fc, Misiak:2015xwa, Misiak:2017bgg}, \textit{e.g.} from
$b\rightarrow s \gamma$ interactions. With a type I Yukawa symmetry, the bound 
is heavily $\tan\beta$ dependent and becomes irrelevant for us when
 $\tan\beta \gtrsim 2$. For a type II 2HDM the 
bound is largely $\tan\beta$ independent; with a conservative lower value of
580 GeV \cite{Misiak:2017bgg}. We will, however, not impose these constraints 
for the charged boson mass in this paper.

\subsection*{Consistency}

On the theoretical side, we make basic checks to ensure tree-level stability of
the scalar potential \cite{Ivanov:2006yq,Ivanov:2007de} and that the VEV is in a
global minimum \cite{Ivanov:2015nea}.  We also check the unitarity of the
scattering matrix for scalar particles at high energies \cite{Ginzburg:2005dt}.
These tests are implemented in \code{2HDME} \cite{Oredsson:2018vio}.

\subsection*{Collider data}

The first check of each parameter point is that the lightest Higgs scalar falls
in the range $m_{h_1} \in [120, 130]$ GeV.

To check whether a parameter point is allowed by the current collider data from
LEP, the Tevatron and LHC, we make use of the codes \code{HiggsBounds}
\cite{Bechtle:2008jh,Bechtle:2011sb,Bechtle:2013wla} and
\code{HiggsSignals} \cite{Bechtle:2013xfa}.  \code{HiggsBounds} excludes models at a 95 \% confidence
level by comparing to experimental cross section limits and \code{HiggsSignals}
ensures that the 125 GeV Higgs boson in the model resembles the one observed at
the LHC.  These codes require the calculations of the decay rates for each
scalar particle, which we compute with \code{2HDMC}
\cite{Eriksson:2009ws}\footnote{Although a modified version that generalizes the
original \code{2HDMC} code to the complex scenario with \CP~violation.}.

\subsection*{Precision measurements}

The oblique electroweak corrections to precision measurements, involving the $W$
and $Z$ bosons, are tightly constrained and simultaneously sensitive to
additional scalar particles.  We calculate the parameters $S$, $T$ and $U$
\cite{Peskin:1990zt,Peskin:1991sw} using the formulas for the 2HDM in
\mycite{Haber:2010bw} and make sure they are within the allowed 68 \% confidence
region of \mycite{Baak:2014ora}.

\subsection*{Electric dipole moment of the electron} 

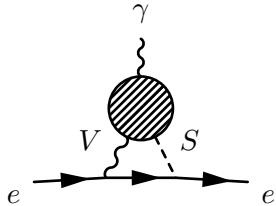
\begin{figure}[h!]                                                               
  \begin{align*}                                                                 
    \vcenter{\hbox{                                                              
      \begin{fmffile}{images/barrZee}                                     
        \begin{fmfgraph*}(80,60)                                                 
        \fmfleft{i1}                                                             
        \fmfright{o1}                                                            
        \fmfbottom{b1,b2}                                                        
        \fmftop{t1}                                                              
        \fmf{fermion,tension=10}{b1,v1,v2,b2}                                    
        \fmf{photon,tension=0.6,label=$V$}{m1,v1}                         
		\fmfblob{0.3w}{m1}
        \fmf{photon,tension=1.2}{m1,t1}                                          
        \fmf{dashes,tension=0.6,label=$S$}{m1,v2}                              
        \fmflabel{$\gamma$}{t1}                                                  
        \fmflabel{$e$}{b1}                                                       
        \fmflabel{$e$}{b2}                                                       
        \end{fmfgraph*}                                                          
      \end{fmffile}                                                              
  }}                                                                             
\end{align*}                                                                     
\caption{The structure of Barr-Zee diagrams, where $V =\{\gamma,Z,W^\pm\}$ and
$S=\{h_i, H^\pm\}$. Each diagram is denoted by $(d_e)^{VS}_l$, where $l$ is the
    loop particle in the blob.}
\label{fig:BarrZee}                                                              
\end{figure}

There is currently no direct evidence of an EDM for a fundamental particle and
an observation would indicate a violation of \CP.  It is a very difficult task
to perform an experiment to measure the EDM of a charged particle.  The electron
serves as the easiest particle to try to measure the EDM of and the current
limit in \eq{ACMEII} are set by the ACMEII collaboration \cite{Andreev:2018ayy} 
using a system of ThO molecules.  Even though the SM predicts a non-zero eEDM,
it is many orders of magnitude below the current limit.  New scalar particles
and sources of \CP~violating phases in the 2HDM can quite easily generate EDMs of the
order of $10^{-30}$ to $10^{-26}$ e cm; making a check vital for the
survivability of any model.

There have been many studies of EDMs in the 2HDM
\cite{Shu:2013uua, Jung:2013hka,Cheung:2014oaa, Ipek:2013iba, Bian:2014zka,
Chen:2015gaa,Egana-Ugrinovic:2018fpy}.
It is a well known phenomenon, that while in general there can be a large contribution to
the eEDM for a single Higgs boson, there are regions in parameter
space which exhibit cancellations among all the contributions; thus, making
regions with large \CP~violating phases allowed. Therefore we take into
consideration all the Higgs bosons in the calculation of the EDM.  The largest
contributions, and the only ones relevant for this study, are the 2-loop
Barr-Zee diagrams \cite{Barr:1990vd} illustrated in \fig{BarrZee}. We have
collected the necessary formulas and details of the calculation in
\app{BarrZee}. A numerical implementation of the computation is also available
in \code{2HDME}.

\subsection*{Renormalization group evolution}

By evolving the 2HDM in energy, we investigate the energy range where the model
is valid and thus probe the stability of the model.  If the model is not
complete and consequently breaks down in the evolution, it would signal the
need for new physics at a higher energy scale. Sensitivity to starting
conditions is also an indication of fine tuning in choosing the parameters.

The RG evolution is performed at 2-loop order using \code{2HDME}.  For technical
details, we refer to \mycites{Oredsson:2018vio, Oredsson:2018yho}.

In the RG evolution, we look for a breakdown of tree-level stability and
unitarity as mentioned above.  We also check for the presence of Landau poles
where a parameter of the model goes to infinity; which we will refer to as a
violation of perturbativity.  This is most effectively imposed as a limit of
$|\lambda_i|<4\pi$. It should, however, not be interpreted as an exact perturbativity limit, but simply a
numerical cut-off; evolving beyond this limit is more computationally demanding
and yield no additional information.  The true Landau pole will lie at a
slightly higher energy scale.

%%%%%%%%%%%%%%%%%%%%%%%%%%%%%%%%%%%%%%%%%%%%%%%%%%%%%%%%%%%%%%%%%%%%%%%%%%%%%%%%
\section{Example of phase dependence}\label{sec:examplePoint}
%%%%%%%%%%%%%%%%%%%%%%%%%%%%%%%%%%%%%%%%%%%%%%%%%%%%%%%%%%%%%%%%%%%%%%%%%%%%%%%%

\begin{figure}[h!]
\begin{center}
\begin{tabular}{cc}
%\hspace{-1cm}
\includegraphics[trim=0cm 0.5cm 0.5cm 0cm,clip,height=0.3\textwidth]{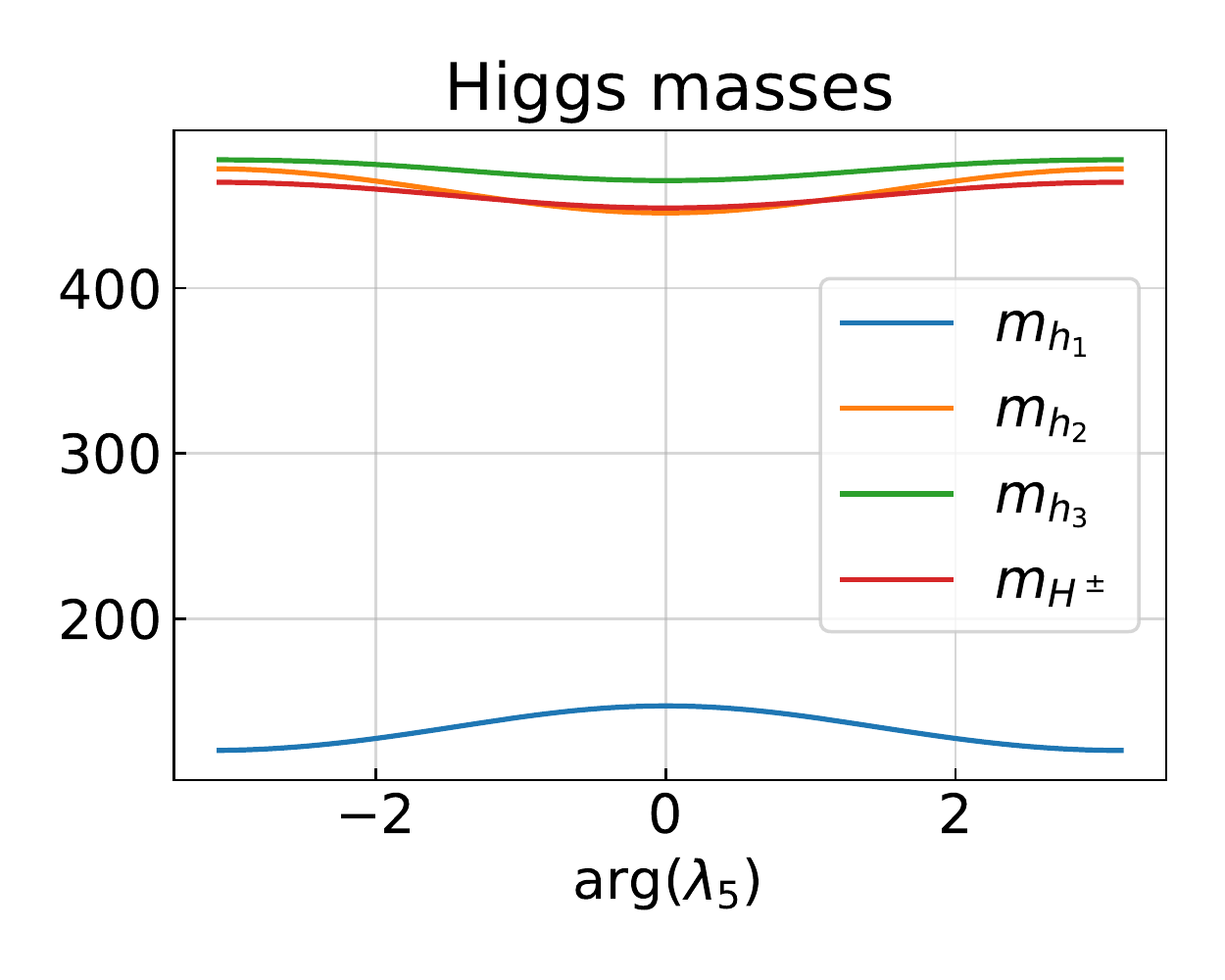} &
\includegraphics[trim=0cm 0.5cm 0.5cm 0cm,clip,height=0.3\textwidth]{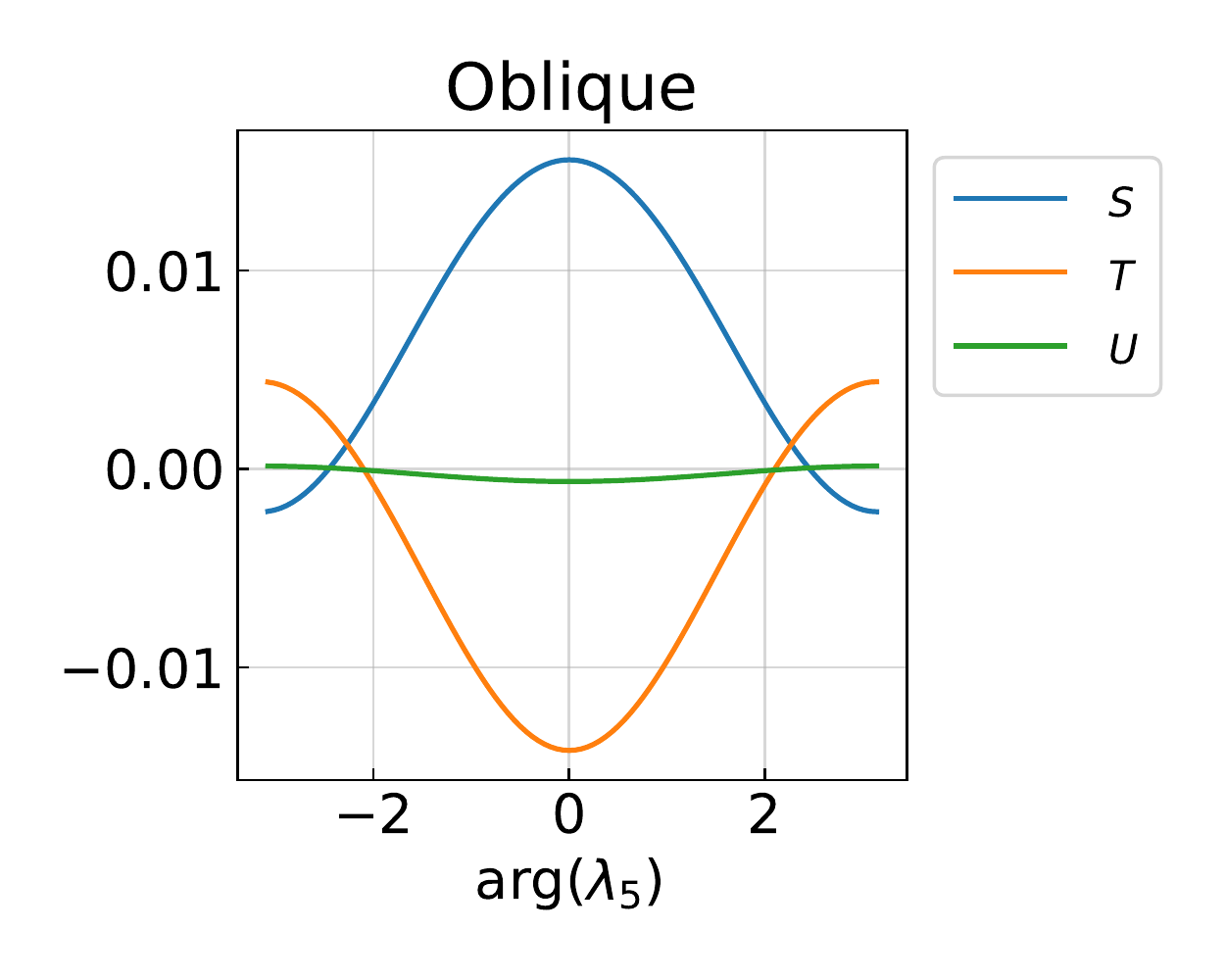} 
\end{tabular}
\caption{Higgs boson masses(left) and oblique parameters (right) for the
parameter point in \eq{genEx} as a function of the phase of $\lambda_5$.}
\label{fig:1Dphase5_2}
\end{center}
\end{figure}

Allowing for \CP~violation can induce effects that are otherwise absent in a
\CP~conserving 2HDM.  The softly broken \Zsym symmetric 2HDM contains only one
phase in its scalar potential.  By a Higgs flavor transformation, one can
therefore fix all parameters to be real except for $\lambda_5$.  Here, we show
the phase dependence of different quantities by varying arg$(\lambda_5)$ in a
softly broken \Zsym symmetric 2HDM of type I, using
the fixed values
\begin{align}\label{eq:genEx}
  \tan\beta =~&  2.5, &&& M_{12}^2 =~&   73~000\text{ GeV}^2,\nn \\
  \lambda_1 =~&  0.47, &&& \lambda_2 =~&  0.40,\nn \\
  \lambda_3 =~&  -0.17, &&& \lambda_4 =~& 0.16,\nn \\
  |\lambda_5| =~&  0.25, &&& \lambda_6 =~& \lambda_7 = 0.
\end{align}
The Higgs boson masses are dependent on the phase of $\lambda_5$ as can be seen
in the left plot in \fig{1Dphase5_2}.

\begin{figure}[h!]
\begin{center}
\includegraphics[trim=0cm 0.5cm 0.5cm 0cm,clip,height=0.45\textwidth]{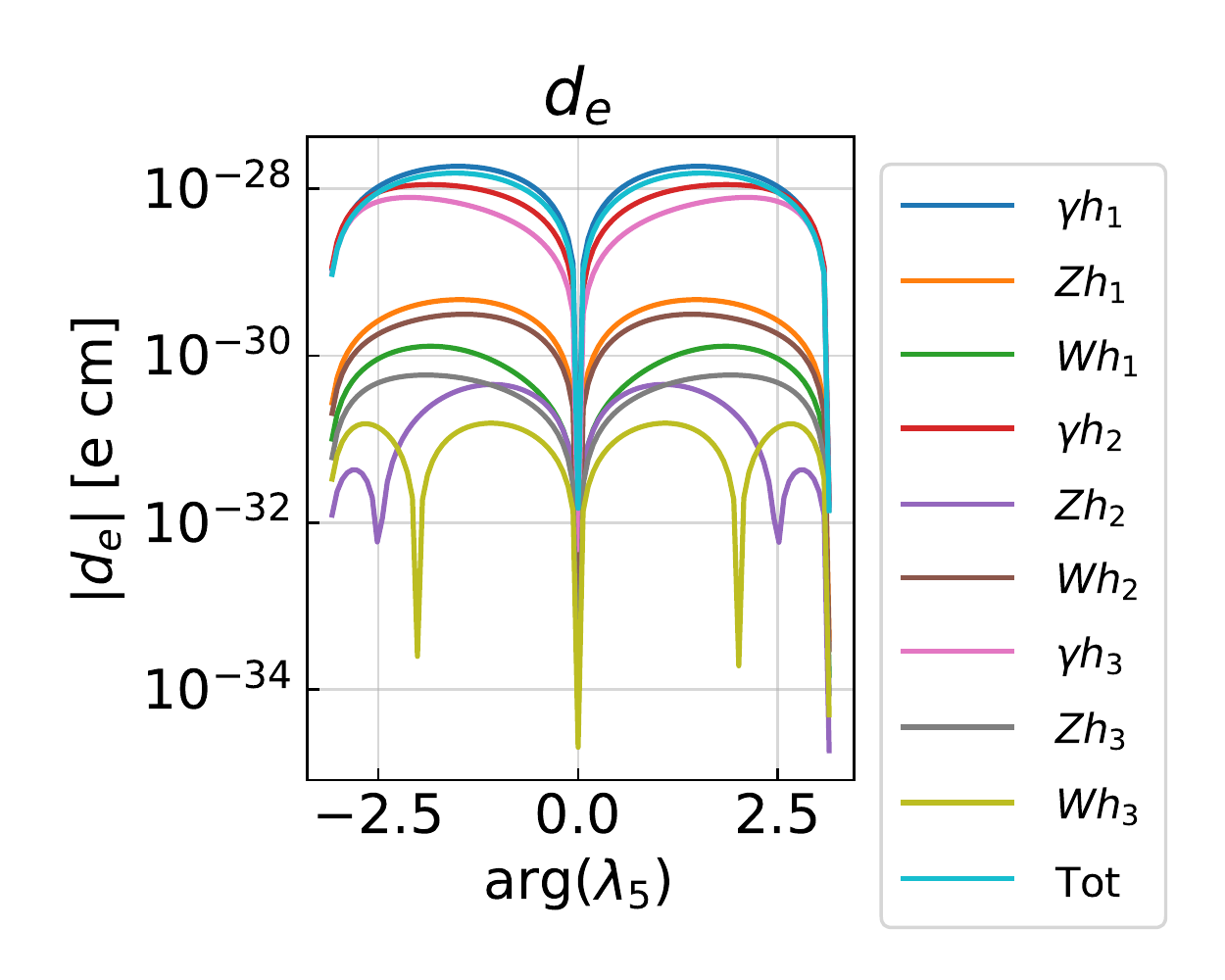}
\caption{The eEDM as a function of arg$\lambda_5$ for the parameter point in
\eq{genEx}. The different contributions, $VS$, refers to the individual Barr-Zee
diagrams, as in \fig{BarrZee}, summed over all different loop particles.}
\label{fig:1Dphase5_1}
\end{center}
\end{figure}

Varying arg$\lambda_5$ has a large effect on the oblique parameters $S$, $T$
and $U$, as can be seen in the right plot in \fig{1Dphase5_2}. In
\fig{1Dphase5_1}, it is shown how  a non-zero phase quickly induces a large
EDM, $d_e$, for the electron.  There can be non-trivial cancellations among the
many contributions to $d_e$; with the lightest Higgs boson usually dominating.
Including the $(d_e)^{\gamma h_{2,3}}$ diagram is however important since it is
at the same order of magnitude.

\begin{figure}[h!]
\begin{center}
\begin{tabular}{cc}
\hspace{-1cm}
\includegraphics[trim=0cm 0.5cm 0.5cm 0cm,clip,height=0.35\textwidth]{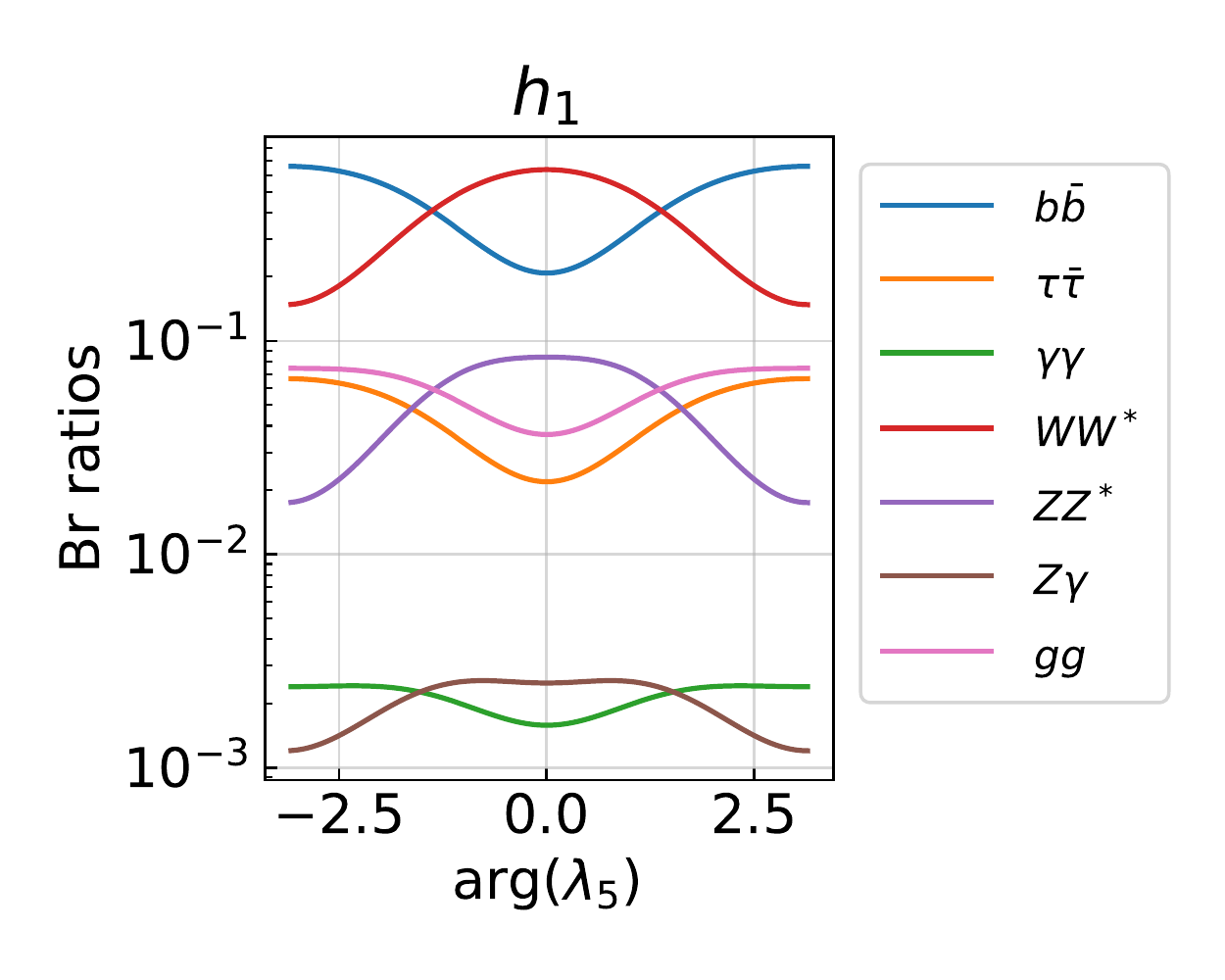} &
\includegraphics[trim=0cm 0.5cm 0.5cm 0cm,clip,height=0.35\textwidth]{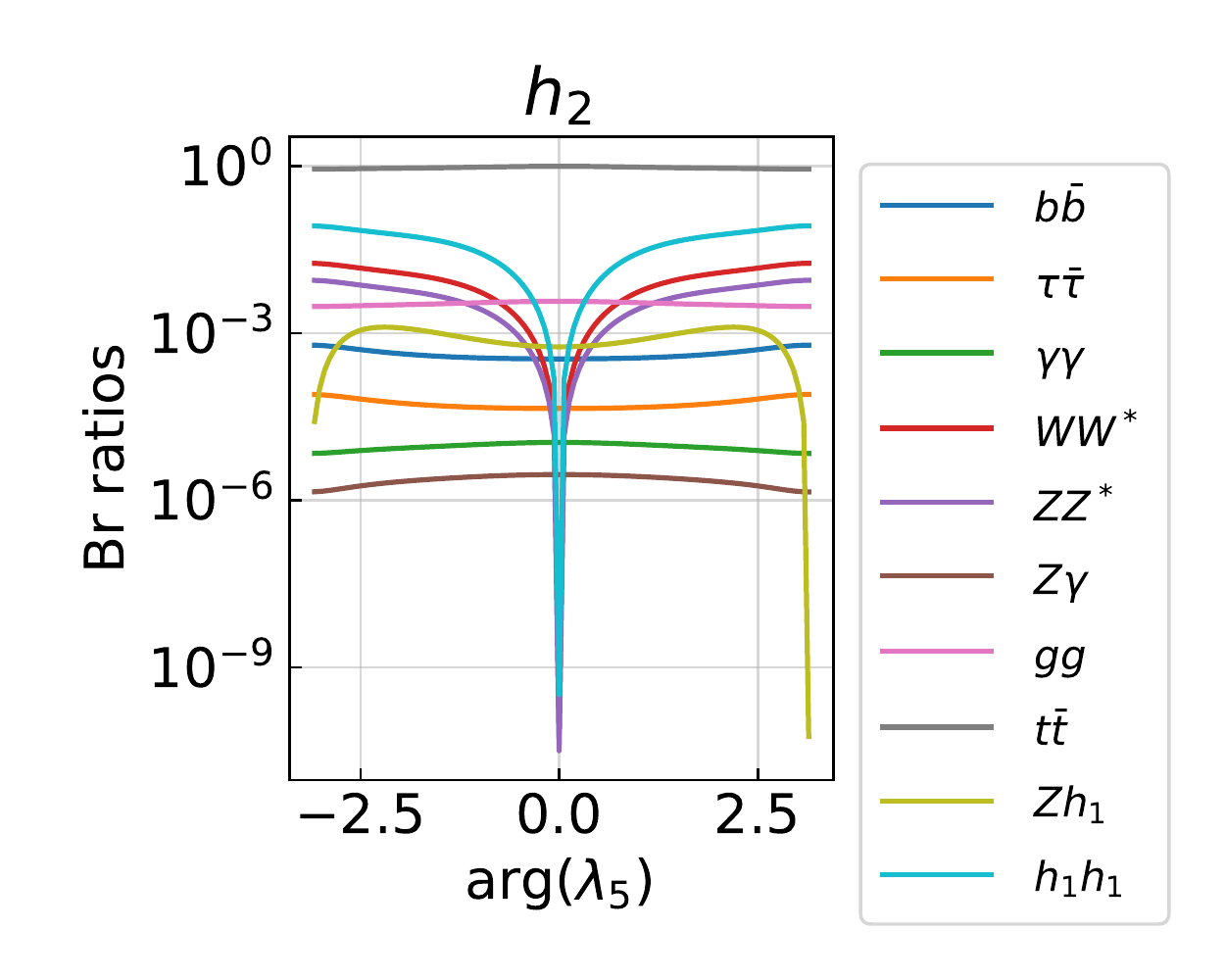} \\
\includegraphics[trim=0cm 0cm 0.5cm 0cm,clip,height=0.35\textwidth]{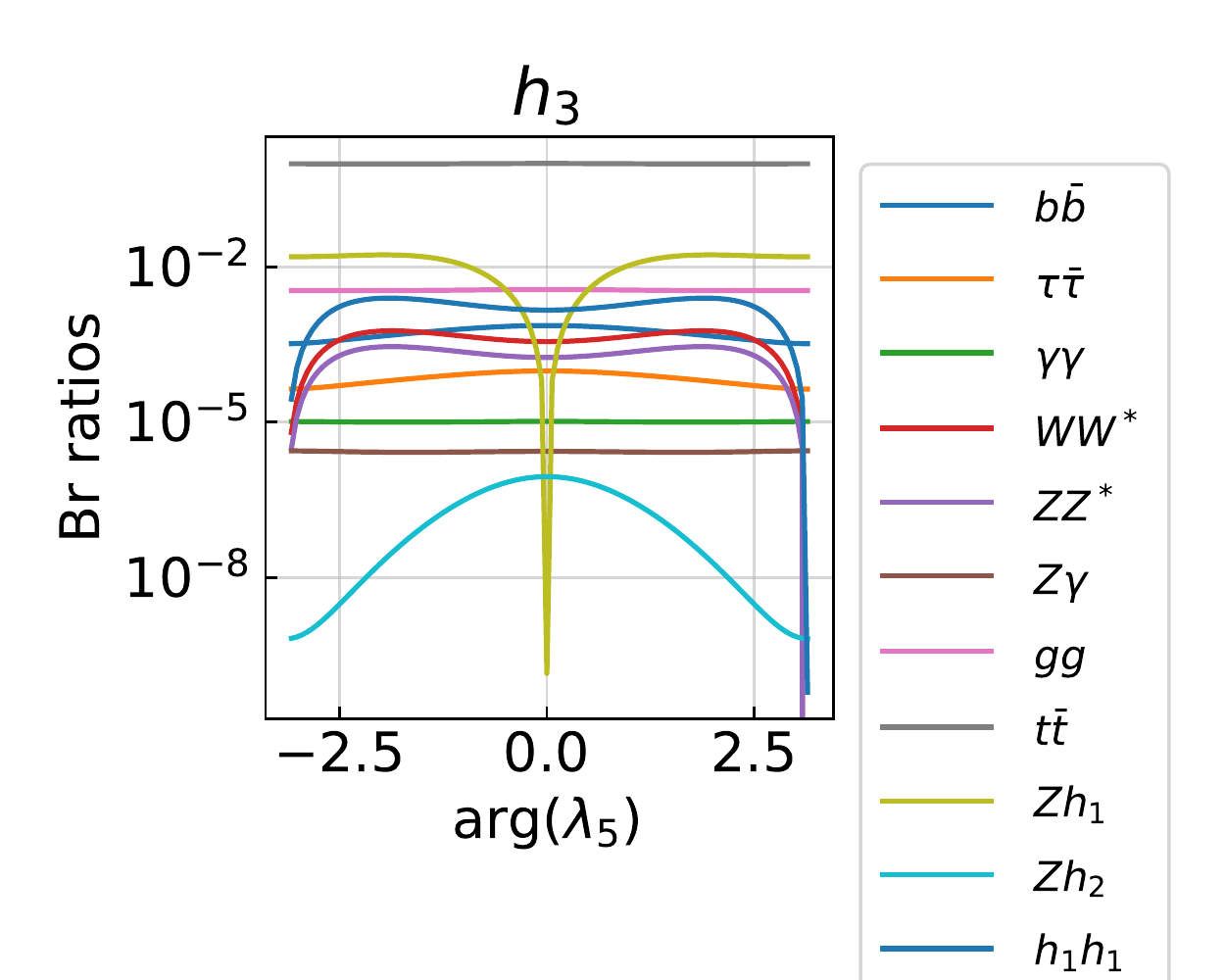} &
\includegraphics[trim=0cm 0cm 0.5cm 0cm,clip,height=0.35\textwidth]{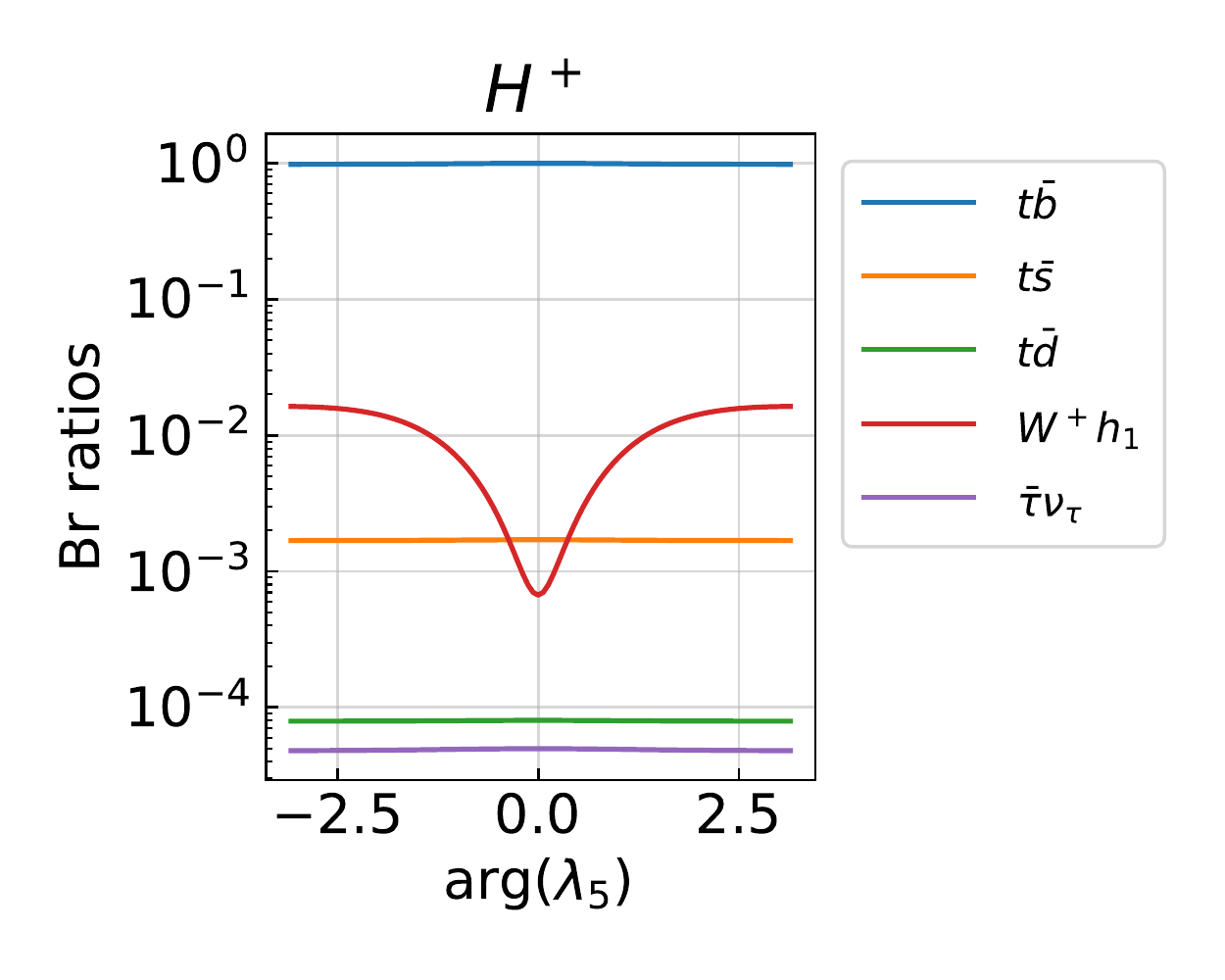}
\end{tabular}
\caption{Branching ratios for the most important decays of the Higgs particles.}
\label{fig:1Dphase5_3}
\end{center}
\end{figure}

In \fig{1Dphase5_3} we show the branching ratios of the Higgs bosons as a
function of arg$\lambda_5$.  There are a number of new possible decays opening
up when going to a \CP~violating 2HDM since the neutral Higgs bosons all mix
together.  For example, one can have $h_2$ and $h_3$ simultaneously decaying
into $Zh_1$ as well as to $h_1h_1$. 

%%%%%%%%%%%%%%%%%%%%%%%%%%%%%%%%%%%%%%%%%%%%%%%%%%%%%%%%%%%%%%%%%%%%%%%%%%%%%%%%
\section{Results}\label{sec:results}
%%%%%%%%%%%%%%%%%%%%%%%%%%%%%%%%%%%%%%%%%%%%%%%%%%%%%%%%%%%%%%%%%%%%%%%%%%%%%%%%

In the parameter scan, we perform the RG evolution from the top mass scale
until the evolution breaks down. We define the energy scale $\Lambda$ as
the breakdown energy scale; meaning the energy where either perturbativity,
stability or unitarity is violated.

If nothing else if mentioned, the figures presented below are constructed from
the parameter points that are allowed by \code{HiggsBounds} and
\code{HiggsSignals} as well as within the limits of $S$, $T$ and $U$. 

\begin{table}[h!]
\centering
    \begin{tabular}{|c|cccccc|}\hline
        Scenario	 & \Zsym	& Pass HB & Pass HS &  Pass ST & Pass eEDM & Pass
        all\\
\hline 
        I & type I & 43\% & 9\% & 80\%  & 9\% & 1.3\%\\
        I & type II & 39\% & 7\%  & 80\% & 6\% & 0.5\% \\
        II & type I & 38\% & 8\% & 74\%  & 5\% & 0.7\%\\
        II & type II & 36\% & 6\%  & 74\% & 2\% & 0.2\% \\
        III & type I & 44\% & 9\% & 80\%  & 3\% & 0.4\%\\
        III & type II & 44\% & 8\% & 79\%  & 1\% & 0.01\%\\
        III & type X & 43\% & 8\% & 79\%  & 1\% & 0.01\%\\
        \hline
\end{tabular}
\caption{Statistics of the parameter scans of scenario I-III. HB (HS) refers to
\code{HiggsBounds} (\code{HiggSignals}). There is a total of 50~000 points in
each scenario. The \Zsym symmetry for scenario III sets the magnitude for the
complex $a_F$ coefficients.}
\label{tab:stats}
\end{table}

The fraction of points that survives the different constraints at the starting
scale of the parameter scans is shown in \Tab{stats} for all the scenarios. In
\Tab{eEDMStats}, we also list the three largest contributions to the eEDM.

\begin{table}[h!]
\centering
    \begin{tabular}{|c|cccc|}\hline
        Scenario	 & \Zsym	& 1st & 2nd &  3rd \\
\hline 
        I & type I & $\gamma h_1 (W)$: 85 \%  & $\gamma h_2 (W)$: 10 \% & $\gamma h_2 (t)$: 3 \% \\
        I & type II & $\gamma h_1 (W)$: 60 \%  & $\gamma h_1 (t)$: 24 \% & $\gamma h_2 (t)$: 5 \% \\
        II & type I & $\gamma h_1 (W)$: 79 \%  & $\gamma h_2 (W)$: 12 \% & $\gamma h_2 (t)$: 5 \% \\
        II & type II & $\gamma h_1 (W)$: 58 \%  & $\gamma h_1 (t)$: 20 \% & $\gamma h_2 (t)$: 11 \% \\
        III & type I & $\gamma h_2 (t)$: 31 \%  & $\gamma h_3 (t)$: 22 \% & $\gamma h_1 (W)$: 19 \% \\
        III & type II & $\gamma h_2 (t)$: 31 \%  & $\gamma h_1 (W)$: 25 \% &
        $\gamma h_3 (t)$: 22 \% \\
        III & type X & $\gamma h_2 (t)$: 31 \%  & $\gamma h_2 (W)$: 25 \% &
        $\gamma h_3 (t)$: 23 \% \\
        \hline
\end{tabular}
\caption{Statistics of the top three largest contributions to the eEDM for all
50~000 parameter points in each scenario. The notation is according to
\fig{BarrZee}, \textit{i.e.} $VS$(loop particle).}
\label{tab:eEDMStats}
\end{table}
%------------------------------------------------------------------------------
\subsection{Scenario I}
%------------------------------------------------------------------------------

\begin{figure}[h!]
\begin{center}
\textbf{Scenario I}\\
\begin{tabular}{cc}
    type I & type II \\
\includegraphics[trim=1cm 1.5cm 4.5cm 0.5cm, clip, height=0.3\textwidth]
    {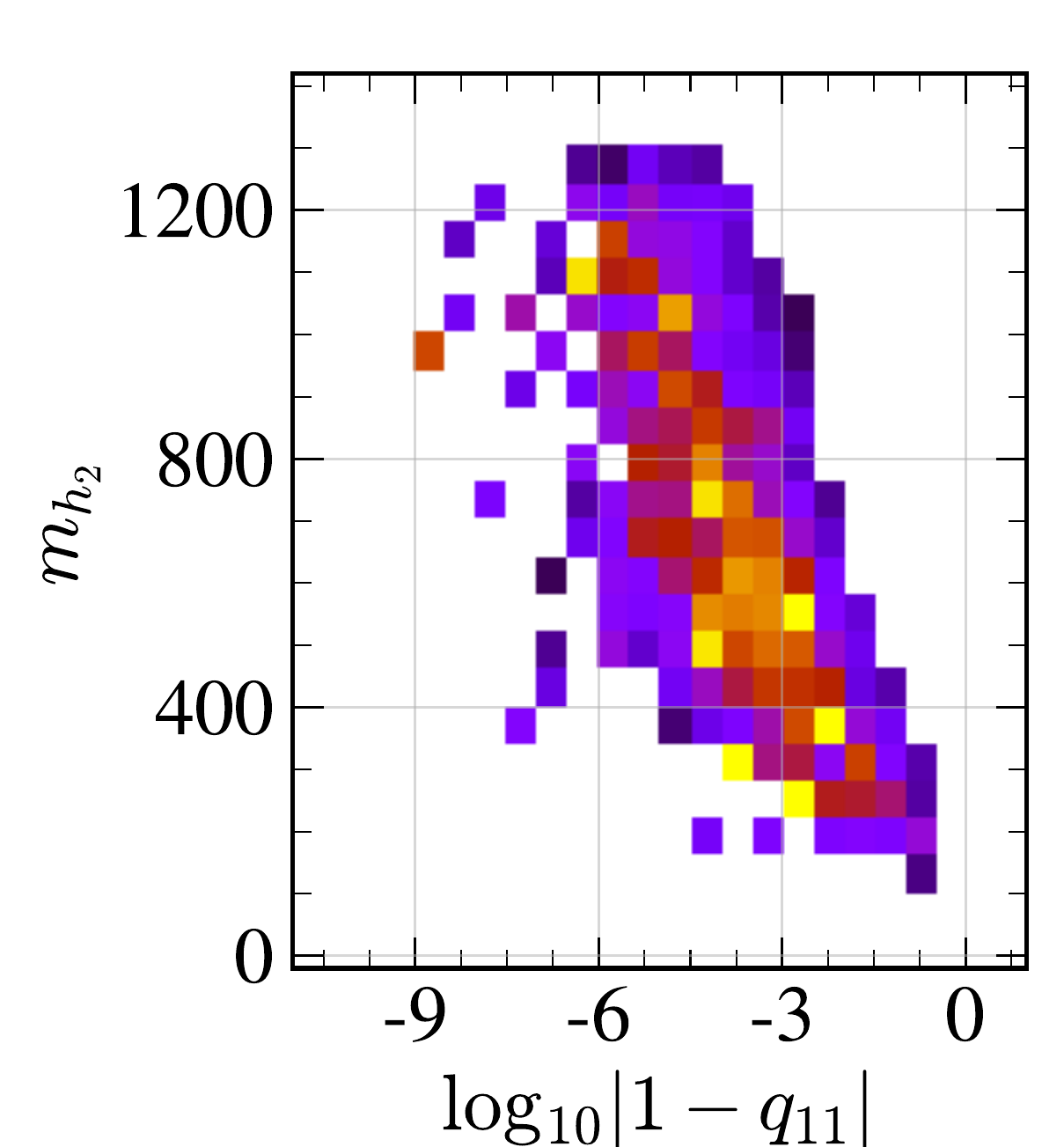} &
\includegraphics[trim=1cm 1.5cm 0.5cm 0.5cm,clip,height=0.3\textwidth]
    {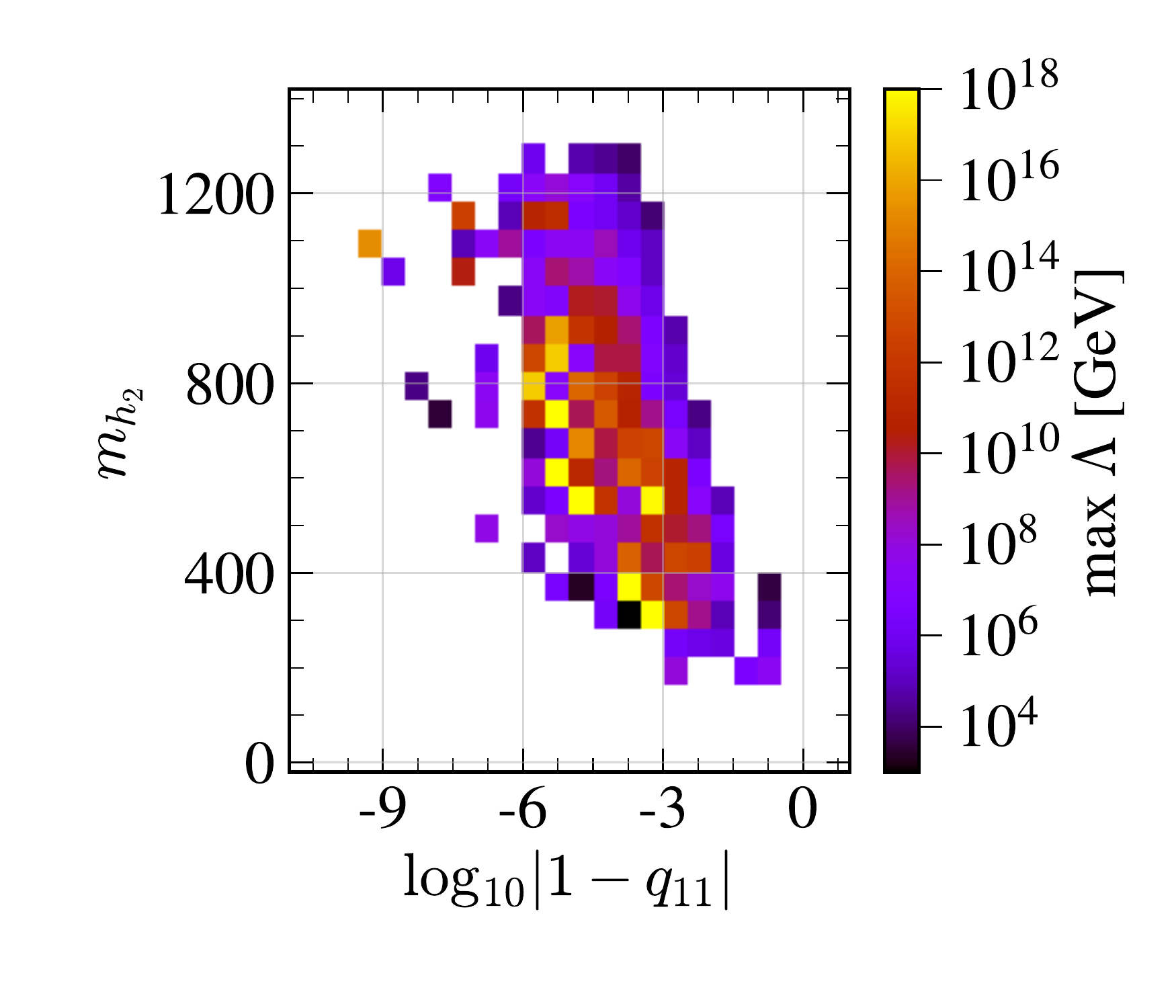} \\
\includegraphics[trim=1cm 1.3cm 0.5cm 0.5cm, clip, height=0.3\textwidth]
    {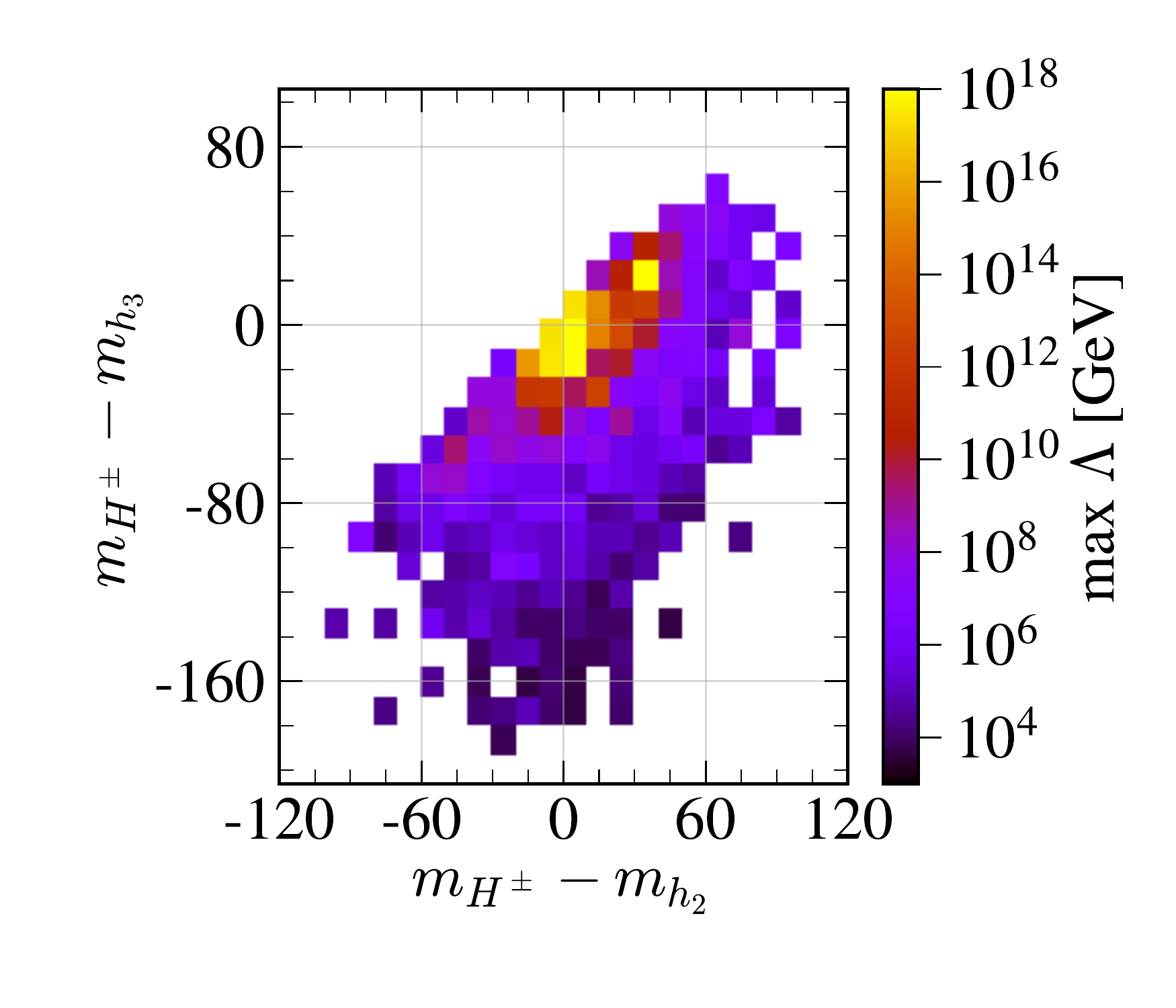} &
\includegraphics[trim=1cm 1.3cm 0.5cm 0.5cm,clip,height=0.3\textwidth]
    {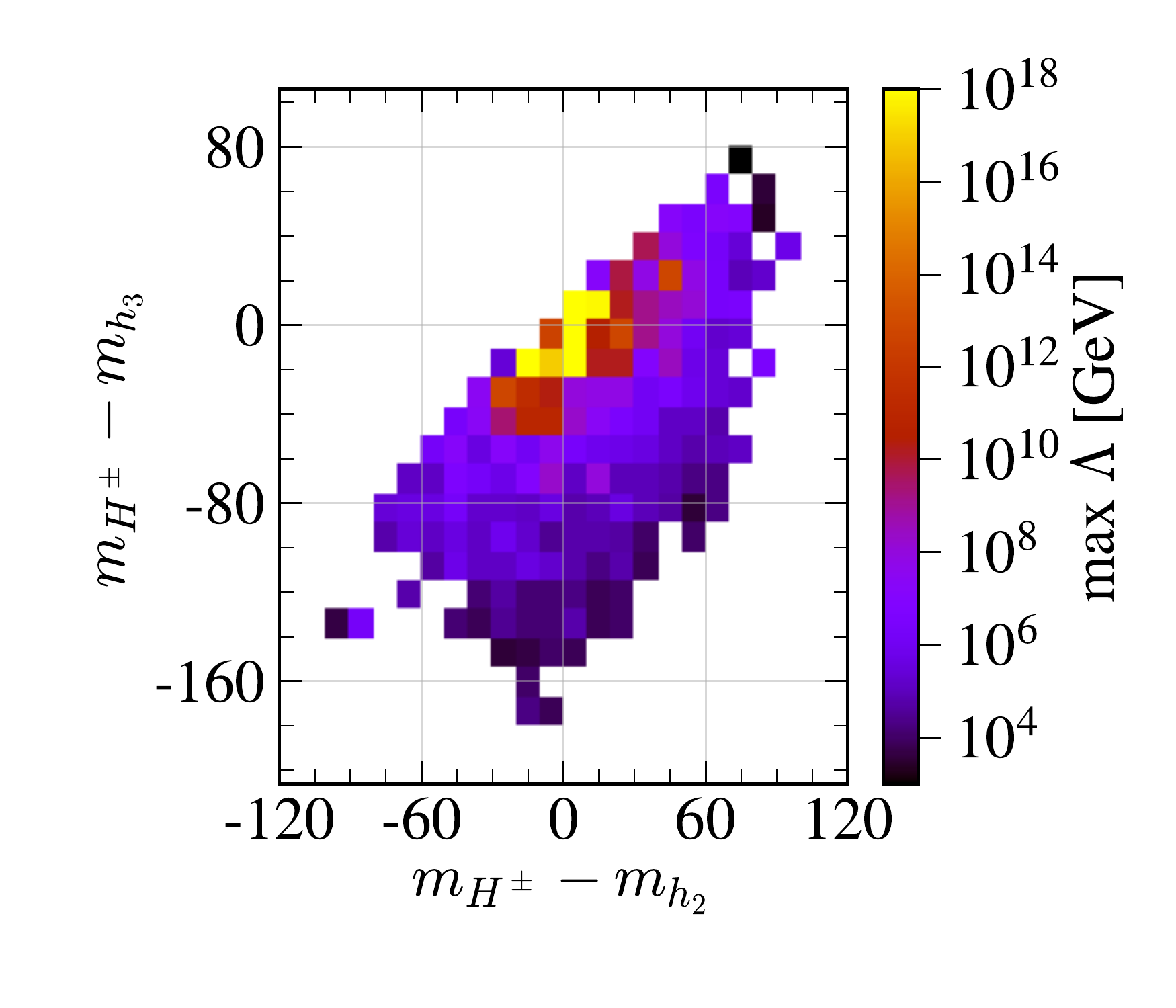} 
\end{tabular}
    \caption{The maximum breakdown energy $\Lambda$ as a function of masses, mass
    differences and the $q_{11}$ parameter for type I (left) and type II
    (right).}
\label{fig:CPVI1}
\end{center}
\end{figure}

\begin{figure}[h!]
\begin{center}
\textbf{Scenario I}\\
\begin{tabular}{cc}
    type I & type II \\
\includegraphics[trim=1cm 1.2cm 5cm 0.5cm, clip, height=0.3\textwidth]
    {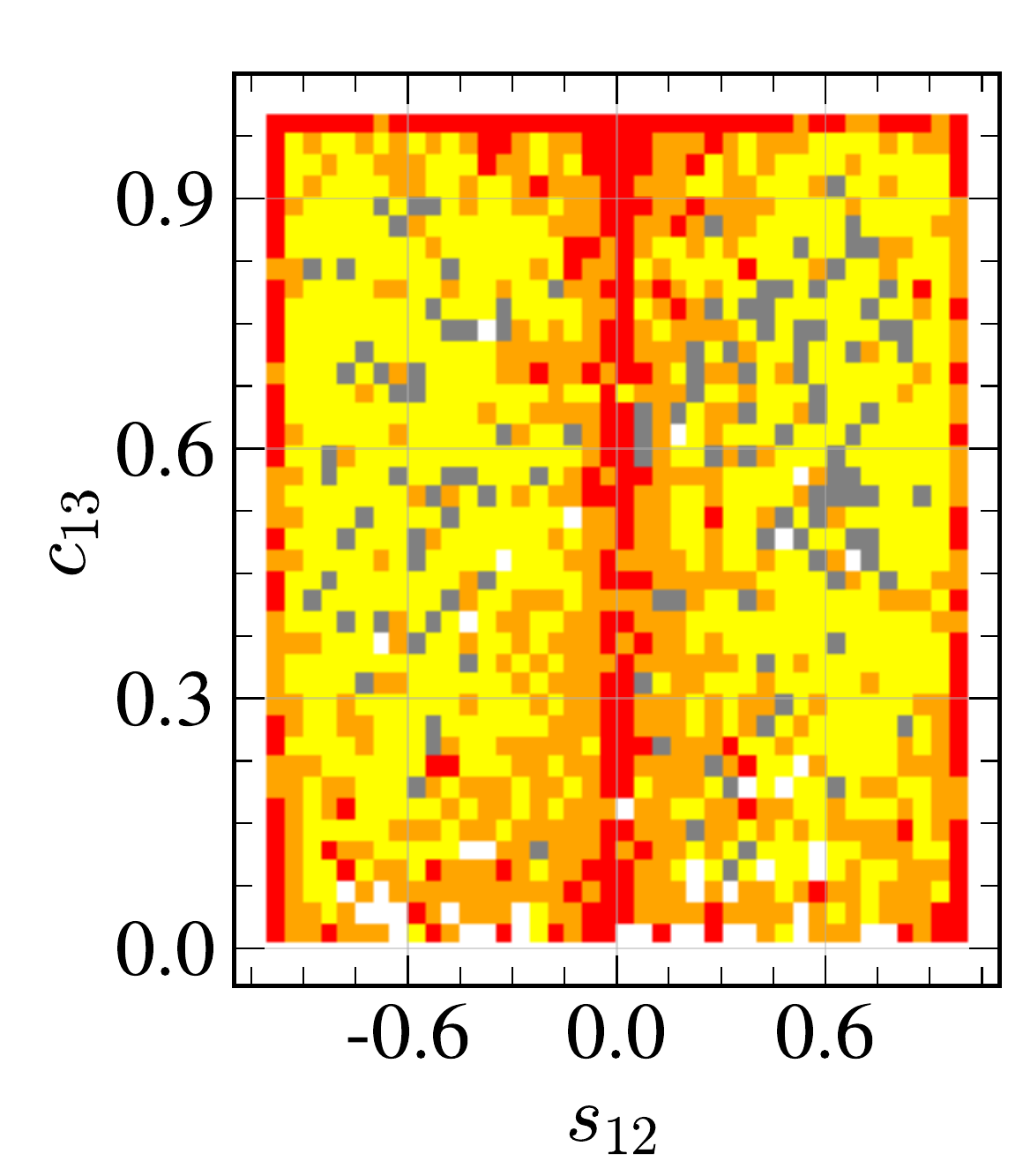} &
\includegraphics[trim=0.5cm 1.2cm 0.5cm 0.5cm,clip,height=0.3\textwidth]
    {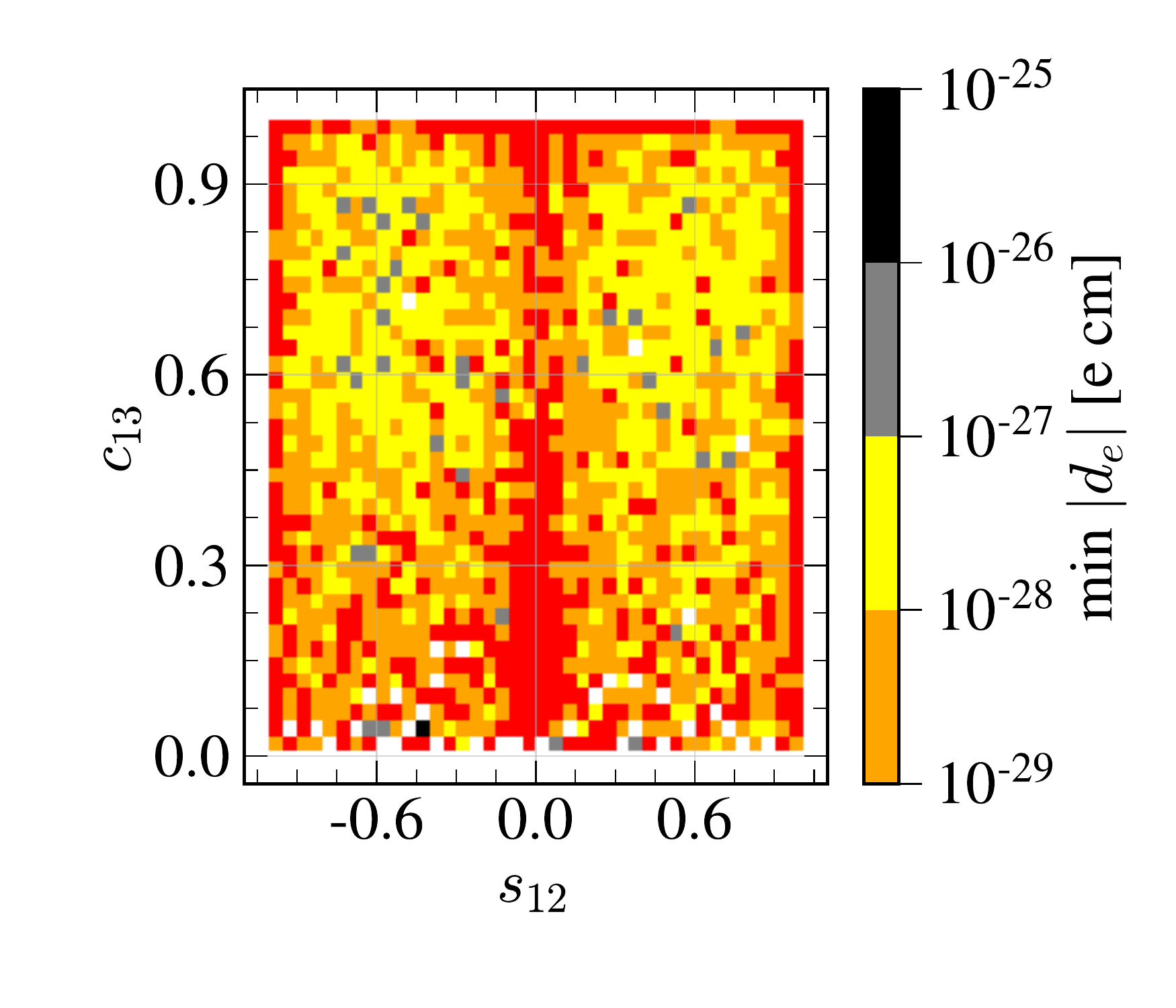} \\
\includegraphics[trim=1cm 1.5cm 5cm 0cm, clip, height=0.3\textwidth]
    {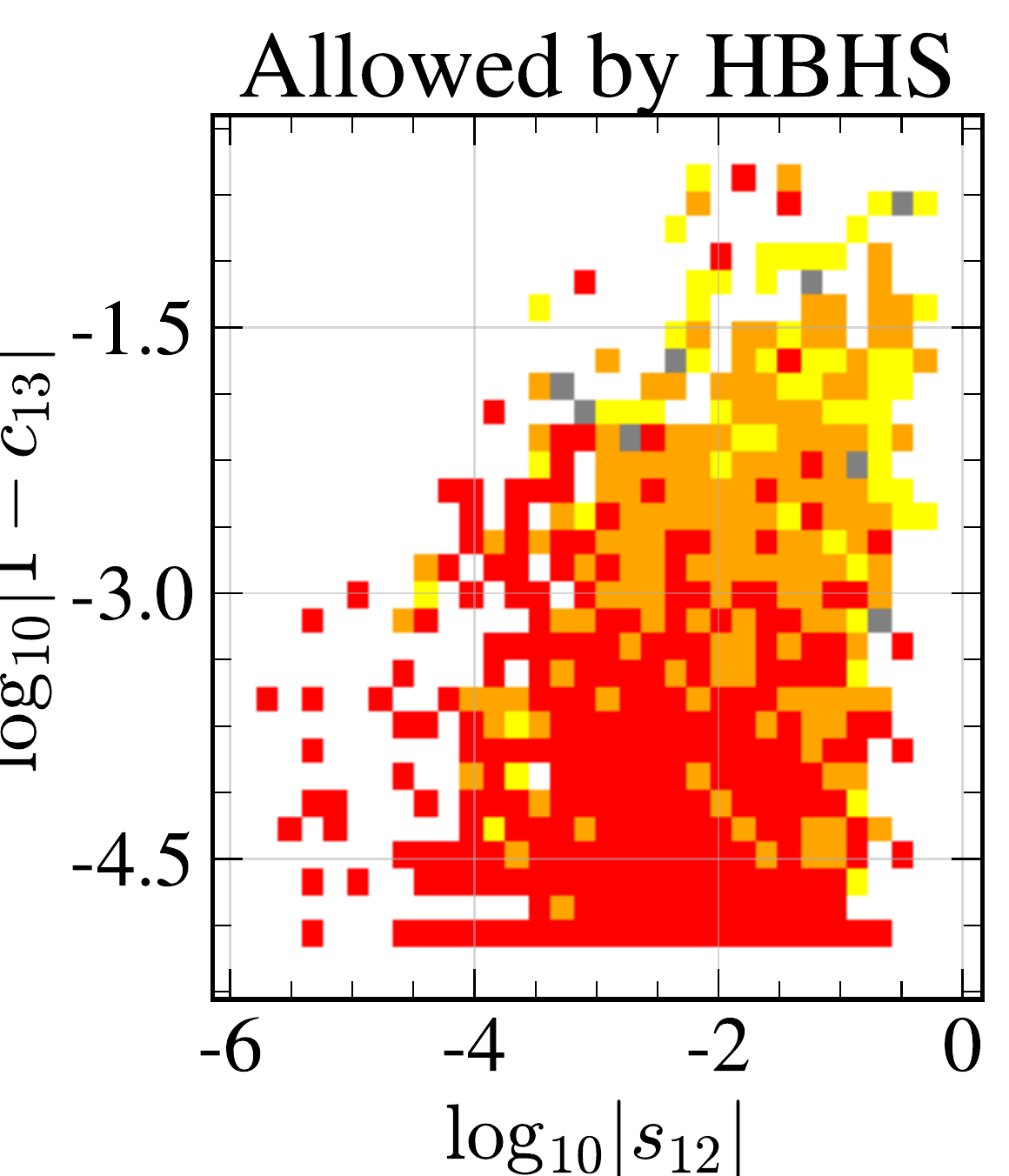} &
\includegraphics[trim=0.5cm 1.5cm 0.5cm 0cm,clip,height=0.3\textwidth]
    {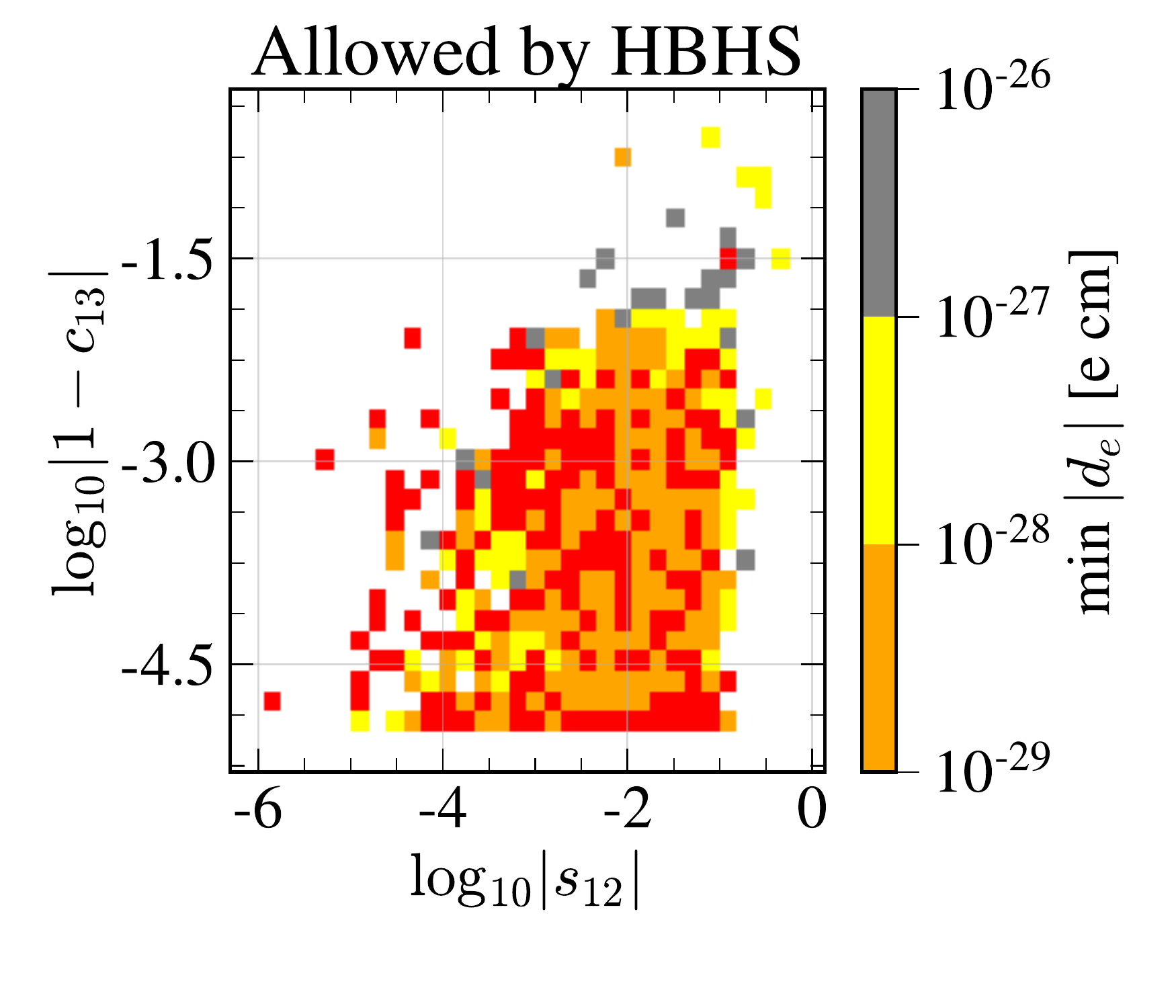}
\end{tabular}
    \caption{The minimum eEDM as a function of the angles $s_{12}$ and $c_{13}$
    for type I (left) and type II (right). The
    top figures are all parameter points, while the bottom figures contain only
    points allowed by \code{HiggsBounds} and \code{HiggsSignals}. Red denotes
    eEDM below $10^{-29}$ e cm.}
\label{fig:CPVI2}
\end{center}
\end{figure}

\begin{figure}[h!]
\begin{center}
\textbf{Scenario I}\\
\begin{tabular}{cc}
    type I & type II \\
\includegraphics[trim=1cm 1cm 0cm 0.5cm, clip, height=0.3\textwidth]
    {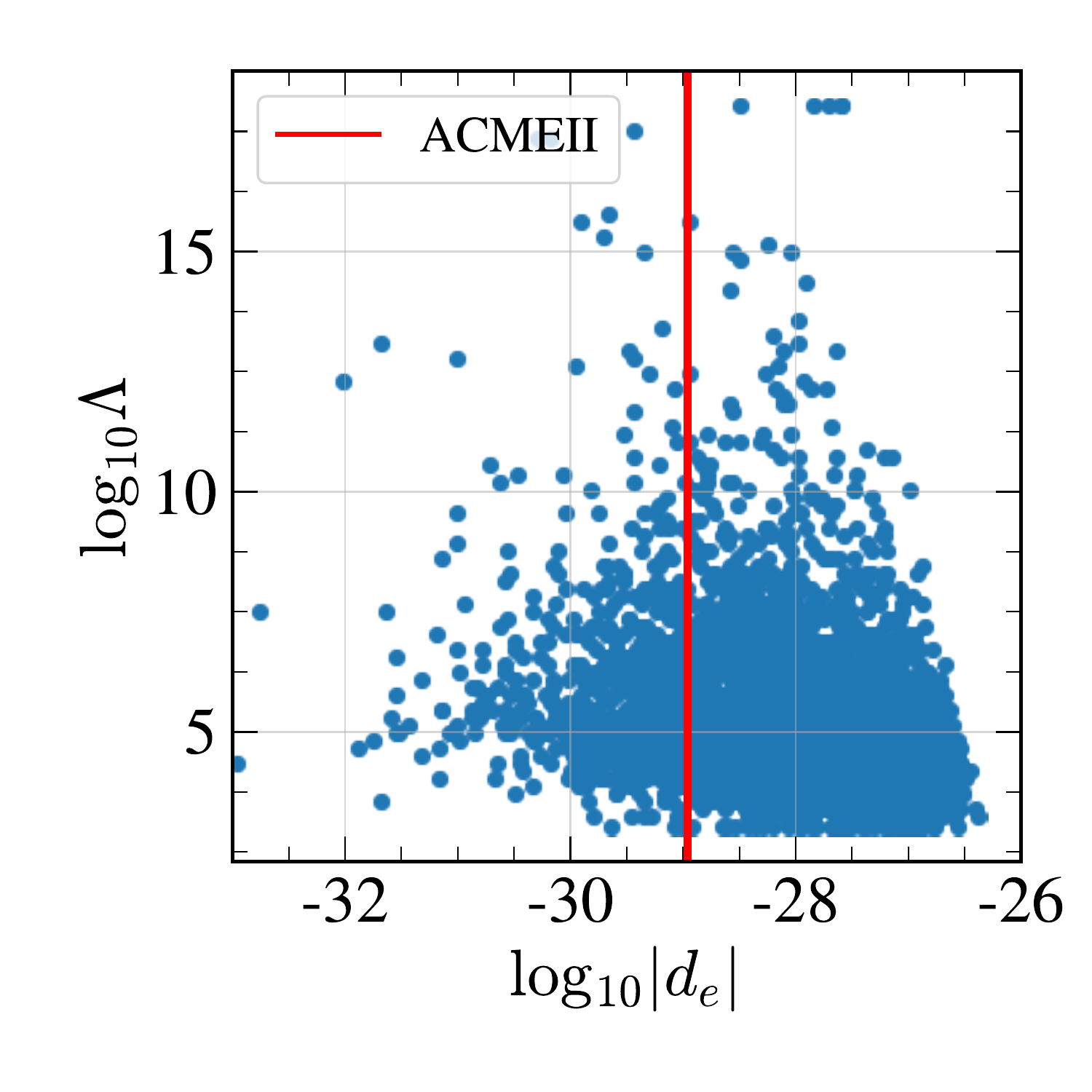} &
\includegraphics[trim=1cm 1cm 0cm 0.5cm, clip, height=0.3\textwidth]
    {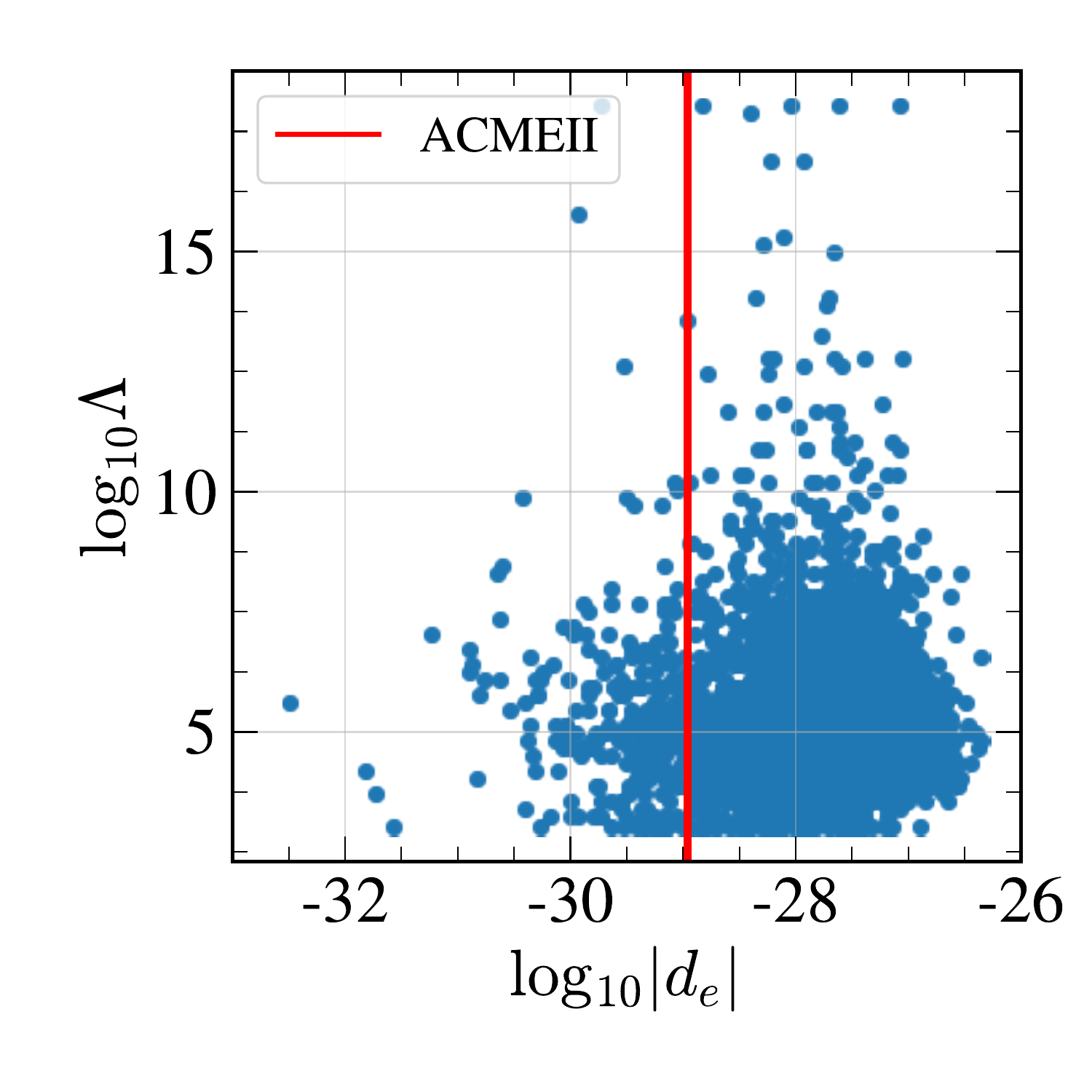} 
\end{tabular}
    \caption{The relation between breakdown energy and eEDM for type I (left)
    and type II (right). The red line is the ACMEII eEDM limit, $1.1\times
    10^{-29}$ e cm.}
\label{fig:CPVI3}
\end{center}
\end{figure}

\begin{figure}[h!]
\begin{center}
\textbf{Scenario I}\\
\begin{tabular}{cc}
    type I & type II \\
\includegraphics[trim=0cm 1cm 5cm 0.5cm, clip, height=0.3\textwidth]
    {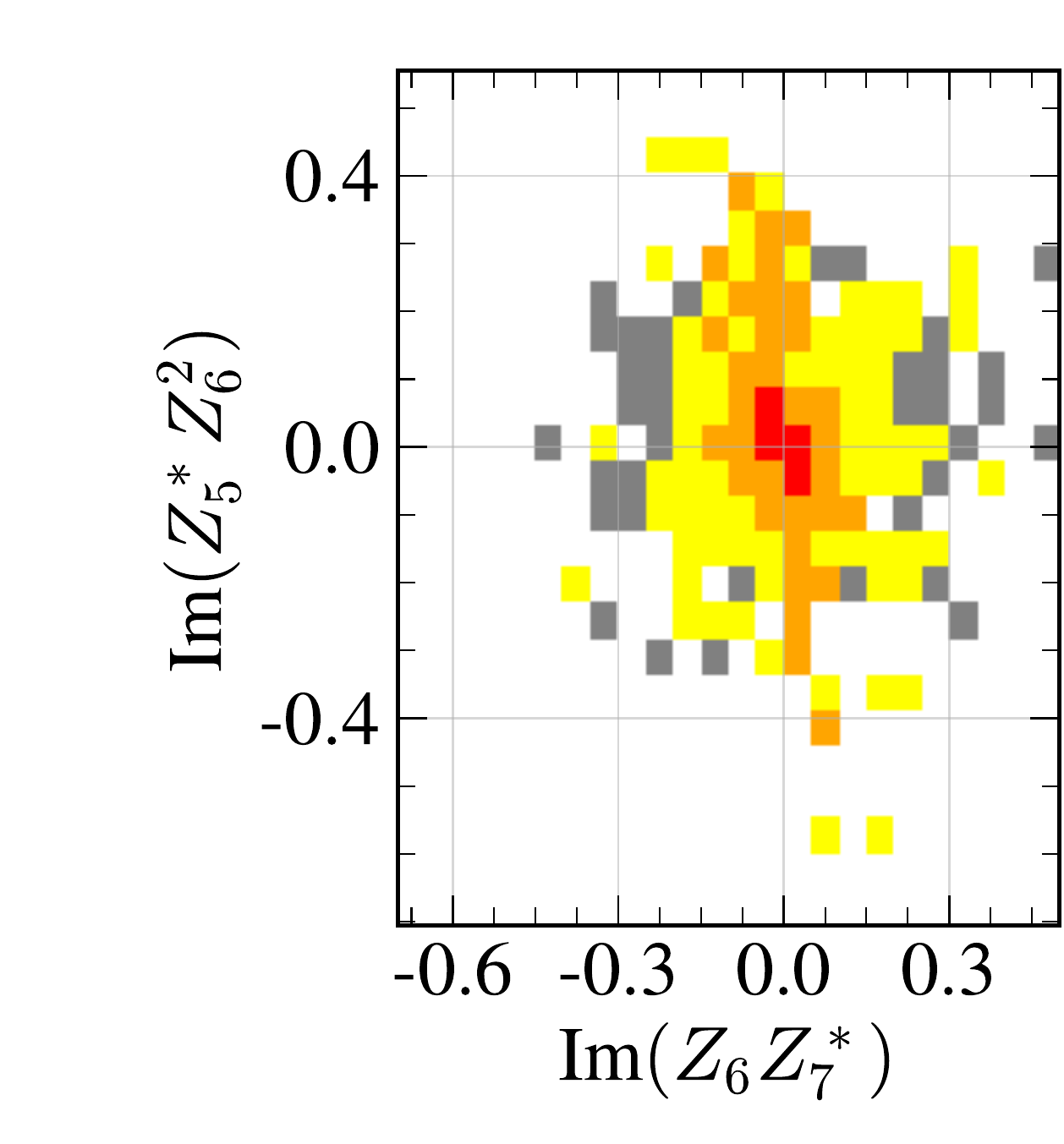} &
\includegraphics[trim=0cm 1cm 0.5cm 0.5cm, clip, height=0.3\textwidth]
    {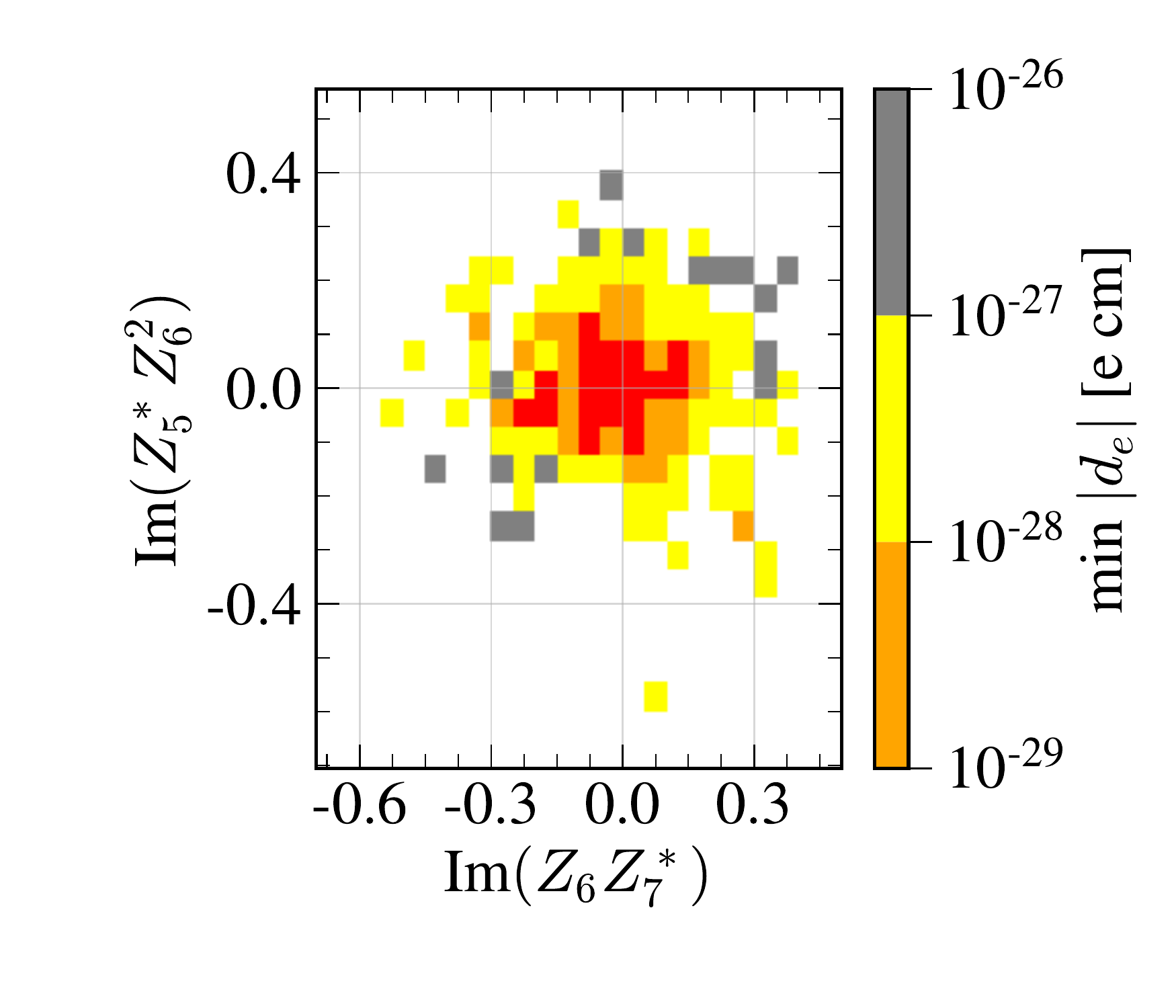} 
\end{tabular}
    \caption{The minimum eEDM as a function of the base invariant
    quantities $Z_5^* Z_6^2$ and $Z_6^*Z_7$ for type I (left)
    and type II (right).}
\label{fig:CPVI4}
\end{center}
\end{figure}

\begin{figure}[h!]
\begin{center}
\textbf{Scenario I}\\
\begin{tabular}{cc}
    type I & type II \\
\includegraphics[trim=0cm 1cm 5cm 0.5cm, clip, height=0.3\textwidth]
    {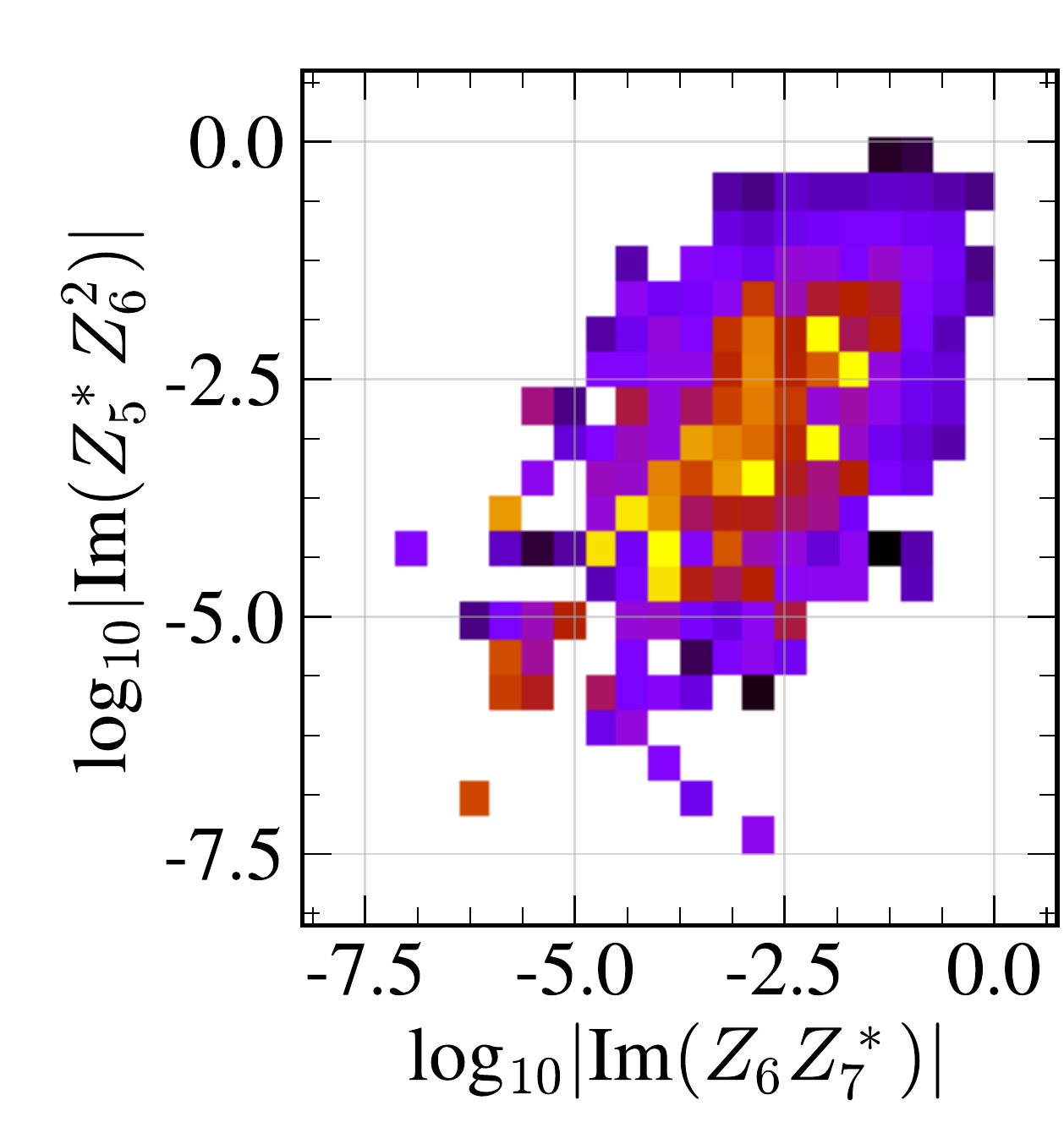} &
\includegraphics[trim=0cm 1cm 0.5cm 0.5cm, clip, height=0.3\textwidth]
    {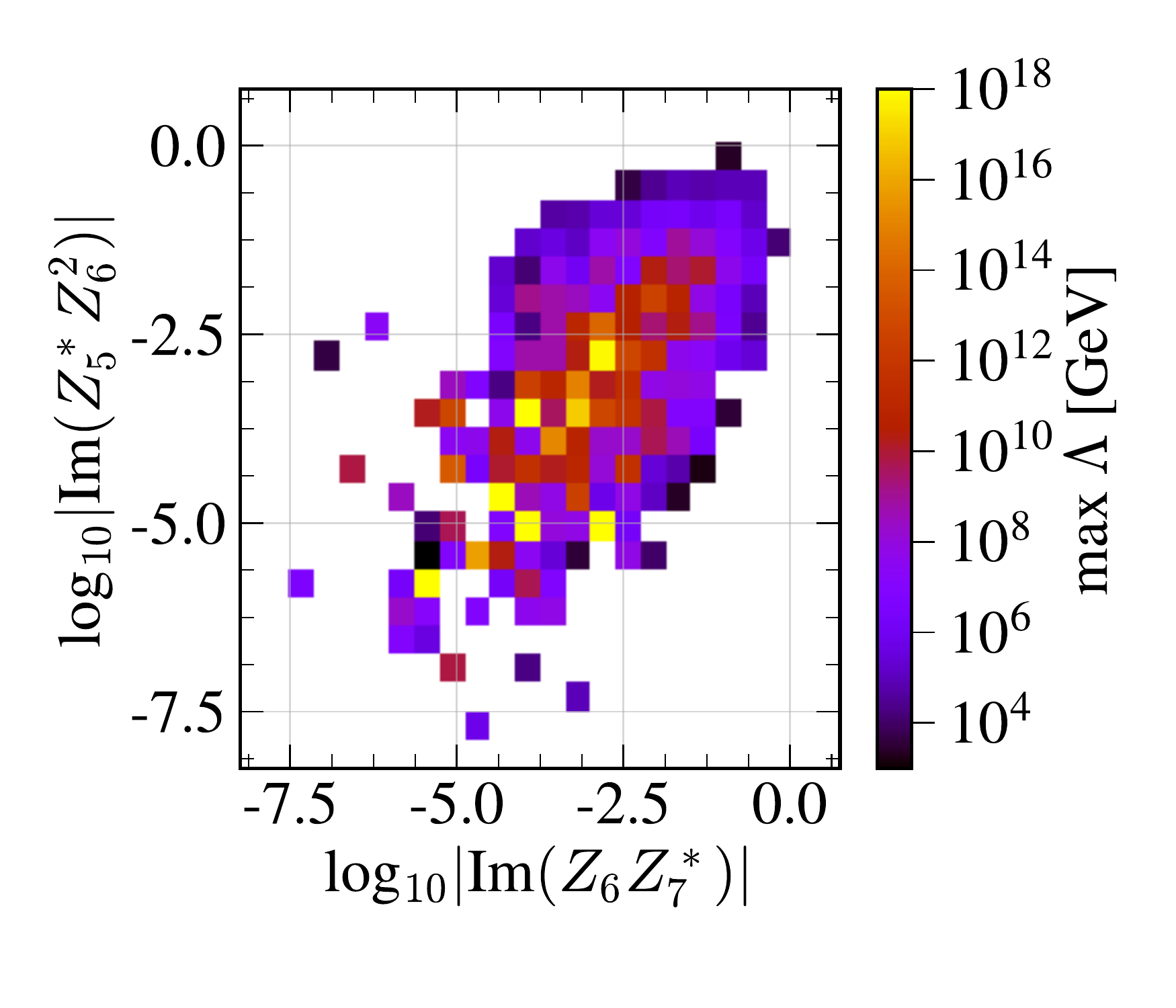} 
\end{tabular}
    \caption{The maximum breakdown energy as a function of the base invariant
    quantities $Z_5^* Z_6^2$ and $Z_6^*Z_7$ for type I (left)
    and type II (right).}
\label{fig:CPVI5}
\end{center}
\end{figure}

Because of the small quartic couplings, the mass spectrum falls easily into an
aligned scenario with $q_{11}\sim 1$. It is easy to find parameter points with
heavy $h_{2,3}$ and $H^\pm$ as can be seen in \fig{CPVI1}; however, heavier
masses implies that the model is more aligned, \textit{i.e.} $q_{11}$ goes to 1
as the masses increase. In the figure, the maximum breakdown energy in each bin
is shown as a function of the masses, mass differences as well as $q_{11}$. From
the figure, it is also clear that only models with small mass differences
between the heavy Higgses can be evolved to high scales. There is also a
preference for $q_{11}$ being very close to 1.

The general property that only very aligned models are viable at the starting
scale is because these are the only ones allowed by \code{HiggsBounds} and
\code{HiggsSignals}. To illustrate this, we show the eEDM as a function of the
angles $s_{12}$ and $c_{13}$  in \fig{CPVI2}. There, two figures for each Yukawa
symmetry are displayed: one before running the parameter points through
\code{HiggsBounds} and \code{HiggsSignals} and one with only the points allowed
by these programs. The allowed points with a small eEDM  all fall in the aligned
region of $s_{12} \sim 0$ and $c_{13} \sim 1$.

To see if a small eEDM also implies a 2HDM that can be evolved to high energies,
we show a scatter plot of the breakdown energy and eEDM in \fig{CPVI3}. As can
be seen from the figure, this is not the case; there is no definite correlation
saying that small eEDM gives a high $\Lambda$. Most points have a too high eEDM
and points that are valid up to the Planck scale exist, presumably, in the
entire region.

Even though there is only one phase in scenario I, $\arg \lambda_5$, that is the
source of \CP~violation in the scalar potential, we find that the base invariant
quantities in \eq{CPVparams} are better measures of the amount of \CP~violation
in the 2HDM. This is because they are also dependent on $\tan\beta$ and other
quartic couplings that for example also influence the eEDM; in addition, it also
simplifies the comparison between different scenarios. All parameter points that
pass the ACMEII bound have these quantities  at the order of $0.1$. The eEDM as
a function of Im$(Z_5^*Z_6^2)$ and Im$(Z_6^*Z_7)$ is shown in \fig{CPVI4}.  In
\fig{CPVI5}, we show the maximum breakdown energy as a function of the same
\CP~violating quantities and there one can see that all parameter points that
are valid all the way to the Planck scale have Im$(Z_5^*Z_6^2) \sim
$Im$(Z_6^*Z_7)\sim 10^{-2}$, thus constraining these parameters even
further.

%------------------------------------------------------------------------------
\FloatBarrier
\subsection{Scenario II}
%------------------------------------------------------------------------------

\begin{figure}[h!]
\begin{center}
\textbf{Scenario II}\\
\begin{tabular}{cc}
    type I & type II \\
\includegraphics[trim=1cm 1.5cm 5cm 0cm, clip, height=0.3\textwidth]
    {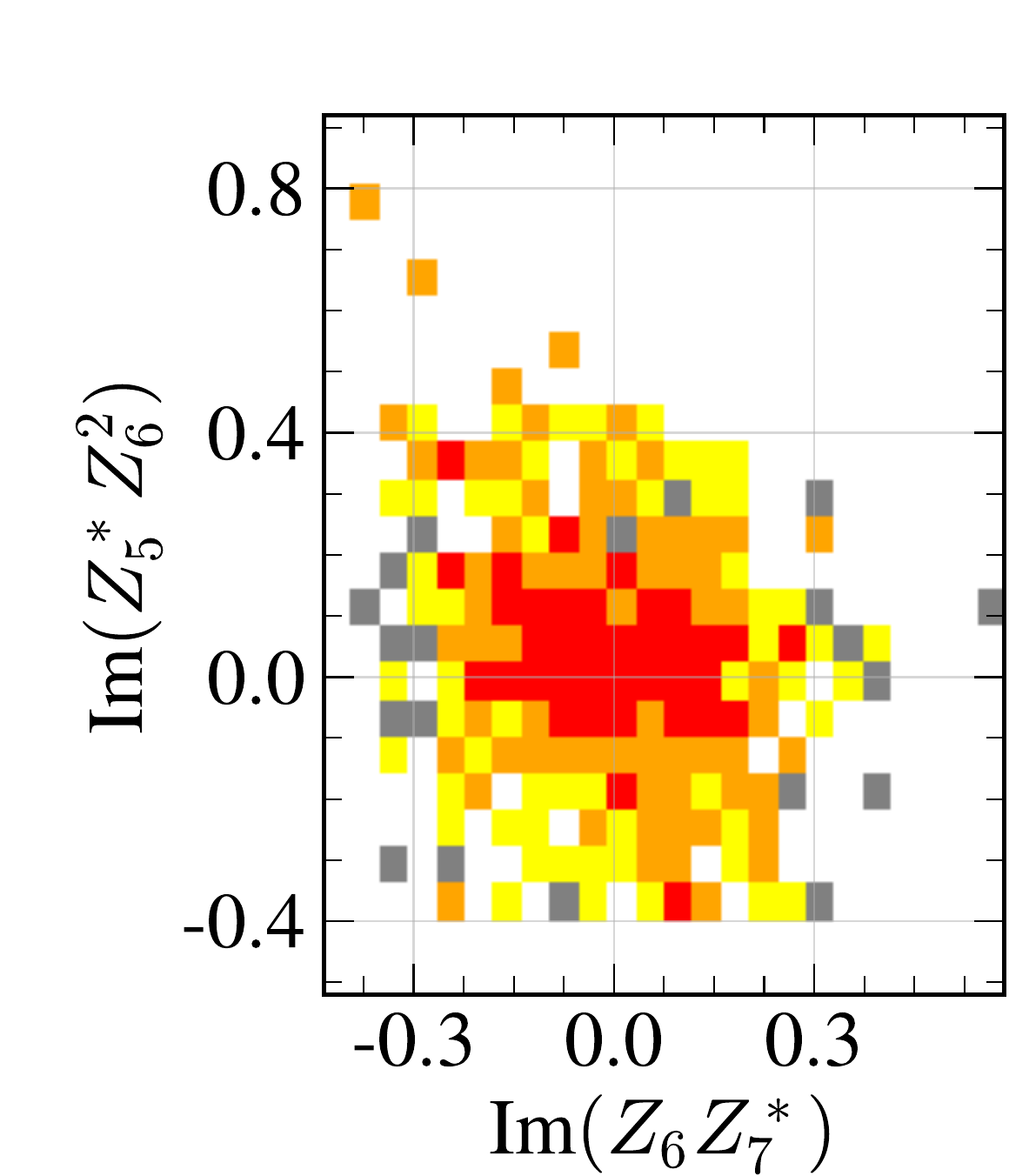} &
\includegraphics[trim=0.5cm 1.5cm 0.5cm 0cm,clip,height=0.3\textwidth]
    {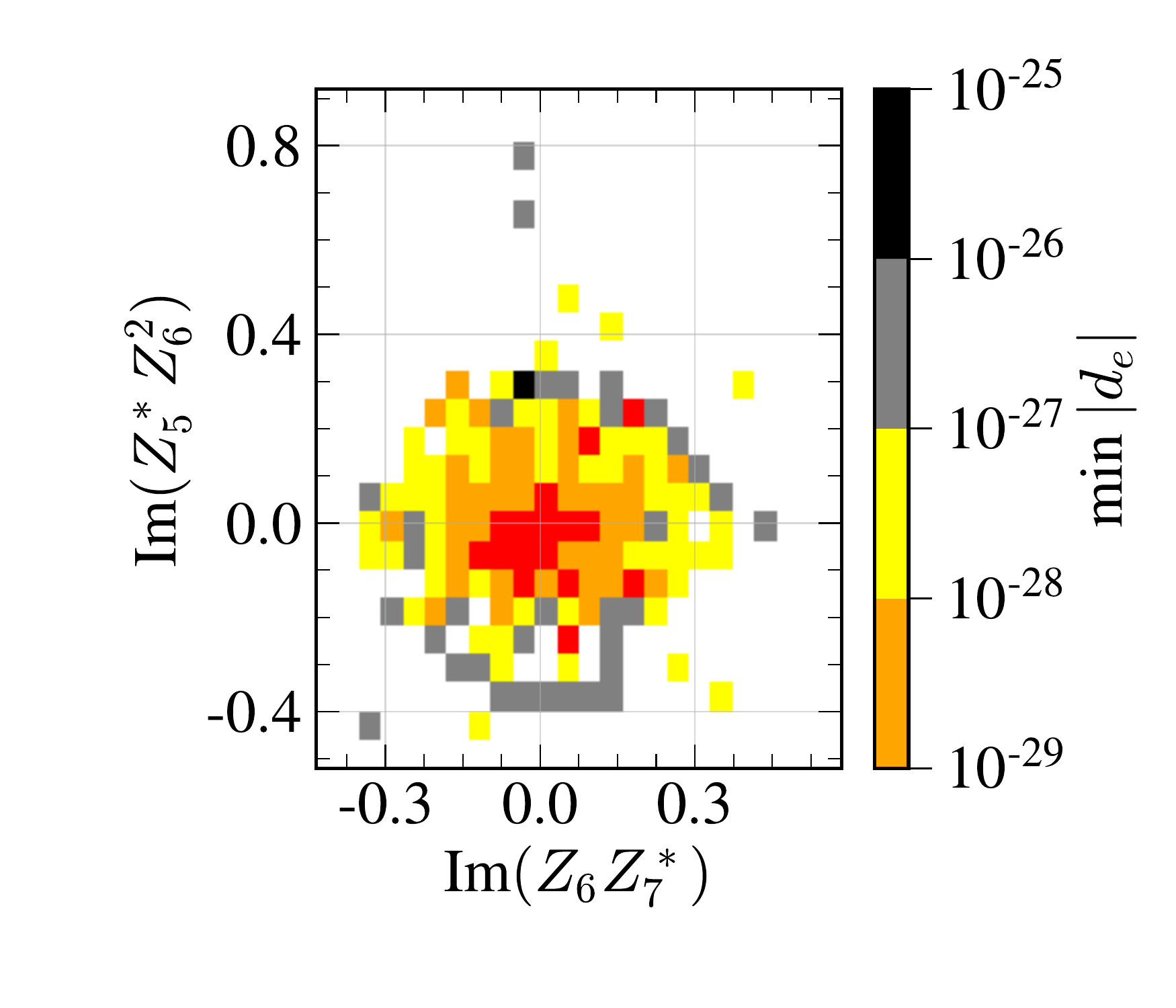}
\end{tabular}
    \caption{The minimum eEDM as a function of the base invariant
    quantities $Z_5^* Z_6^2$ and $Z_6^*Z_7$ for type I (left) and type II (right).}
\label{fig:CPVII1}
\end{center}
\end{figure}

The results of scenario II are largely following that of scenario I; one gets
the same mass spectrum characteristics and the parameter points that survive are
aligned in a similar way as in \fig{CPVI1}. The addition of non-zero
$\lambda_{6,7}$ parameters does, however, have some effects. 

We first note that base-invariant quantities $Z_5^*Z_6^2$ and $Z_6^*Z_7$ get
additional contributions from $\lambda_6$ and $\lambda_7$. In \fig{CPVII1} we
see that this increases the allowed range of the imaginary parts of these
quantities, while still having an allowed eEDM.

\begin{figure}[h!]
\begin{center}
\textbf{Scenario II}\\
\begin{tabular}{cc}
    type I & type II \\
\includegraphics[trim=1cm 1.2cm 5cm 0.5cm, clip, height=0.3\textwidth]
    {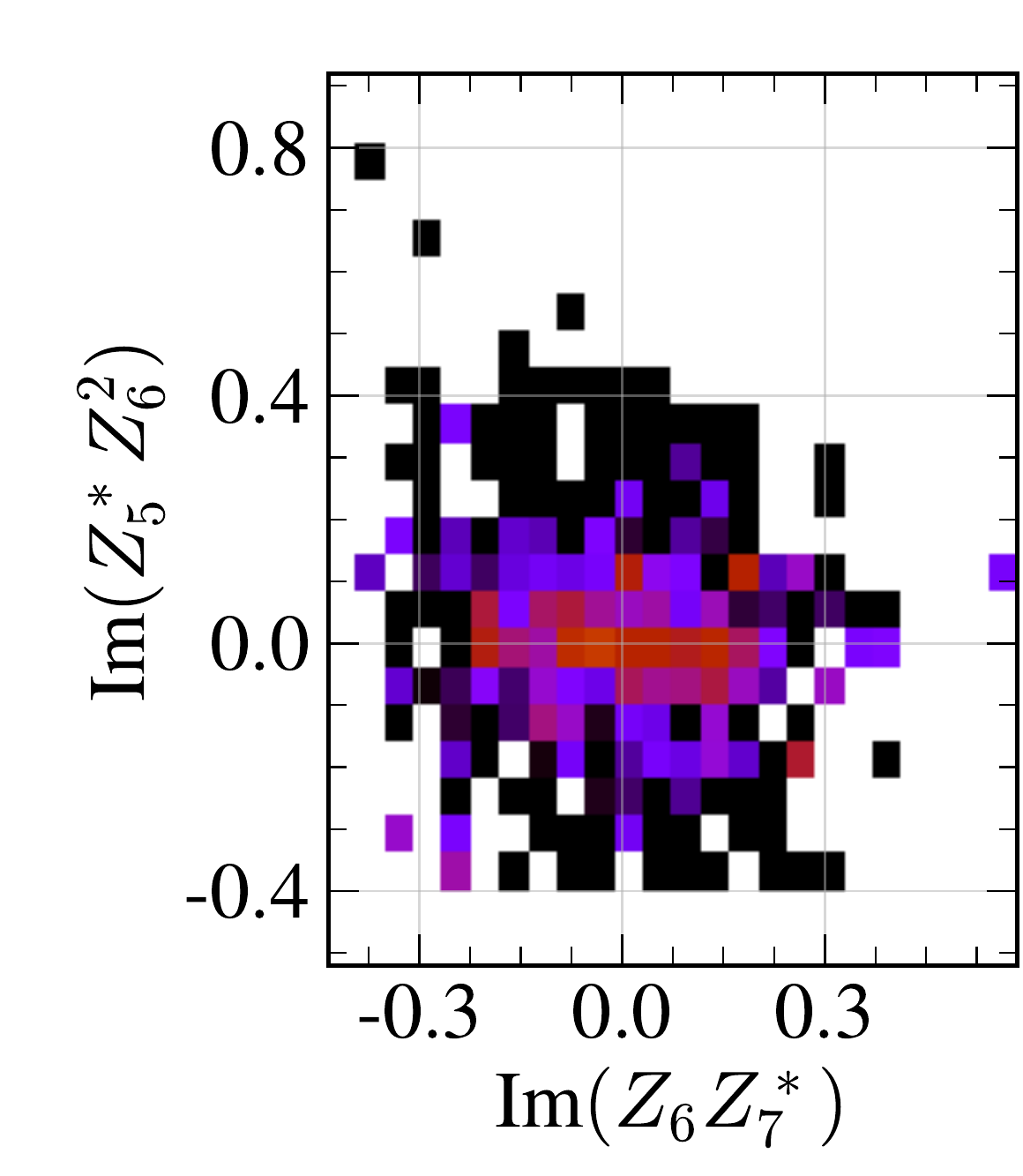} &
\includegraphics[trim=0.5cm 1.2cm 0.5cm 0.5cm,clip,height=0.3\textwidth]
    {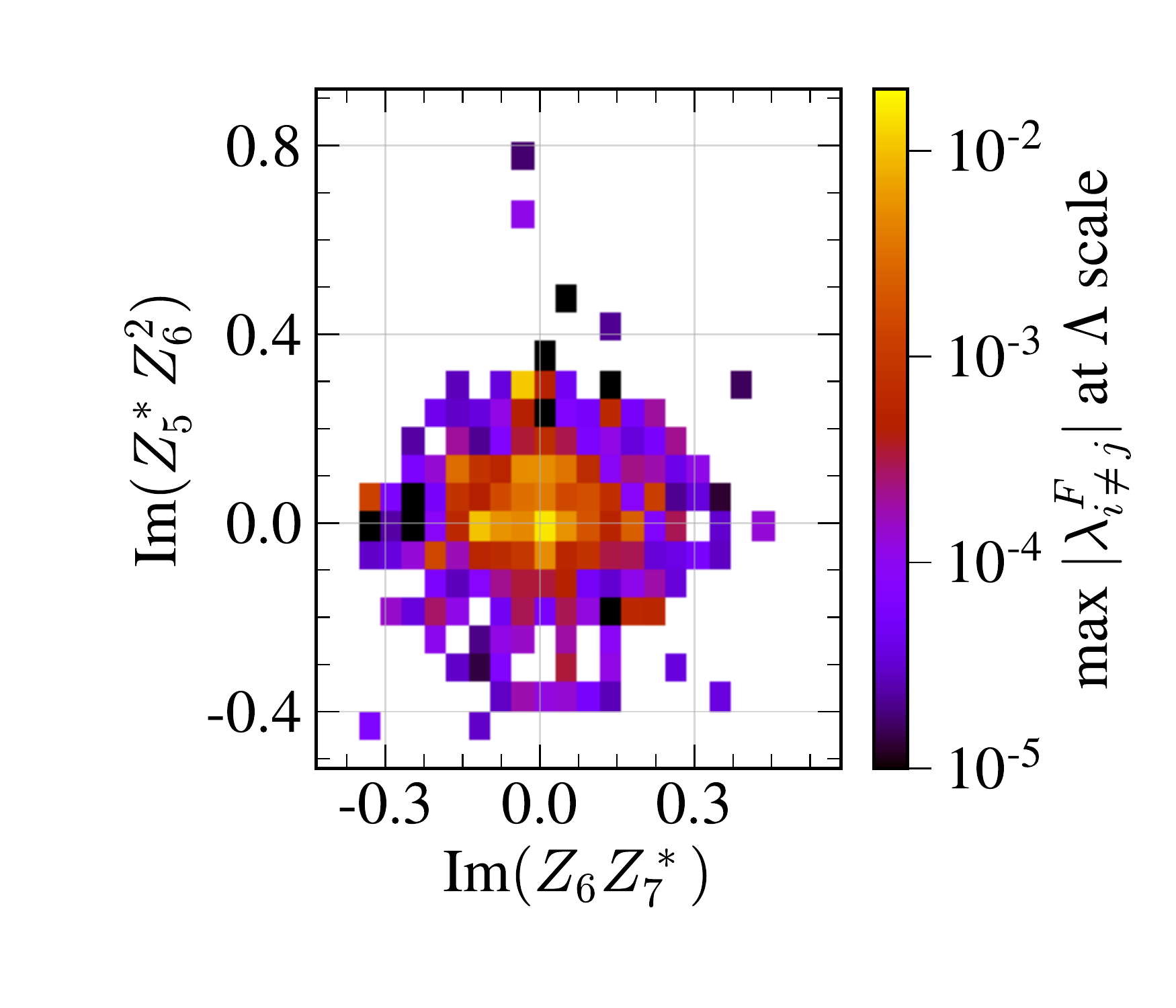} 
\end{tabular}
    \caption{The maximum induced non-diagonal Yukawa element in the Cheng-Sher
    parameterization as a function of the base invariant
    quantities $Z_5^* Z_6^2$ and $Z_6^*Z_7$ for type I (left) and type II (right).}
\label{fig:CPVII2}
\end{center}
\end{figure}

\begin{figure}[h!]
\begin{center}
\textbf{Scenario II}\\
\begin{tabular}{cc}
    type I & type II \\
\includegraphics[trim=1cm 1.2cm 5cm 0.5cm, clip, height=0.3\textwidth]
    {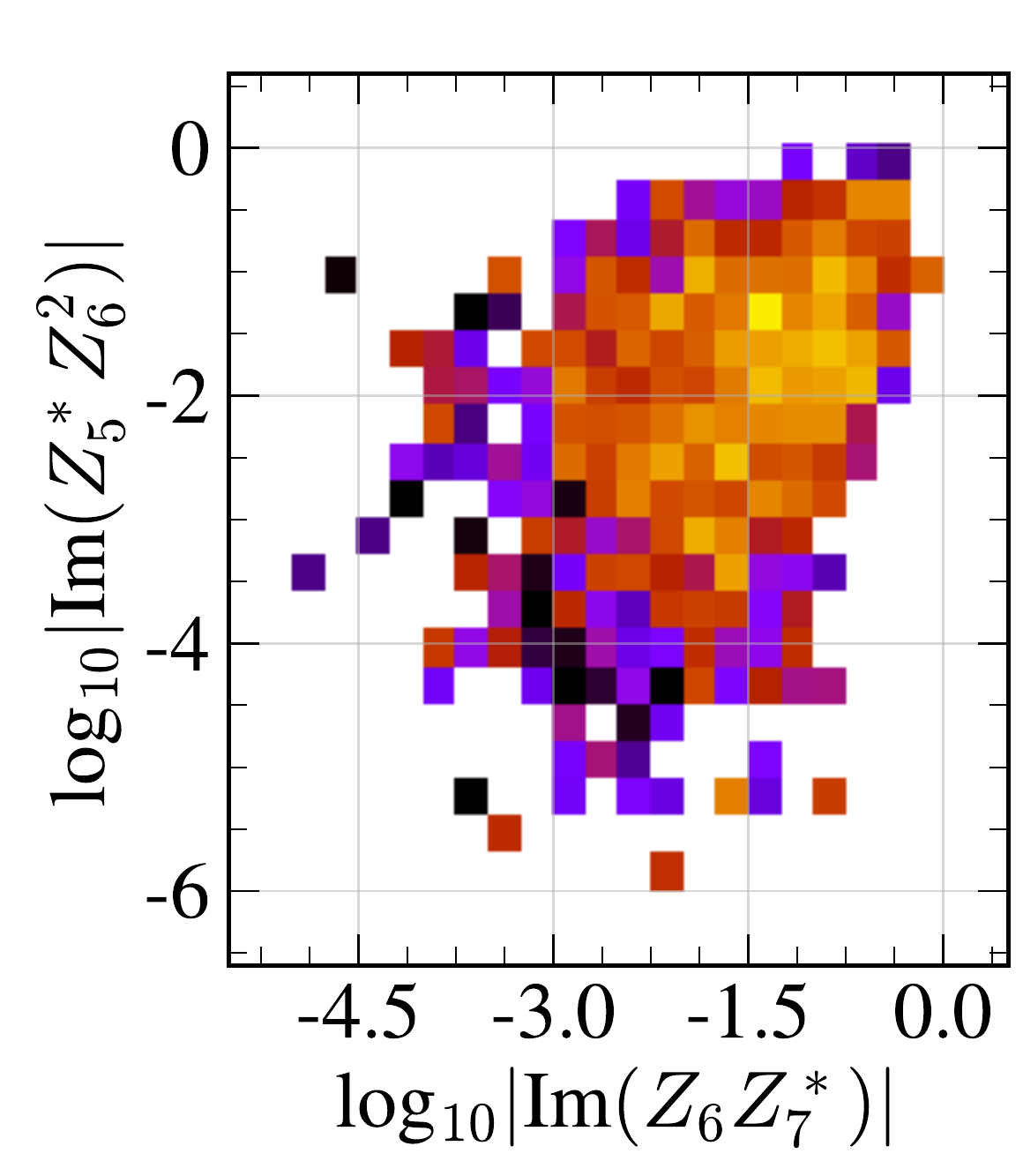} &
\includegraphics[trim=0.5cm 1.2cm 0.5cm 0.5cm,clip,height=0.3\textwidth]
    {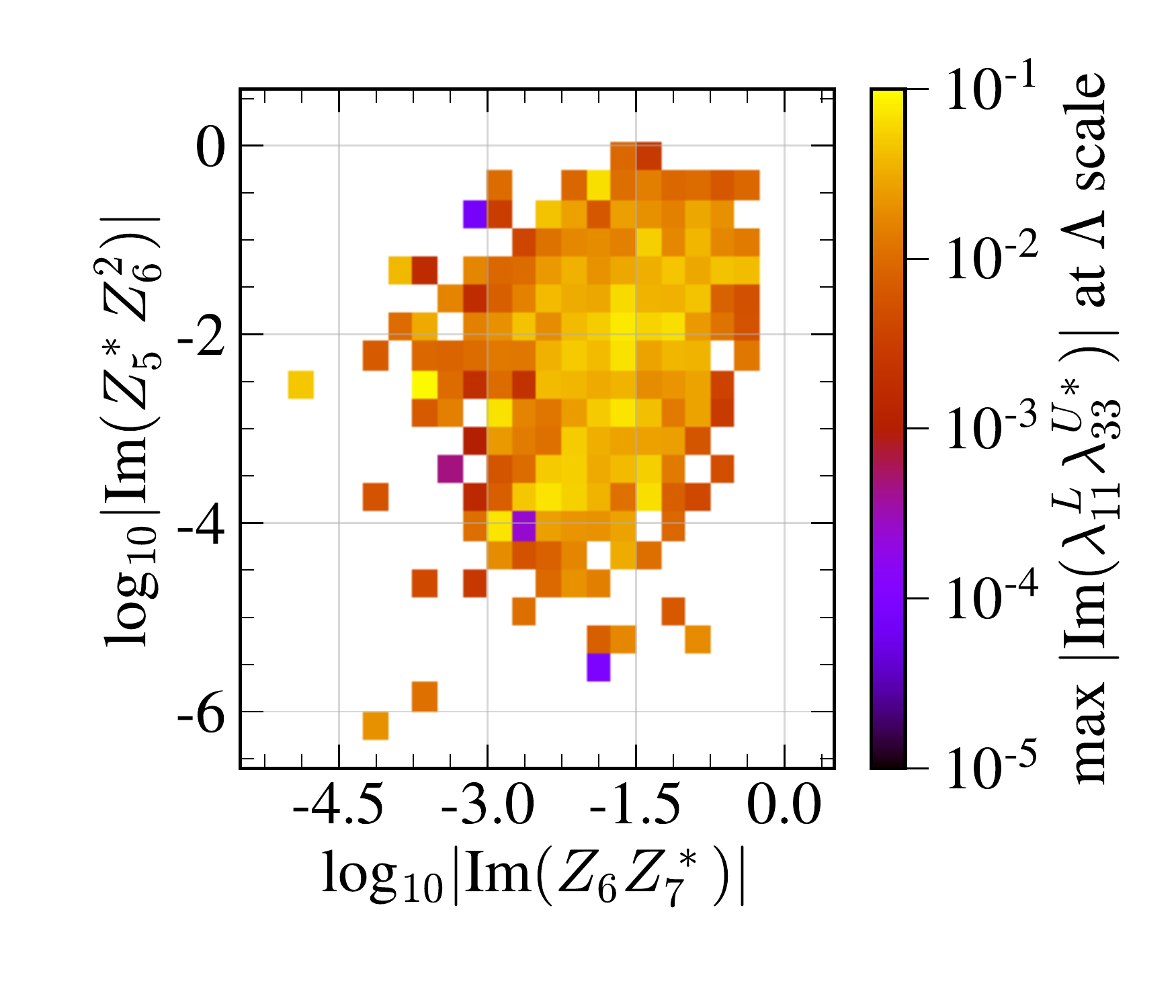}
\end{tabular}
    \caption{The maximum induced imaginary part of
    $\lambda_{33}^L\lambda_{33}^{U*}$ at the breakdown energy scale as a
    function of the base invariant quantities $Z_5^* Z_6^2$ and $Z_6^*Z_7$ for
    type I (left) and type II (right).}
\label{fig:CPVII3}
\end{center}
\end{figure}
Since the $\lambda_{6,7}$ parameters break the \Zsym symmetry hard, the symmetry
breaking spreads in the RG running to the Yukawa sector as well. This does not
have a huge impact, however, since the quartic couplings enter the Yukawa
couplings RGEs first at 2-loop order. The maximum induced non-diagonal Yukawa coupling
as a function of the imaginary parts of $Z_5^* Z_6^2$ and $Z_6^*Z_7$ is shown in
\fig{CPVII2}. There we see that although the effect is much larger with a type II
Yukawa sector, one does not get size-able FCNCs after RG running. For type I
(type II) the maximum generated non-diagonal Yukawa element is $\lambda_{i\neq
j}^F \sim 10^{-3} (10^{-2})$ at the breakdown energy scale.
Similar findings are presented in \mycite{Oredsson:2018yho} in the \CP~conserving
case.
In this \CP~violating case, one can generate a non-trivial amount of
\CP~violation in the Yukawa sector though. To see this, we show the maximum
generated imaginary part of the base invariant quantity
$\lambda_{11}^L\lambda_{33}^{U*}$ in \fig{CPVII3}. This parameter is chosen
because it needs to be small to not yield a too large eEDM; as is discussed in
scenario III.

\begin{figure}[h!]
\begin{center}
\textbf{Scenario II}\\
\begin{tabular}{cc}
    type I & type II \\
\includegraphics[trim=1cm 1.2cm 4.5cm 0.5cm, clip, height=0.3\textwidth]
    {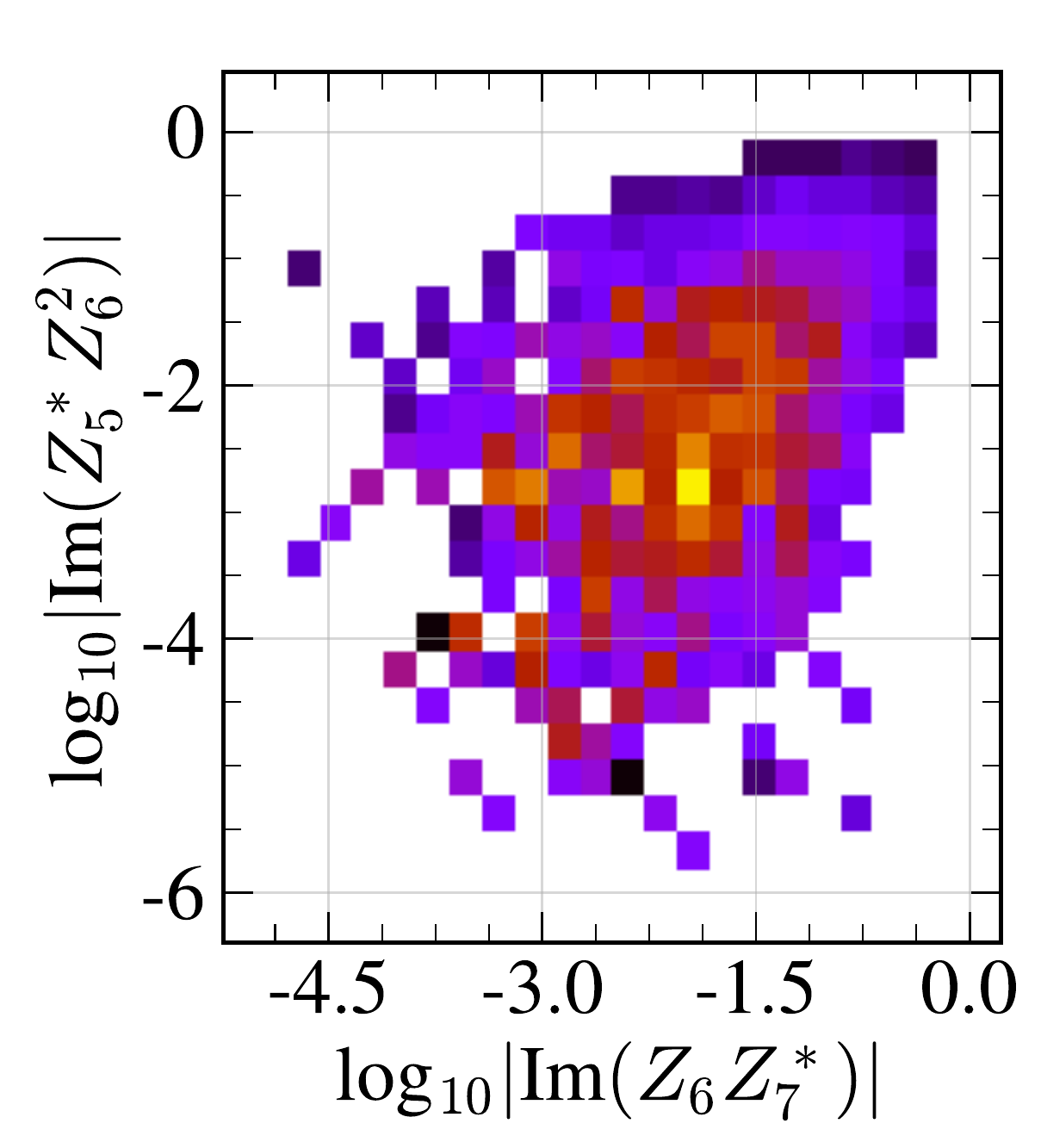} &
\includegraphics[trim=0.5cm 1.2cm 0.5cm 0.5cm,clip,height=0.3\textwidth]
    {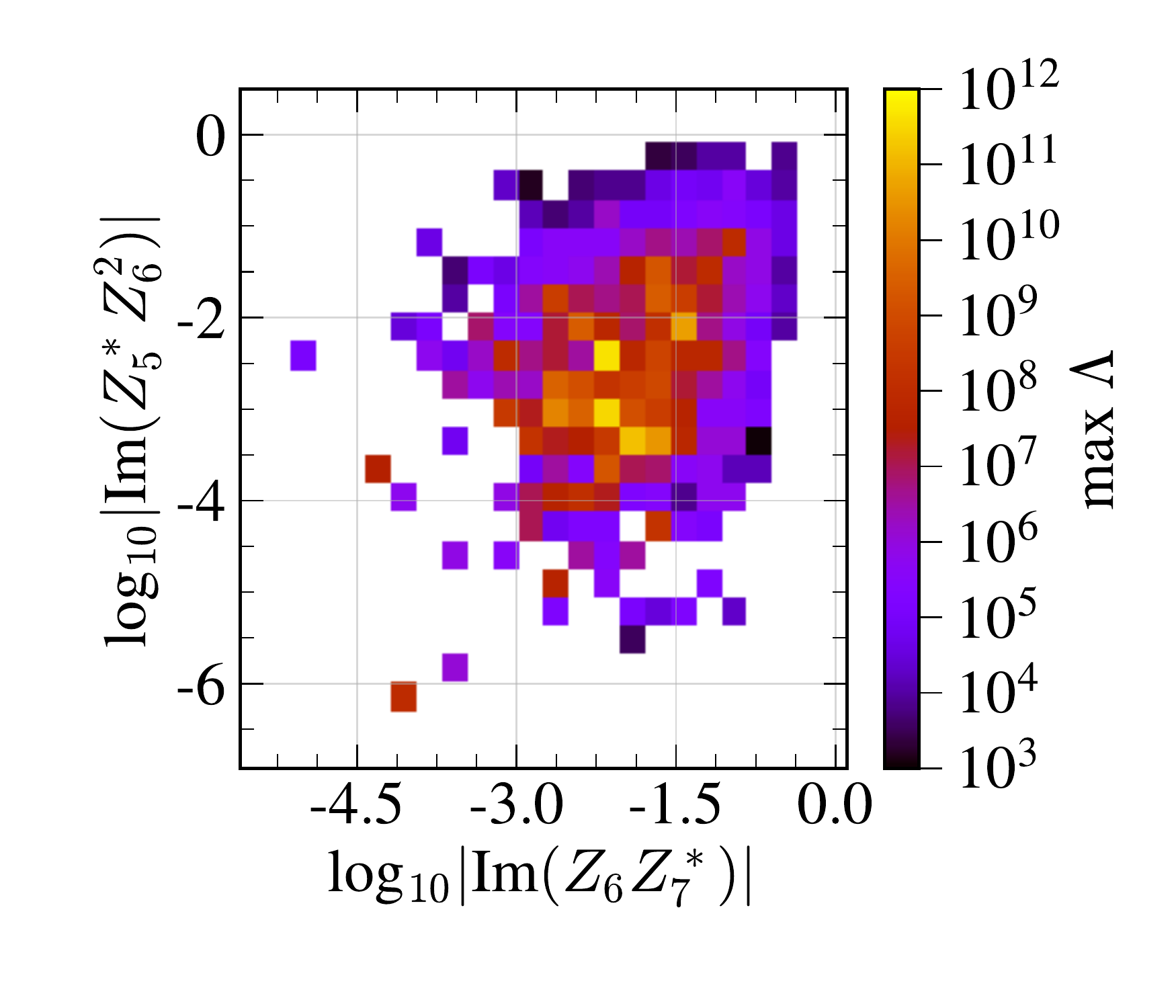}
\end{tabular}
    \caption{The maximum breakdown energy as a function of the base invariant
    quantities $Z_5^* Z_6^2$ and $Z_6^*Z_7$ for type I (left) and type II (right).}
\label{fig:CPVII4}
\end{center}
\end{figure}

The imaginary parts of  $Z_5^* Z_6^2$ and $Z_6^*Z_7$ serves as good measures of
the amount of \CP~violation in the 2HDM as seen in \fig{CPVII1}; all the points
that satisfy the eEDM bound are centered around them being zero. The parameter
points that are valid up to the highest energies also exhibit small Im$(Z_5^*
Z_6^2)$ and Im$(Z_6^*Z_7)$ as can be seen in \fig{CPVII4}.

%------------------------------------------------------------------------------
\FloatBarrier
\subsection{Scenario III}
%------------------------------------------------------------------------------

\begin{figure}[h!]
\begin{center}
\textbf{Scenario III}\\
\begin{tabular}{ccc}
    \footnotesize{type I} & \footnotesize{type II} & \footnotesize{type X} \\
\includegraphics[trim=1cm 1.2cm 5cm 0.5cm, clip, height=0.26\textwidth]
    {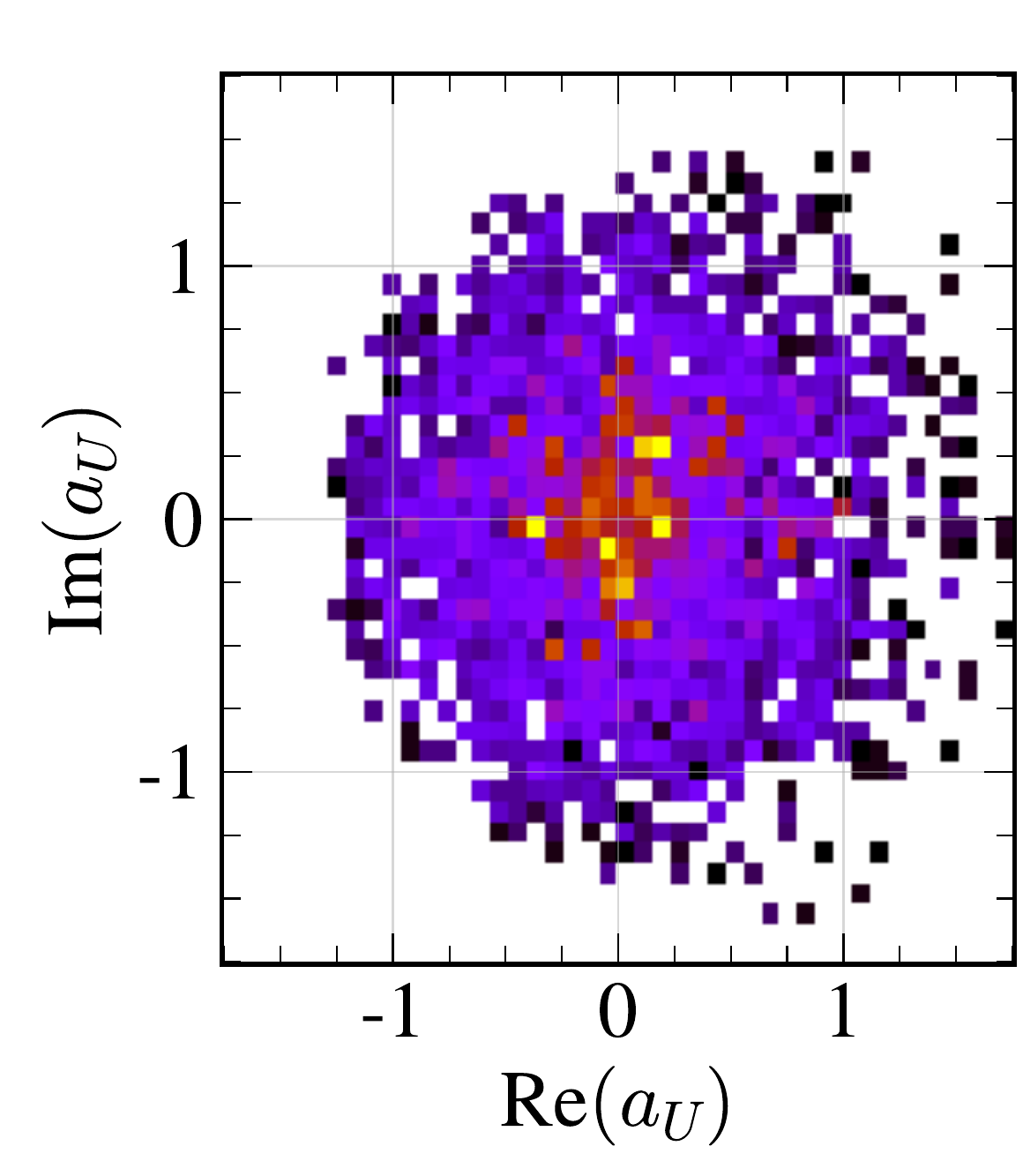} &
\includegraphics[trim=1cm 1.2cm 5cm 0.5cm,clip,height=0.26\textwidth]
    {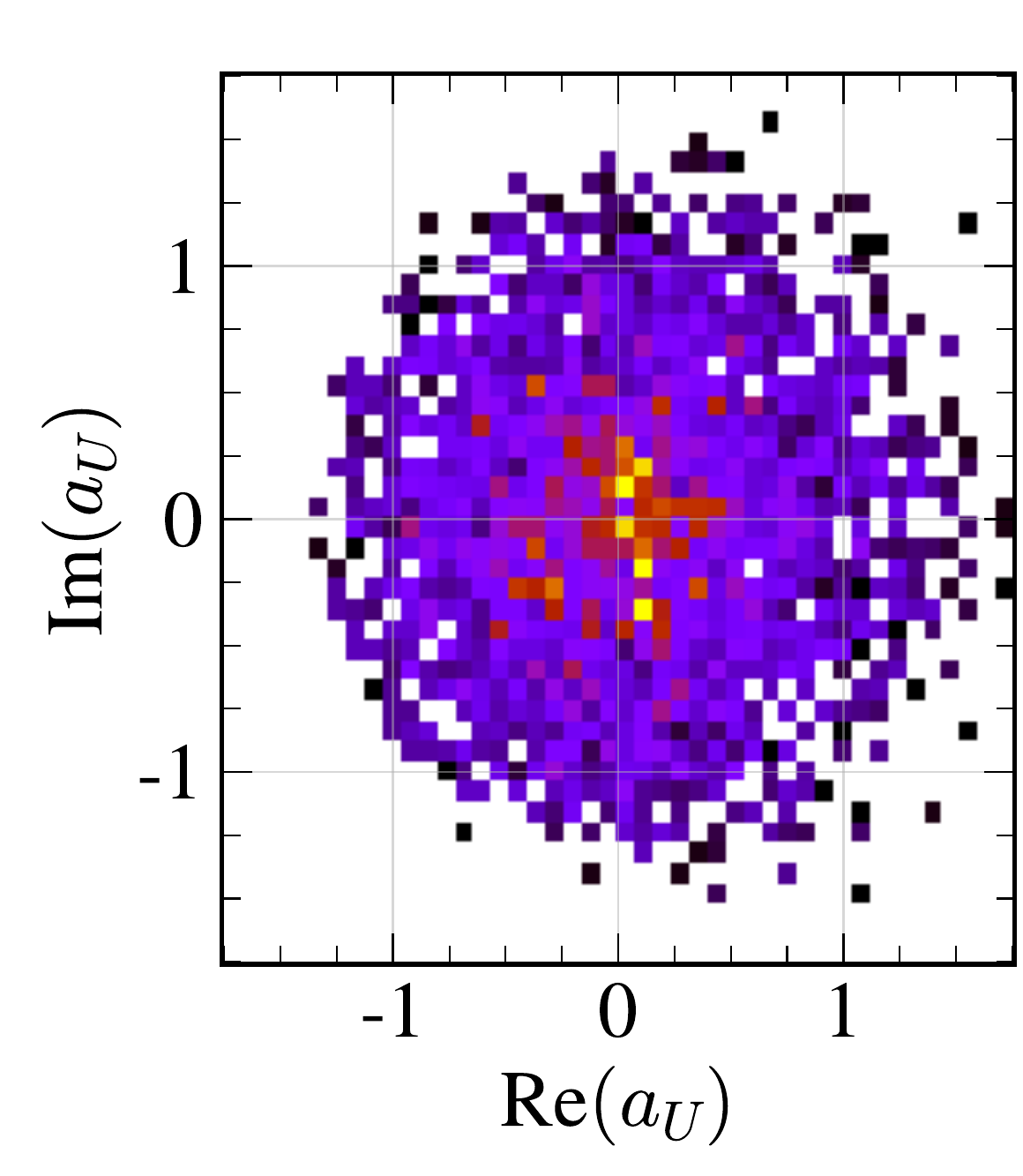} &
\includegraphics[trim=1cm 1.2cm 0.5cm 0.5cm,clip,height=0.26\textwidth]
    {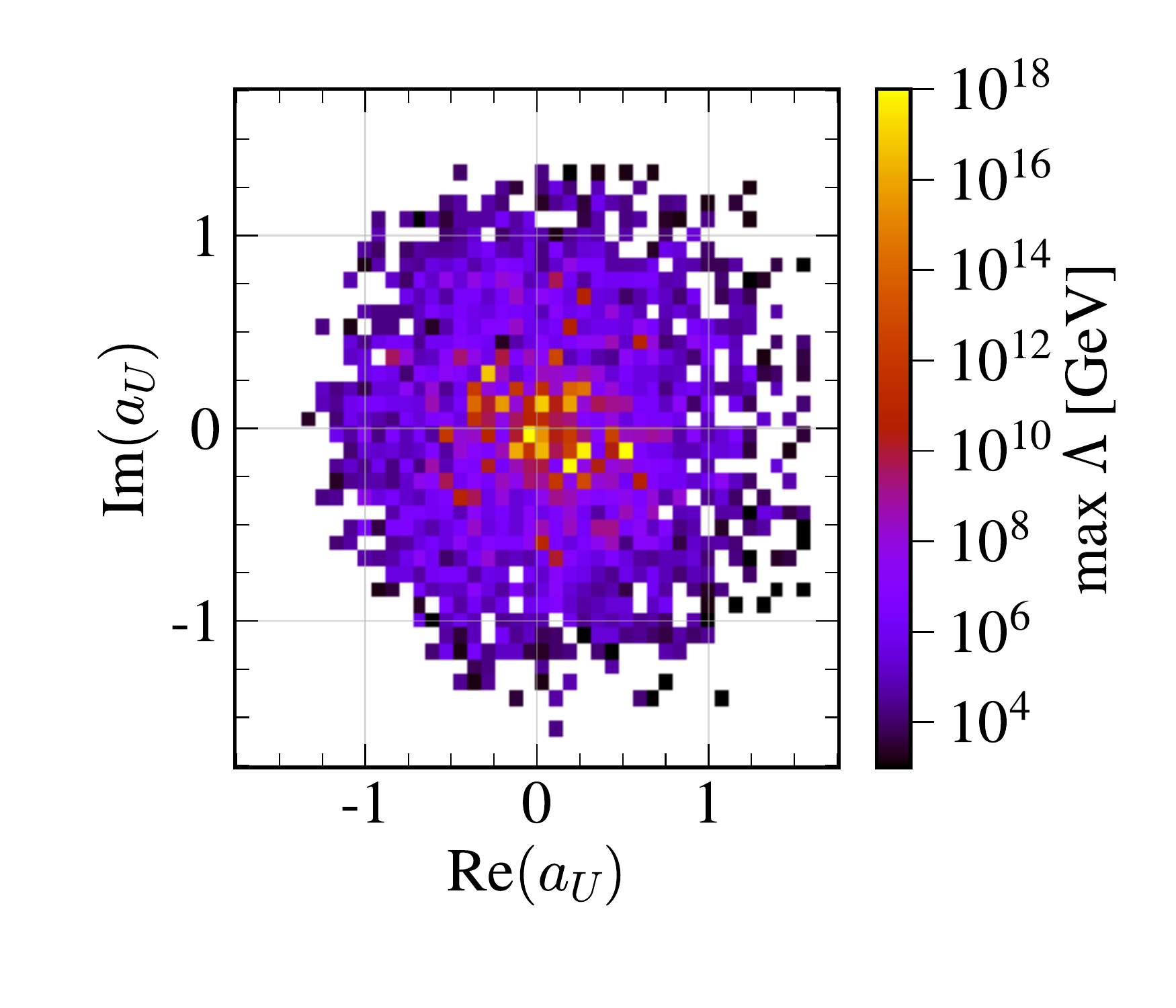}\\
\includegraphics[trim=1cm 1.2cm 5cm 0.5cm, clip, height=0.26\textwidth]
    {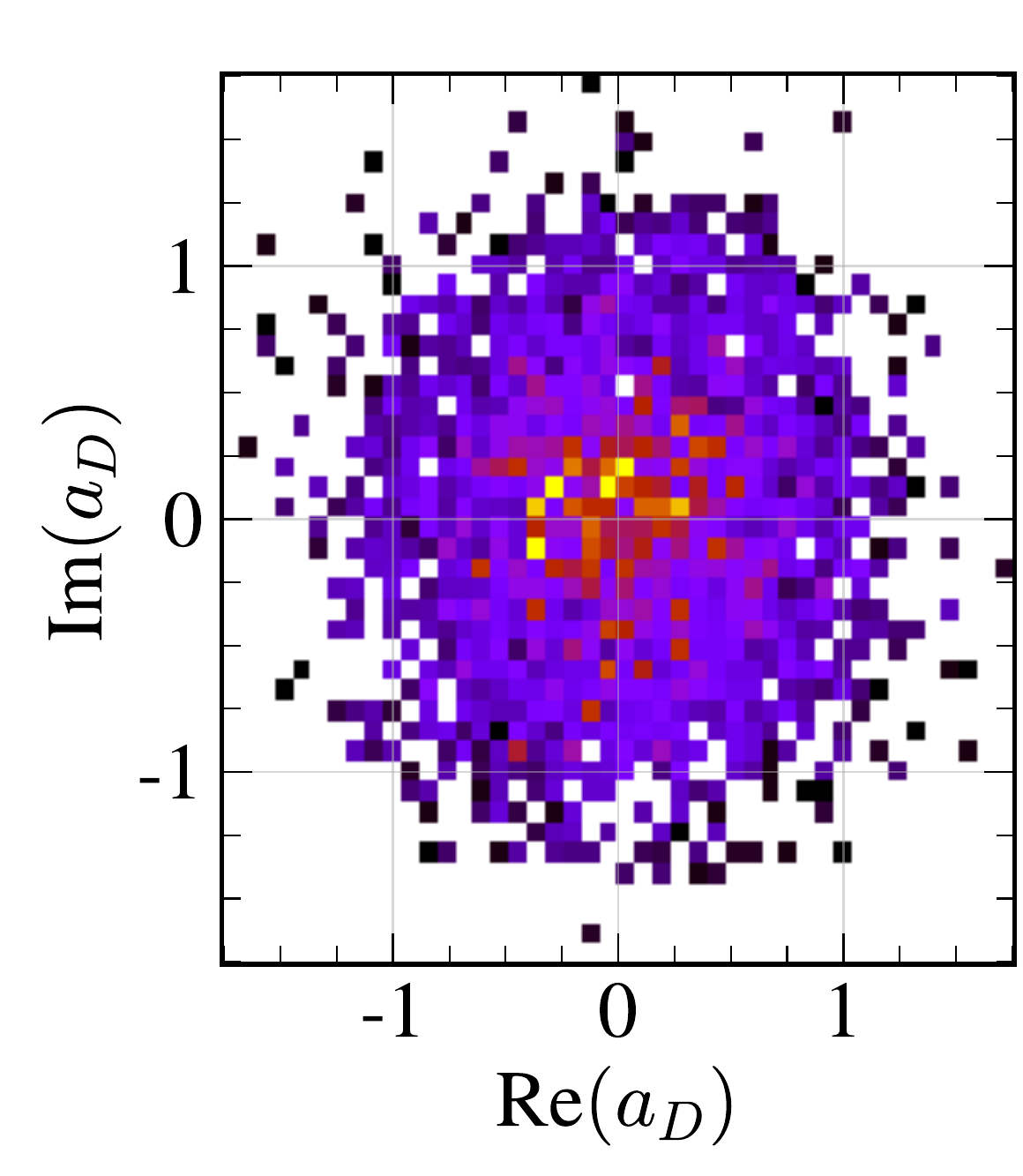} &
\includegraphics[trim=1cm 1.2cm 5cm 0.5cm,clip,height=0.26\textwidth]
    {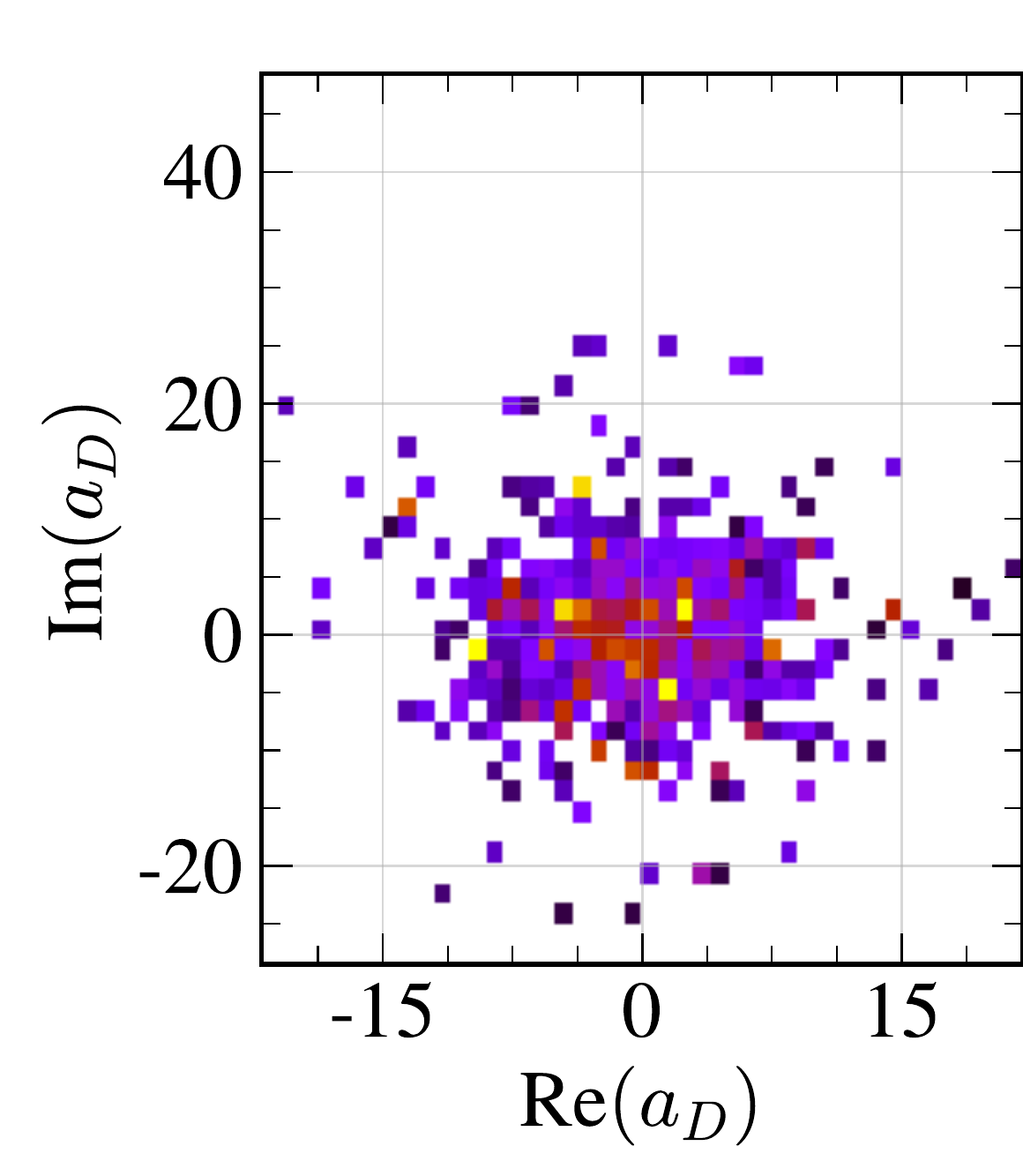} &
\includegraphics[trim=1cm 1.2cm 0.5cm 0.5cm,clip,height=0.26\textwidth]
    {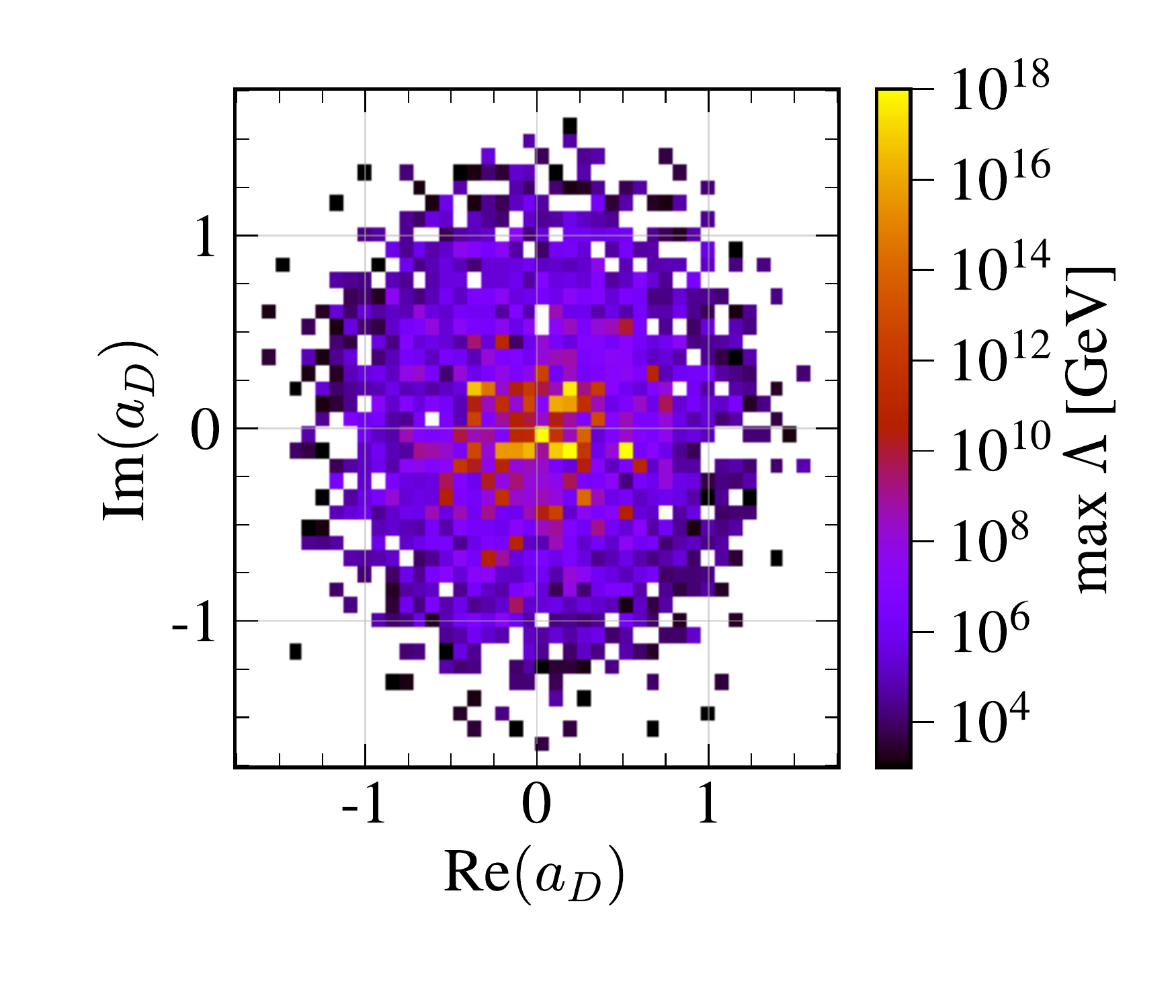}\\
\includegraphics[trim=1cm 1.2cm 5cm 0.5cm, clip, height=0.26\textwidth]
    {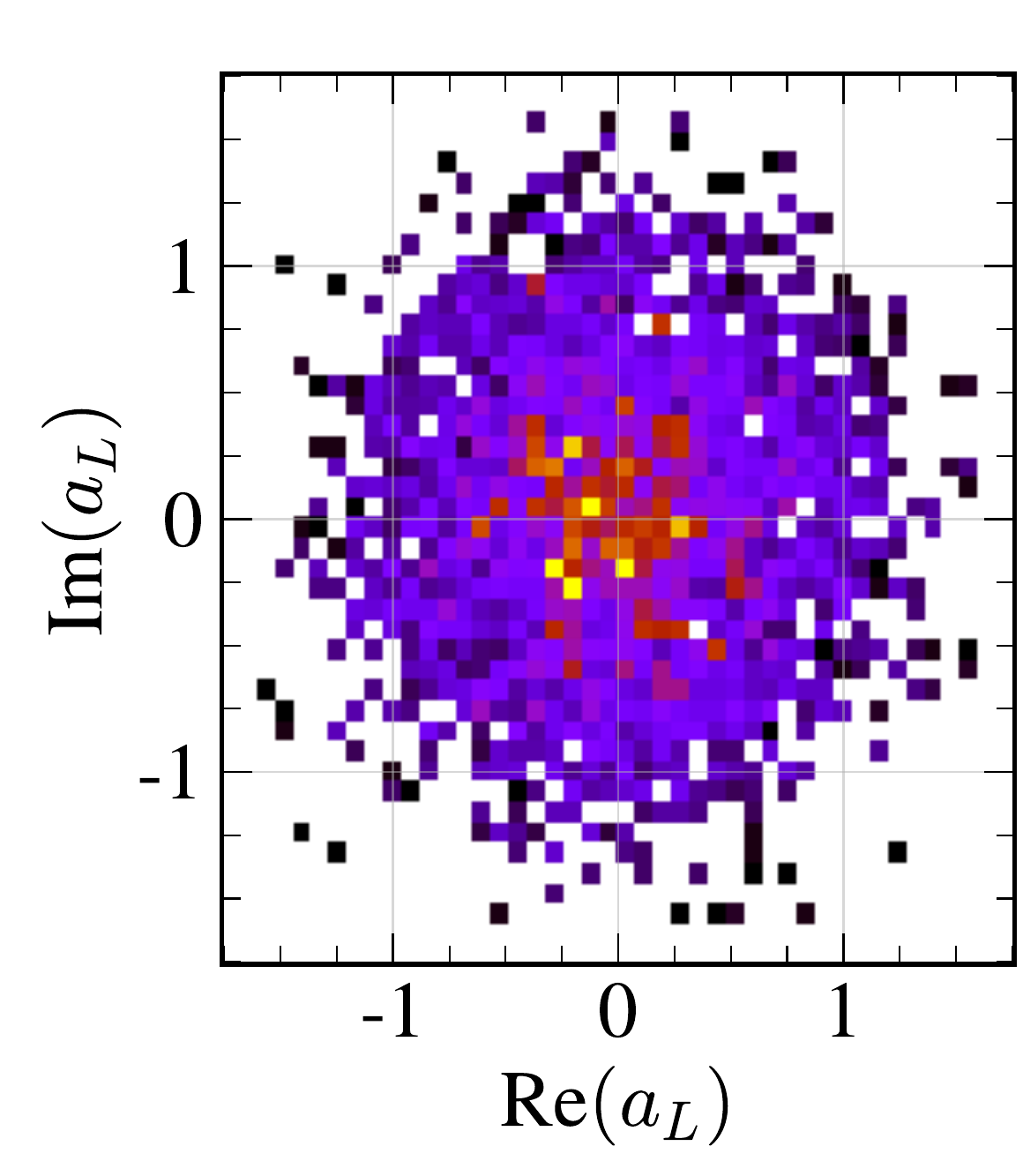} &
\includegraphics[trim=1cm 1.2cm 5cm 0.5cm,clip,height=0.26\textwidth]
    {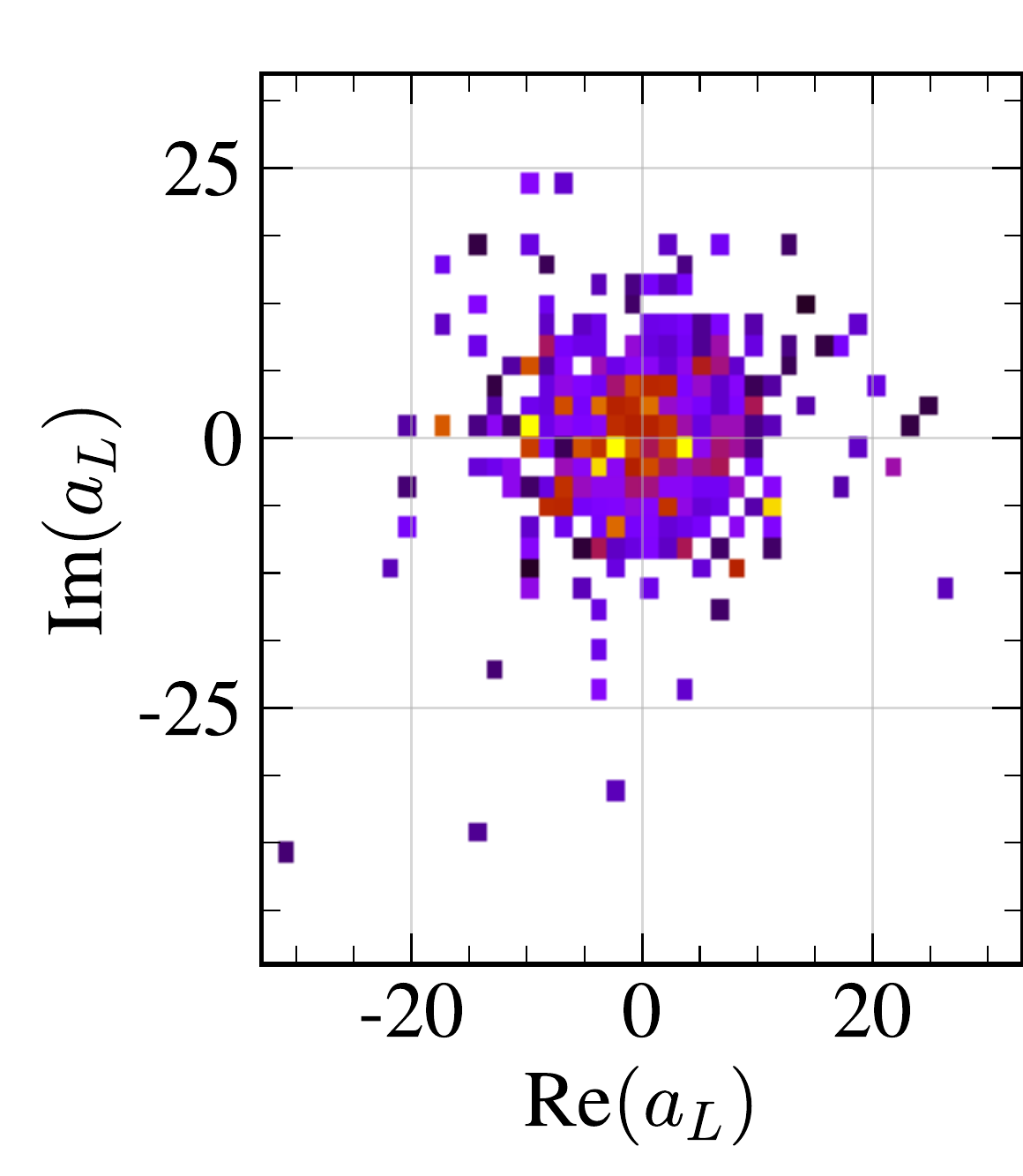} &
\includegraphics[trim=1cm 1.2cm 0.5cm 0.5cm,clip,height=0.26\textwidth]
    {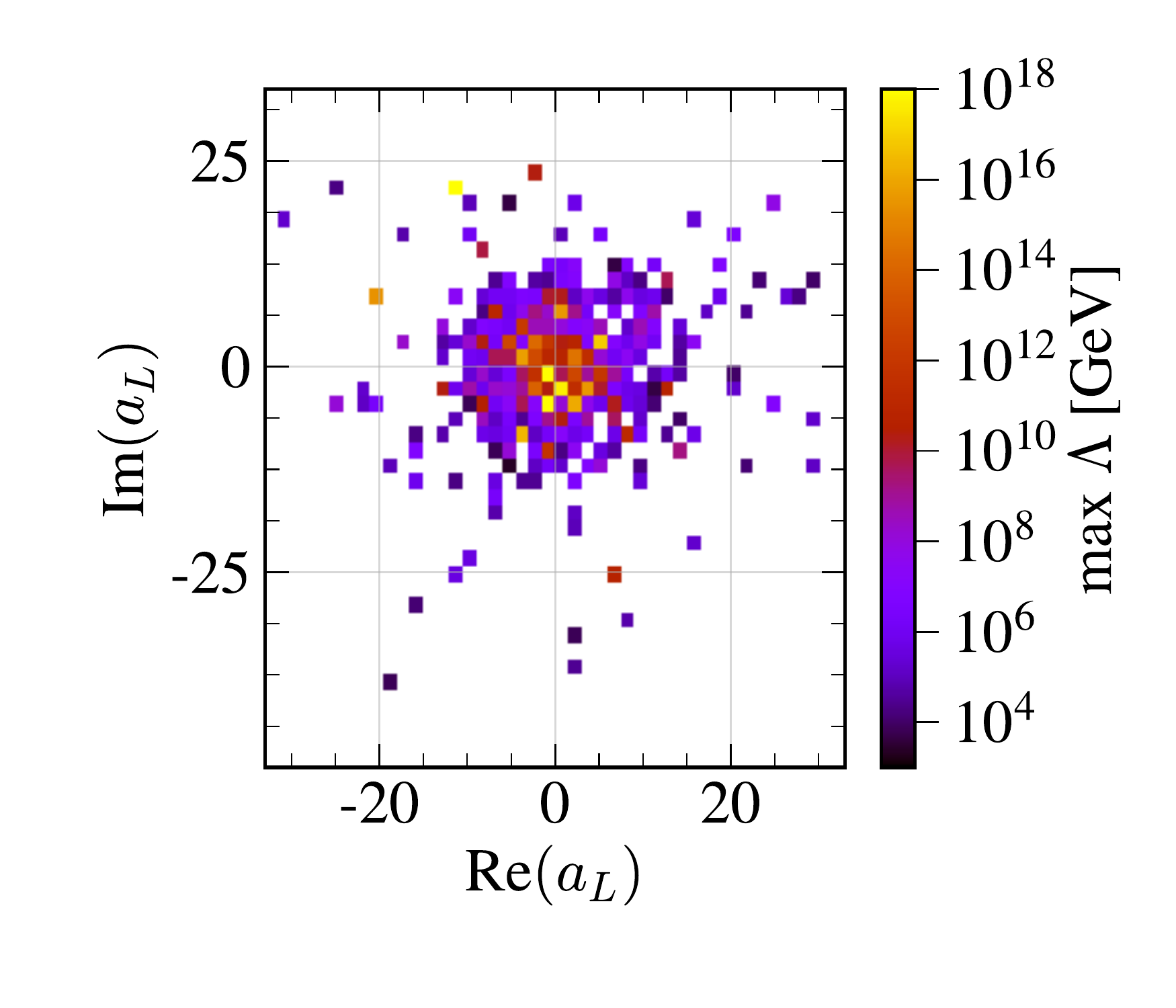}
\end{tabular}
    \caption{The maximum breakdown energy as a function of $a_U$, $a_D$ and
    $a_L$ for type I (left), type II (middle) and type X
    (right).}
\label{fig:CPVIII1}
\end{center}
\end{figure}

\begin{figure}[h!]
\begin{center}
\textbf{Scenario III}\\
\begin{tabular}{ccc}
    \footnotesize{type I} & \footnotesize{type II} & \footnotesize{type X} \\
\includegraphics[trim=0.5cm 1.2cm 5cm 0.5cm,clip,height=0.3\textwidth]
    {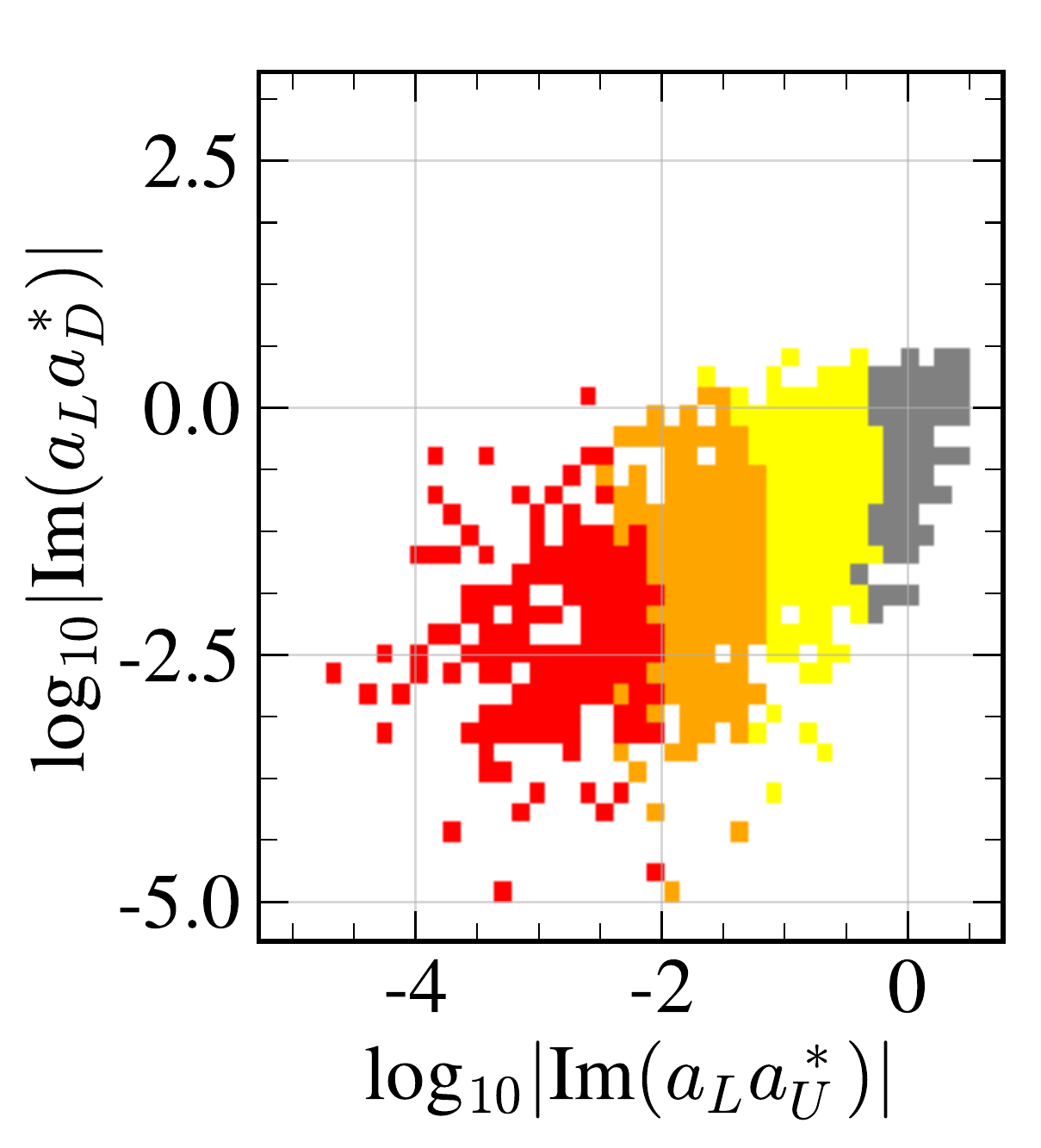} &
\includegraphics[trim=0.5cm 1.2cm 5cm 0.5cm,clip,height=0.3\textwidth]
    {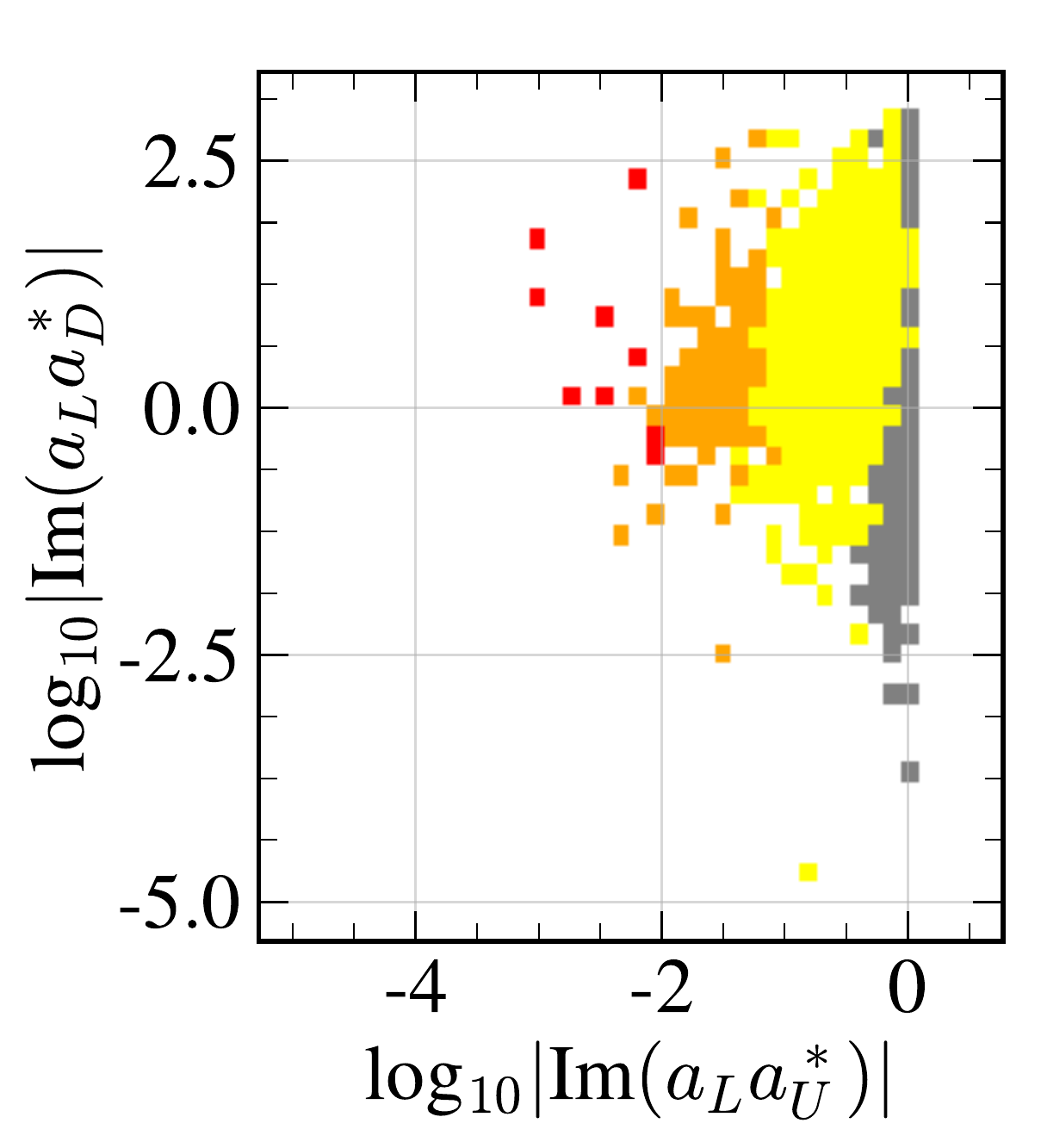} &
\includegraphics[trim=0.5cm 1.2cm 0.5cm 0.5cm,clip,height=0.3\textwidth]
    {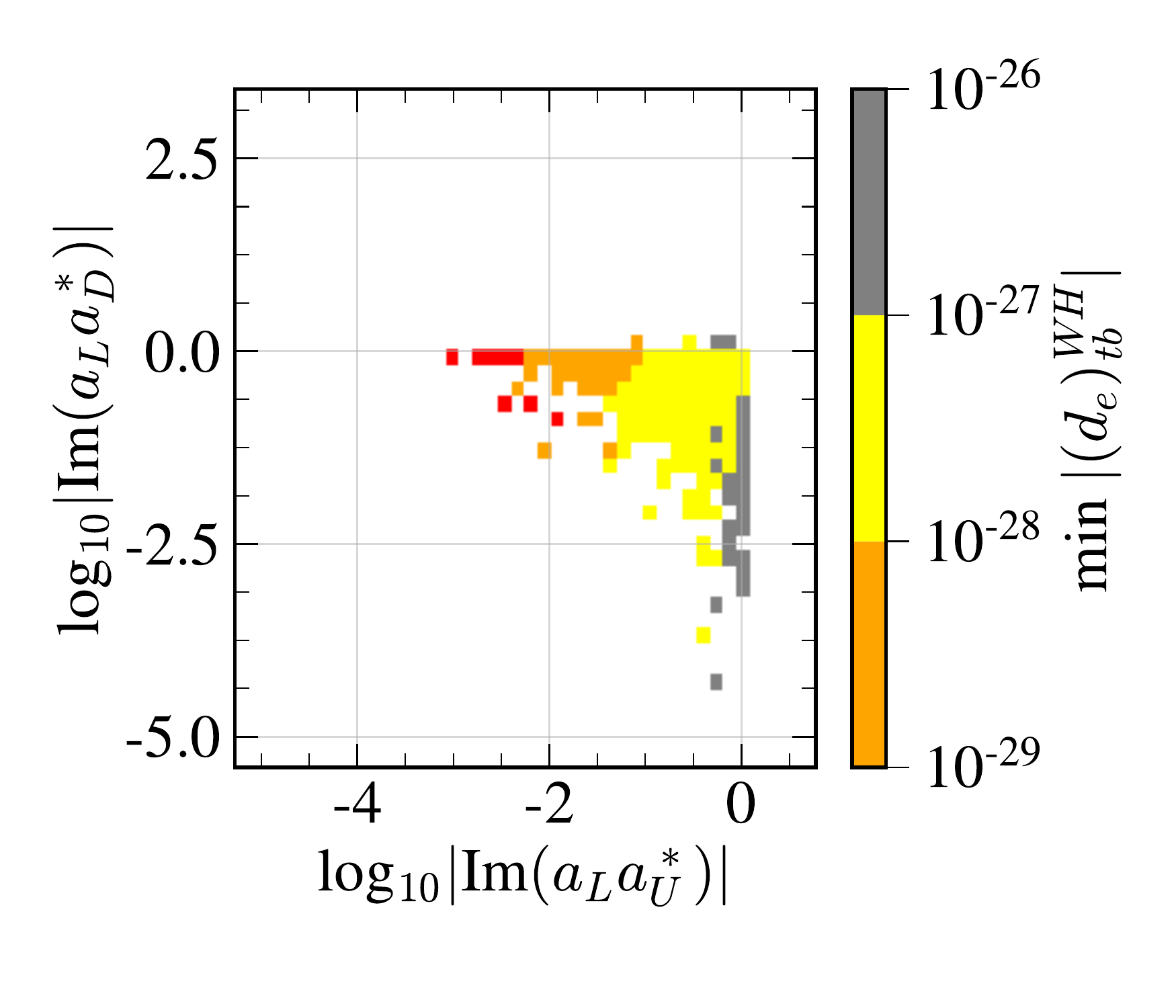}
\end{tabular}
    \caption{The minimum contribution from $(d_e)^{WH}_{tb}$ to the eEDM as a
    function of the imaginary parts of $a_L a_U^*$ and $a_La_D^*$ for type I
    (left), type II (middle) and type X (right).}
\label{fig:CPVIII2}
\end{center}
\end{figure}

\begin{figure}[h!]
\begin{center}
\textbf{Scenario III}\\
\begin{tabular}{ccc}
    \footnotesize{type I} & \footnotesize{type II} & \footnotesize{type X} \\
\includegraphics[trim=0.5cm 1.2cm 5cm 0.5cm, clip, height=0.3\textwidth]
    {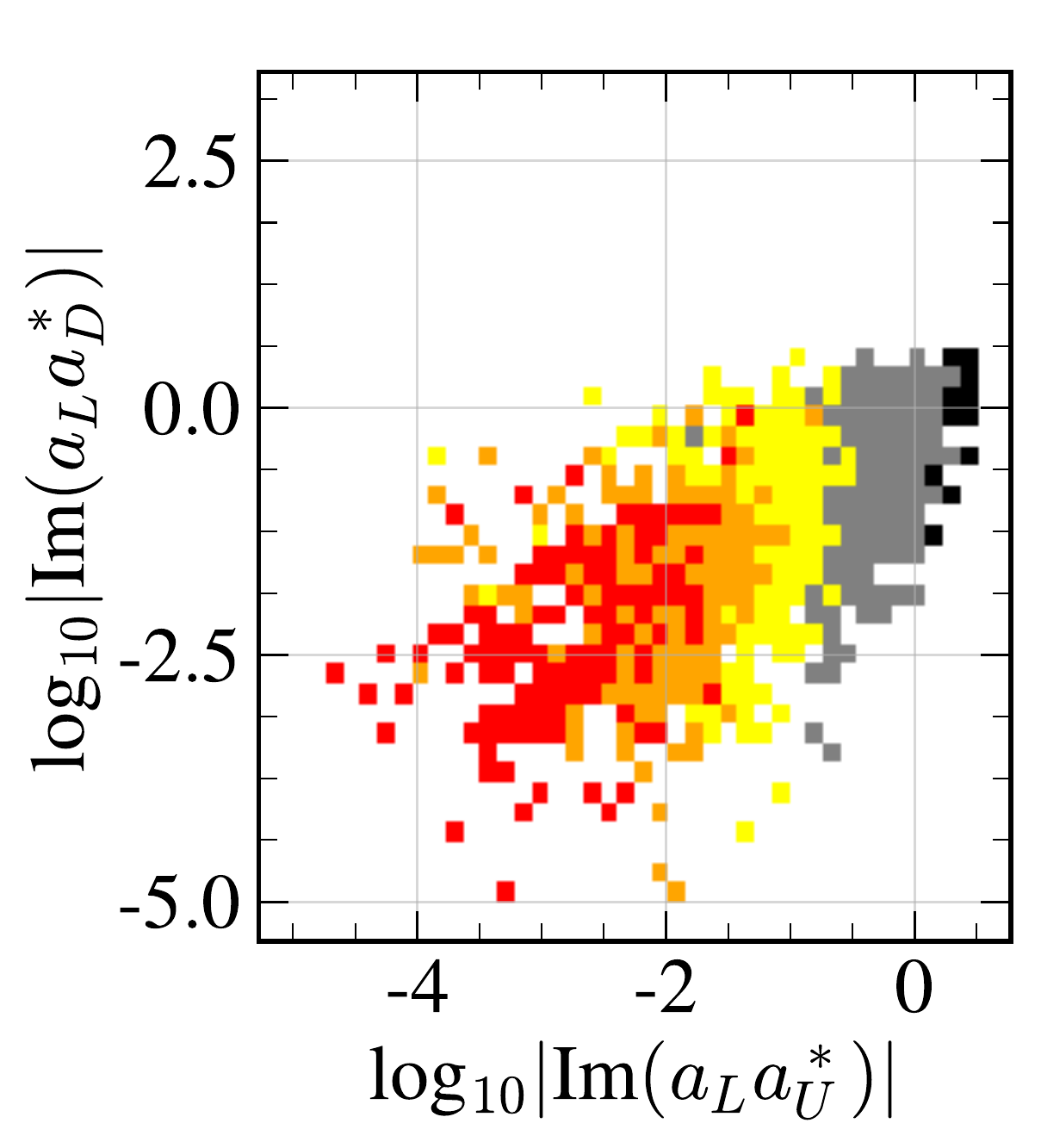} &
\includegraphics[trim=0.5cm 1.2cm 5cm 0.5cm,clip,height=0.3\textwidth]
    {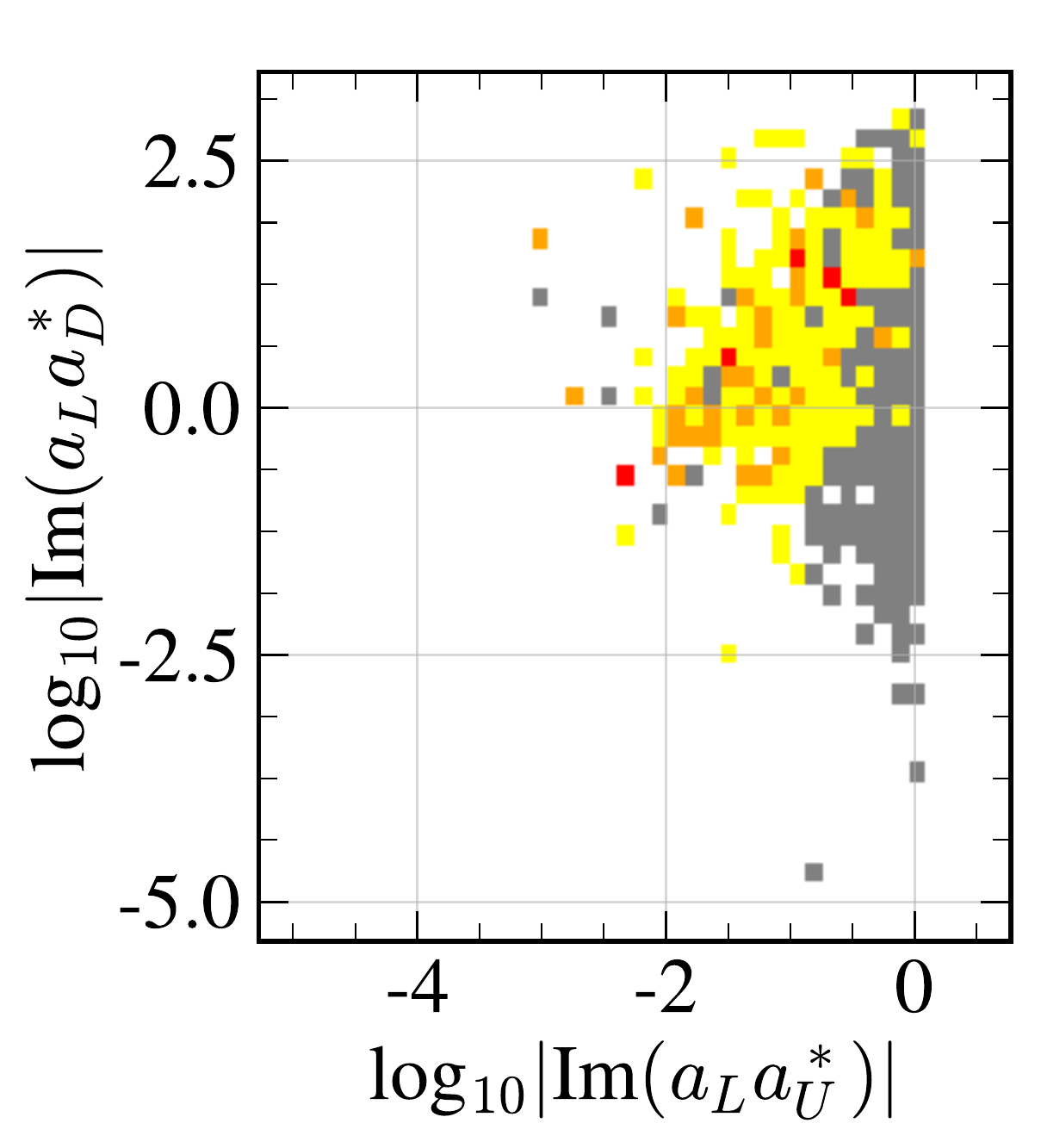}&
\includegraphics[trim=0.5cm 1.2cm 0.5cm 0.5cm,clip,height=0.3\textwidth]
    {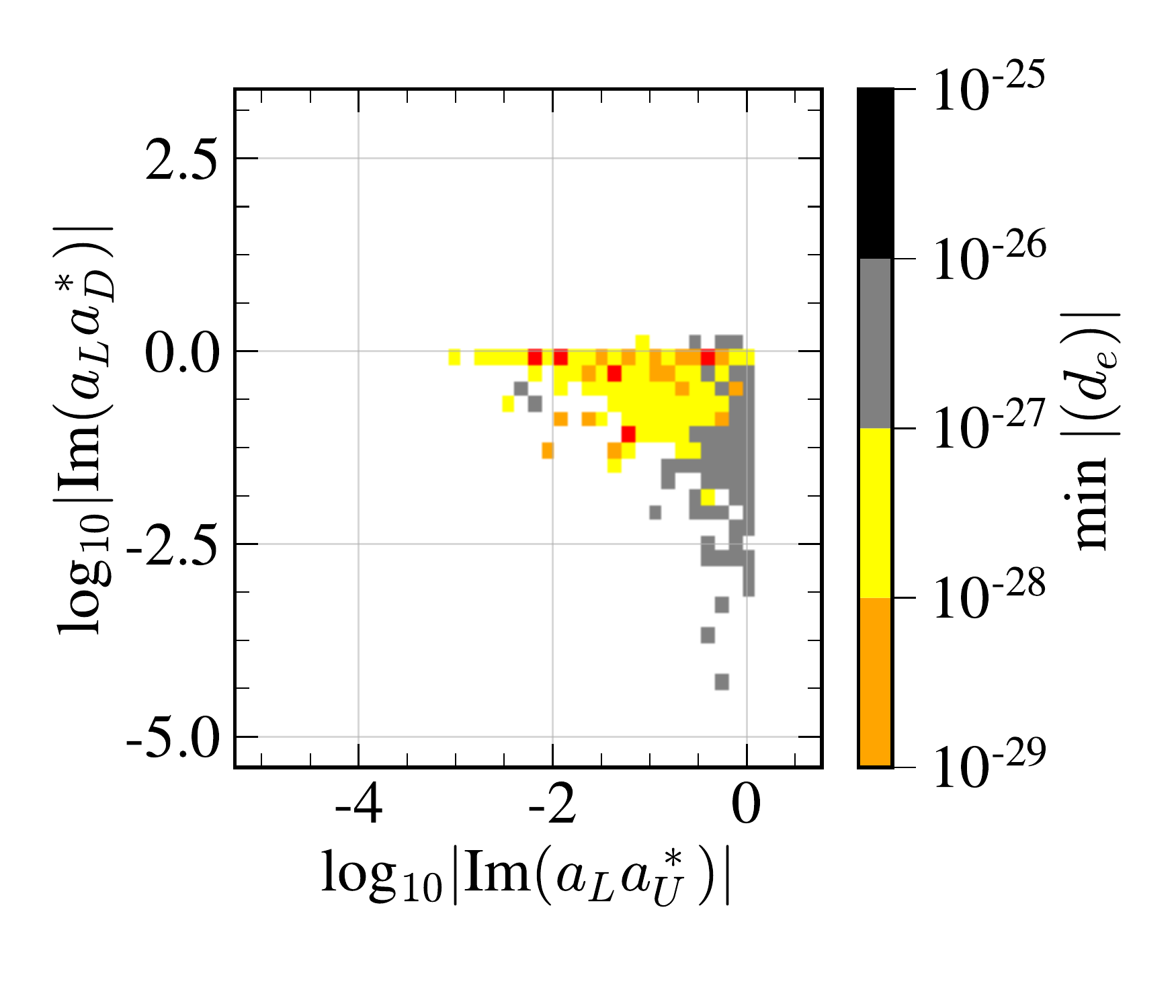} 
\end{tabular}
    \caption{The minimum eEDM as a function of the imaginary parts of $a_L
    a_U^*$ and $a_La_D^*$ for type I (left), type II (middle) and type X
    (right).}
\label{fig:CPVIII3}
\end{center}
\end{figure}

\begin{figure}[h!]
\begin{center}
\textbf{Scenario III}\\
\begin{tabular}{ccc}
    \footnotesize{type I} & \footnotesize{type II} & \footnotesize{type X} \\
\includegraphics[trim=0.5cm 1.2cm 5cm 0.5cm, clip, height=0.3\textwidth]
    {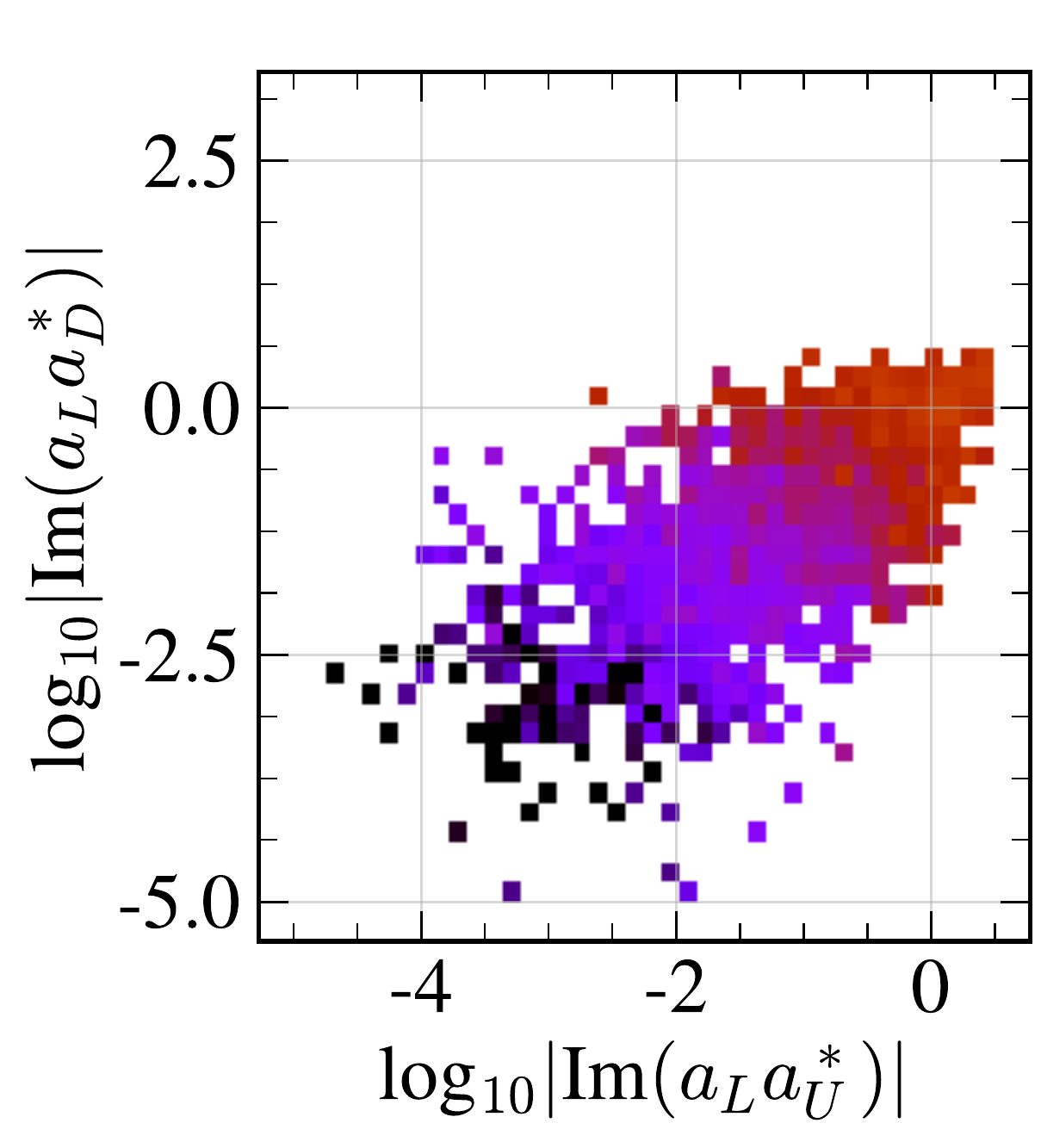} &
\includegraphics[trim=0.5cm 1.2cm 5cm 0.5cm,clip,height=0.3\textwidth]
    {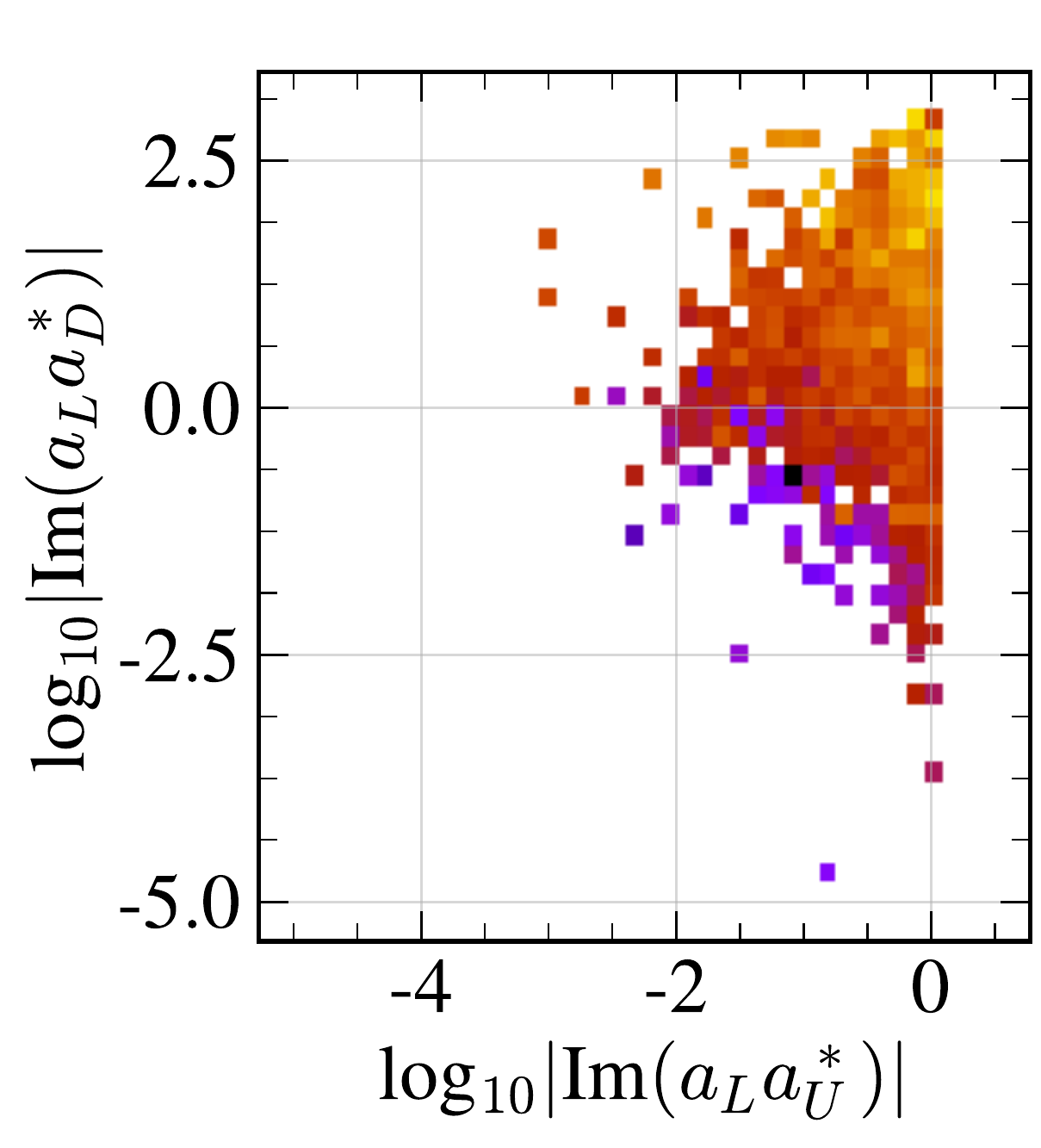}&
\includegraphics[trim=0.5cm 1.2cm 0.5cm 0.5cm,clip,height=0.3\textwidth]
    {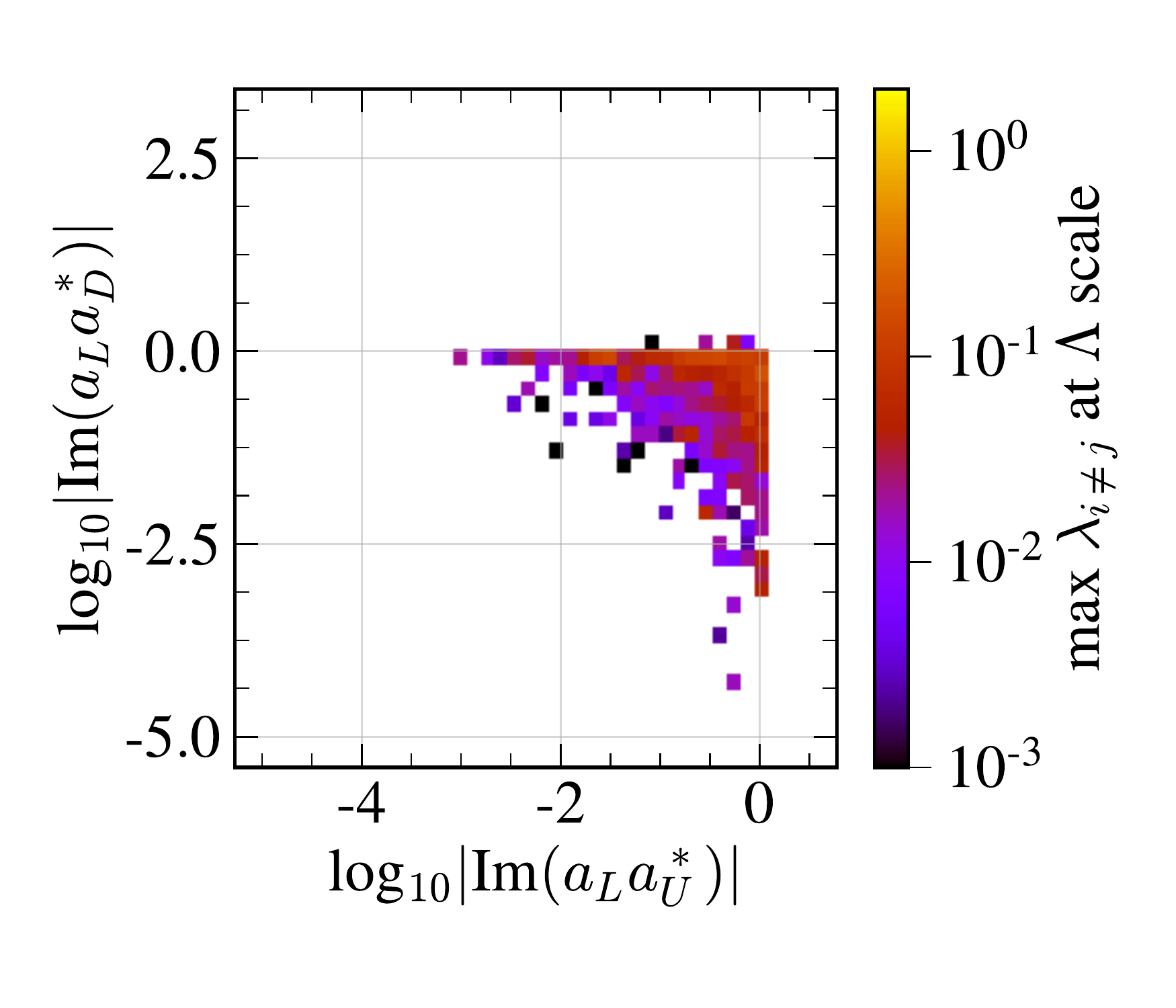} 
\end{tabular}
    \caption{The maximum induced non-diagonal Yukawa couplings in the Cheng-Sher
    ansatz as a function of the imaginary parts of $a_L
    a_U^*$ and $a_La_D^*$ for type I (left), type II (middle) and type X
    (right).}
\label{fig:CPVIII4}
\end{center}
\end{figure}

\begin{figure}[h!]
\begin{center}
\textbf{Scenario III}\\
\begin{tabular}{ccc}
    \footnotesize{type I} & \footnotesize{type II} & \footnotesize{type X} \\
\includegraphics[trim=0.5cm 1.2cm 5cm 0.5cm, clip, height=0.3\textwidth]
    {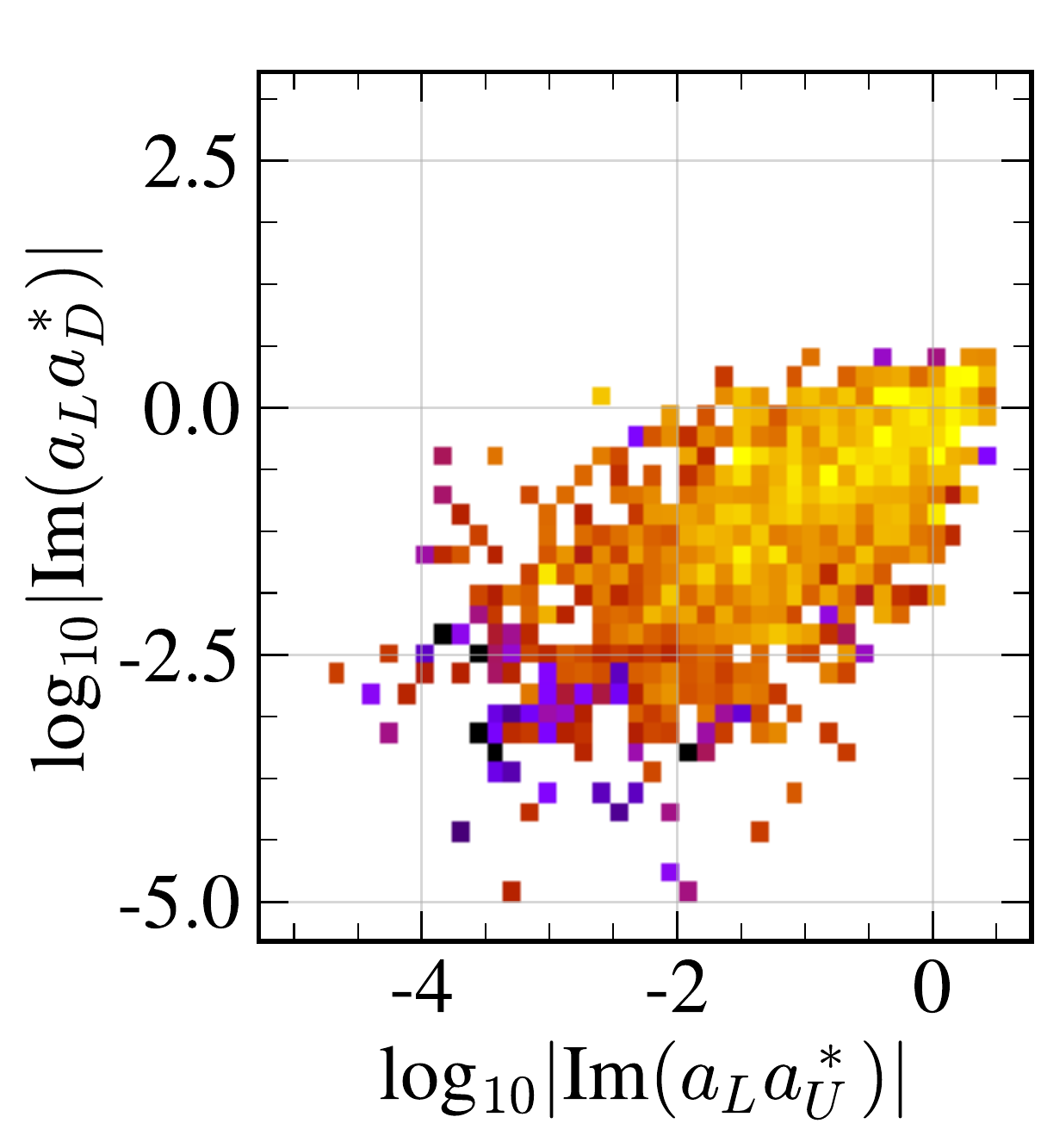} &
\includegraphics[trim=0.5cm 1.2cm 5cm 0.5cm,clip,height=0.3\textwidth]
    {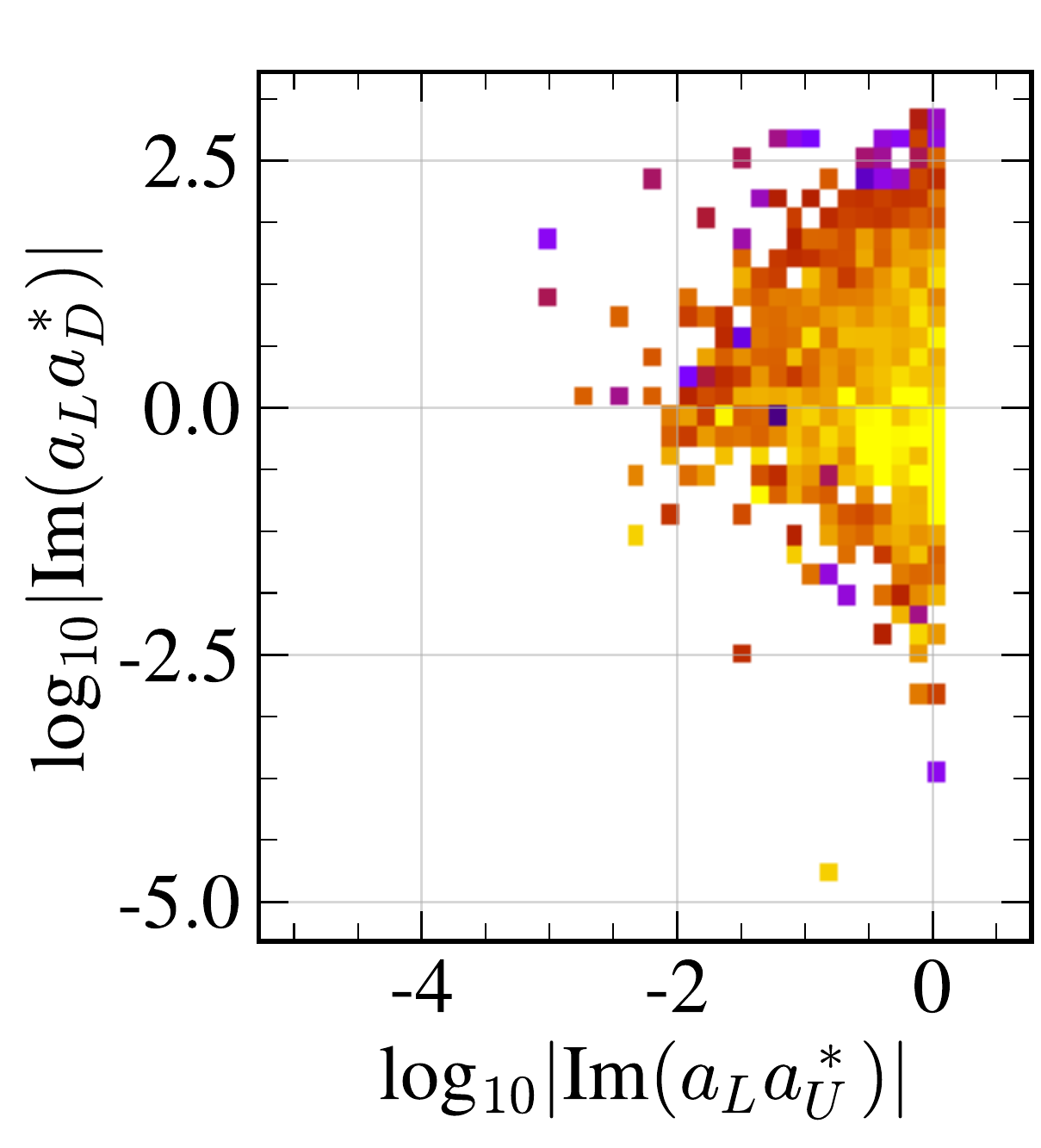}&
\includegraphics[trim=0.5cm 1.2cm 0.5cm 0.5cm,clip,height=0.3\textwidth]
    {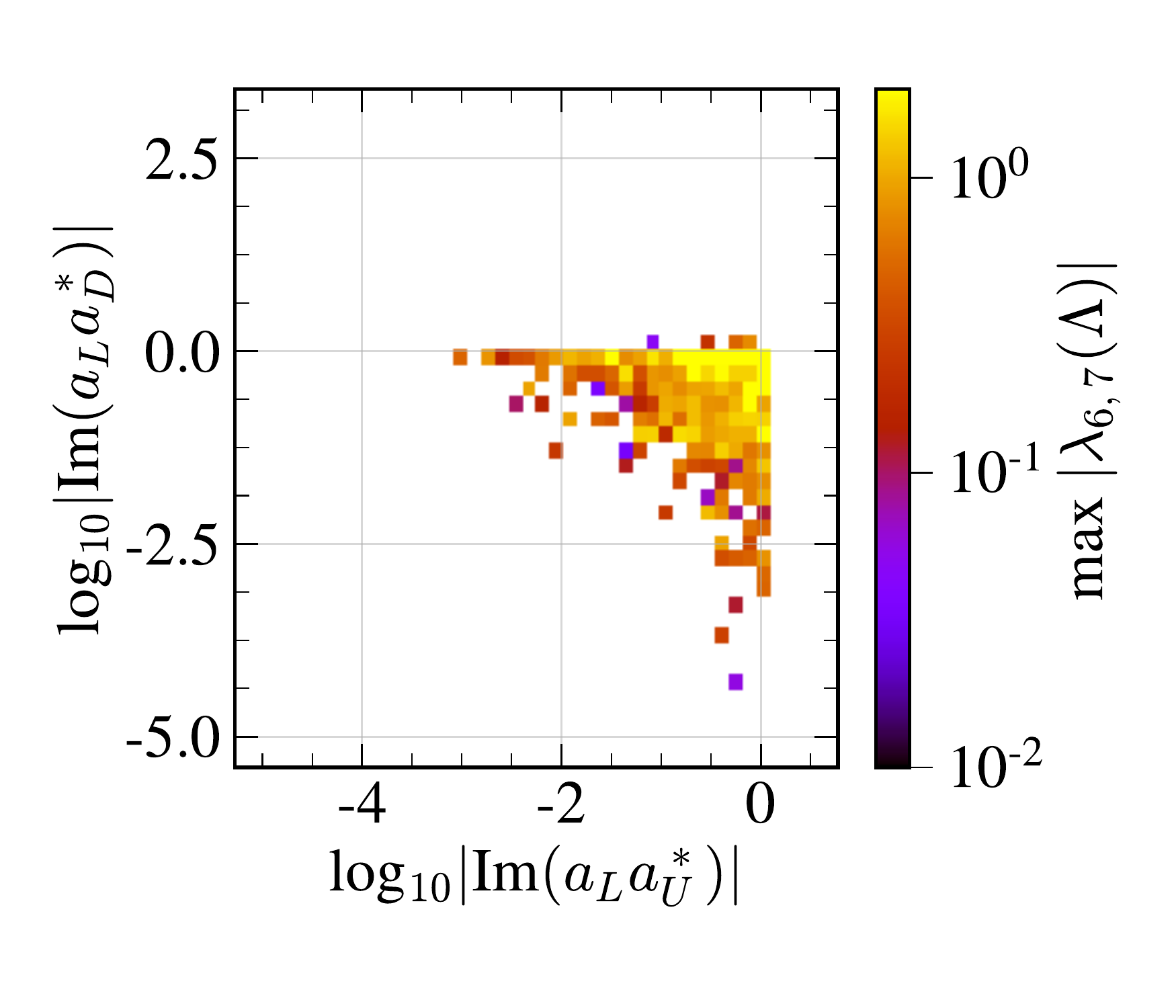} 
\end{tabular}
    \caption{The maximum induced hard \Zsym symmetry breaking parameters
    $\lambda_{6,7}$  as a function of the imaginary parts of $a_L
    a_U^*$ and $a_La_D^*$ for type I (left), type II (middle) and type X
    (right).}
\label{fig:CPVIII5}
\end{center}
\end{figure}

\begin{figure}[h!]
\begin{center}
\textbf{Scenario III}\\
\begin{tabular}{ccc}
    \footnotesize{type I} & \footnotesize{type II} & \footnotesize{type X} \\
\includegraphics[trim=0.5cm 1.2cm 5cm 0.5cm, clip, height=0.3\textwidth]
    {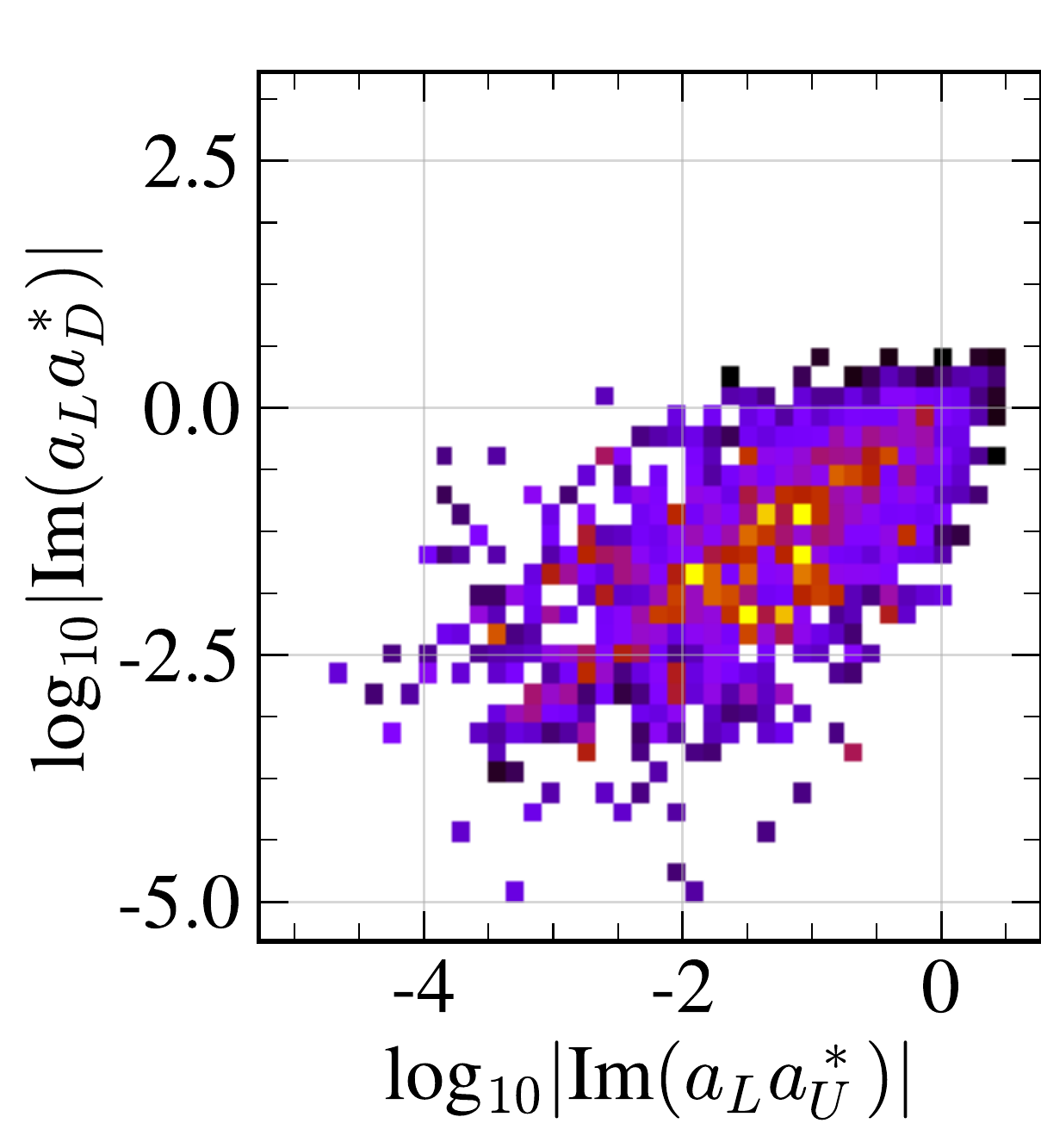} &
\includegraphics[trim=0.5cm 1.2cm 5cm 0.5cm,clip,height=0.3\textwidth]
    {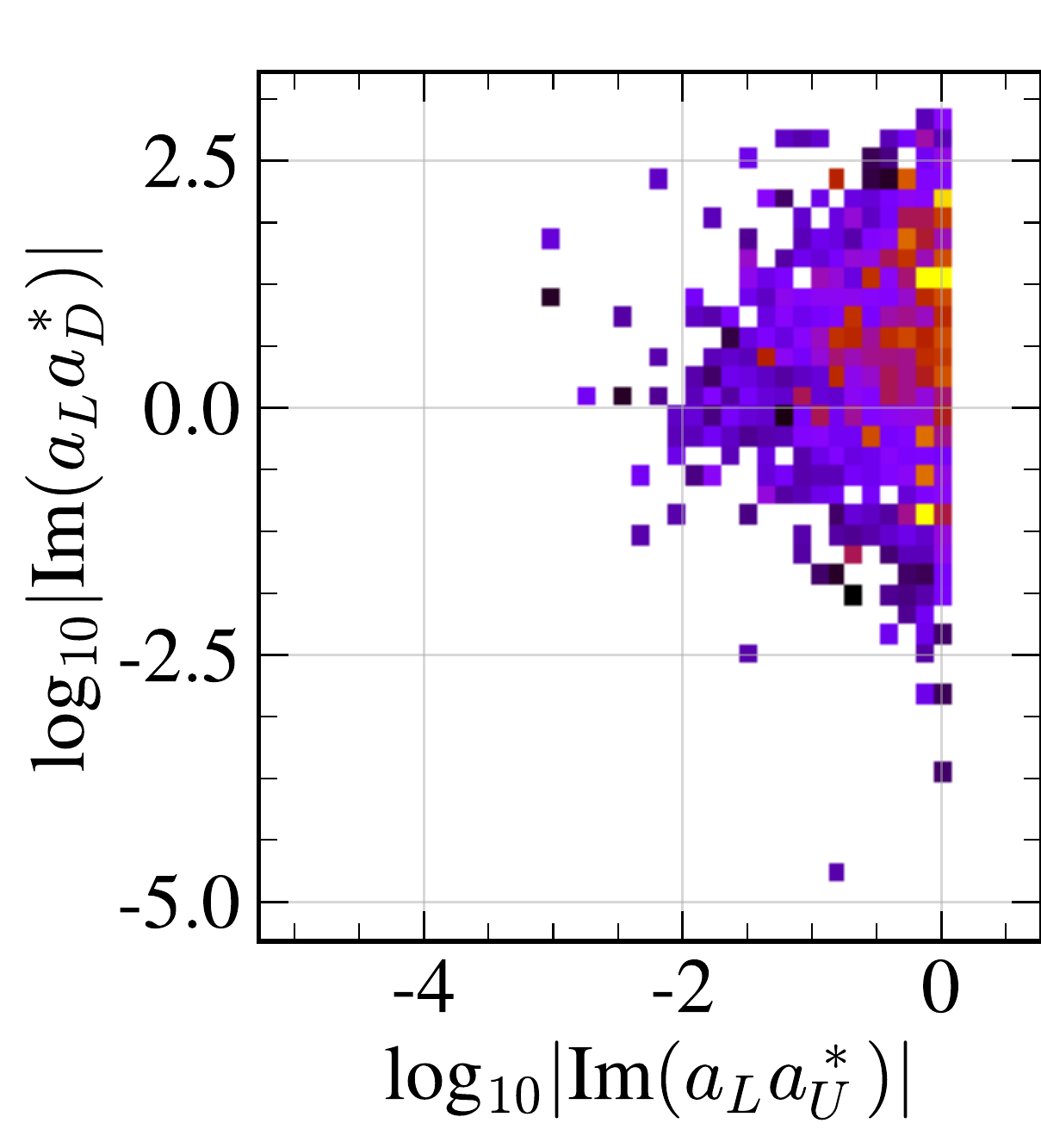}&
\includegraphics[trim=0.5cm 1.2cm 0.5cm 0.5cm,clip,height=0.3\textwidth]
    {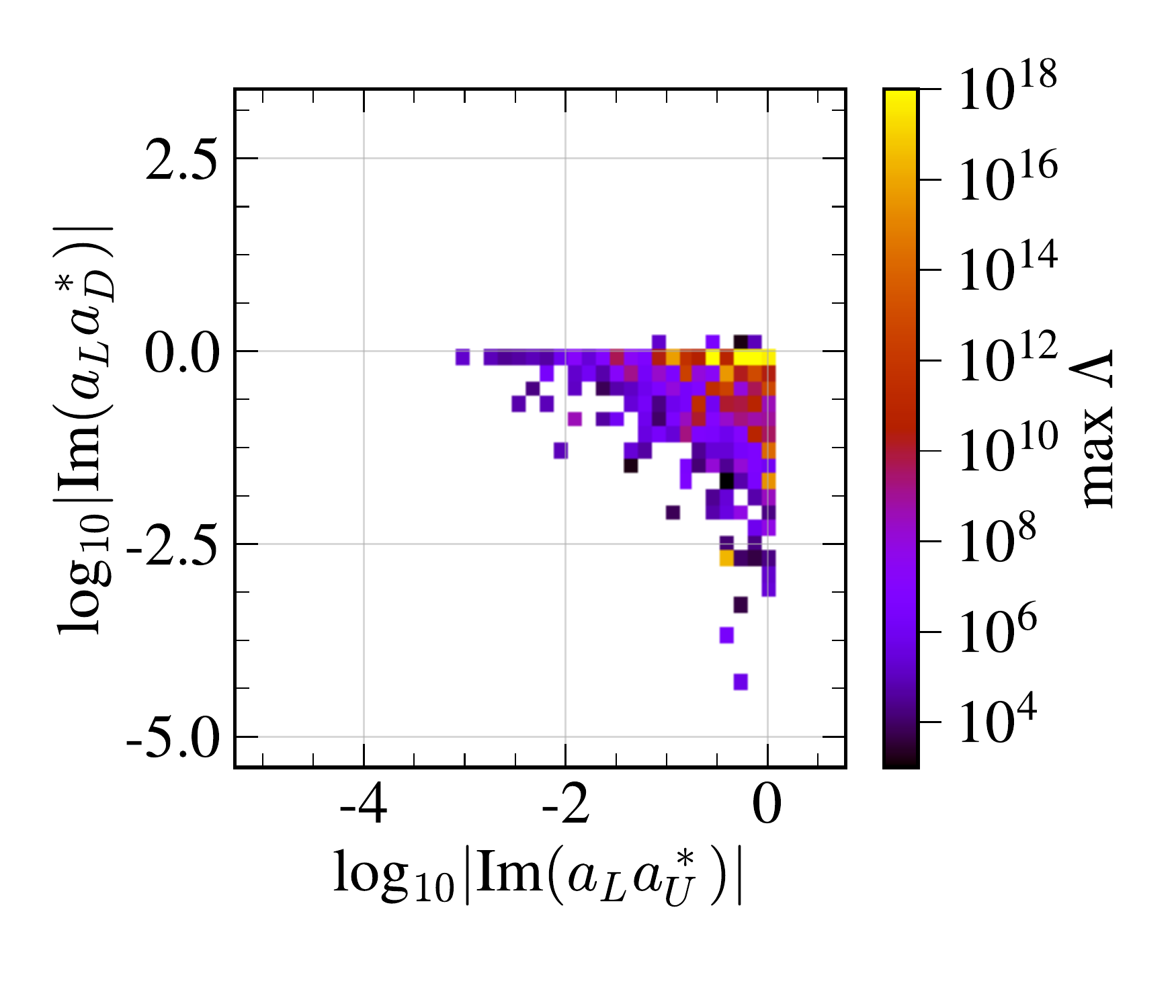} 
\end{tabular}
    \caption{The maximum breakdown energy as a function of the imaginary parts of $a_L
    a_U^*$ and $a_La_D^*$ for type I (left), type II (middle) and type X
    (right).}
\label{fig:CPVIII6}
\end{center}
\end{figure}

The maximum breakdown energy as function of the aligned coefficients $a_F$ is
shown in \fig{CPVIII1} for the three different types. The results are largely
independent on the phase of each $a_F$, while the regions around zero tend to be
better for higher breakdown energies. For type I, this means that
$\tan\beta$ has to be quite large since $|a_F| = 1/\tan\beta$, whereas for type
II and X there is a balance between $|a_U| = 1/\tan\beta$ being small and at the
same time $|a_D| =\tan\beta$ and/or $|a_L| = \tan\beta$ not being too large
($\lesssim 10$).

As measure of the \CP~violation, we use the base invariant quantities Im$(a_L
a_U^*)$ and Im$(a_L a_D^*)$. These give the contribution $(d_e)^{WH}_{tb}$ in
\eq{deWH}; which is the largest contribution in $\sim 10$ \% of the parameter
points. The term involving $a_U$ is the dominant one because of the top mass and
therefore one gets a very clear limit on Im$(a_La_U^*)$ as can be seen in
\fig{CPVIII2}, where the single contribution $(d_e)^{WH}_{tb}$ is shown as a
function of Im$(a_La_U^*)$ and Im$(a_La_D^*)$. There, the Im$(a_La_U^*)$ needs to
be below $\sim 0.01$ for all types to be within the limits of
ACMEII\footnote{The hard limits on $a_La_U^*$ and $a_La_D^*$ for type II and X
arise from the ansatz $|a_U|=1/\tan\beta$ and $|a_{D,L}| = \tan\beta$.}. In \fig{CPVIII3} we show
the total eEDM as a function of the same parameters.  Although, there are some
cancellations that make the limit on Im$(a_La_U^*)$ fuzzier, the eEDM still
gives quite a severe constraint; especially for type I, although for type II and
X we are facing the problem of running out of statistics. For type II and X
there are some points with large Im$(a_L a_U^*)$ and Im$(a_L a_D^*)$ that still
give an allowed eEDM. There could even be regions for these last types that
could pass all constraints and still be valid all the way to the Planck scale;
although this requires more study.

Breaking the \Zsym symmetry in the Yukawa sector can give rise to non-diagonal
Yukawa couplings. In \fig{CPVIII4}, this is shown as a function of 
Im$(a_La_U^*)$ and Im$(a_La_D^*)$. Similarly as in scenario II, type II is
generating the most FCNCs in the RG evolution. The regions in type I that are
within the eEDM limits generate a very low amount of FCNCs.

The hard \Zsym symmetry breaking quartic couplings $\lambda_{6,7}$ are also
generated in general; as seen in \fig{CPVIII5}, where max$(\lambda_{6,7})$ as a
function of Im$(a_La_U^*)$ and Im$(a_La_D^*)$ is plotted.  All scenarios generate
size-able $\lambda_{6,7}$ easily.

%%%%%%%%%%%%%%%%%%%%%%%%%%%%%%%%%%%%%%%%%%%%%%%%%%%%%%%%%%%%%%%%%%%%%%%%%%%%%%%%
\section{Conclusions}\label{sec:conclusions}
%%%%%%%%%%%%%%%%%%%%%%%%%%%%%%%%%%%%%%%%%%%%%%%%%%%%%%%%%%%%%%%%%%%%%%%%%%%%%%%%

We have analyzed the \CP~violating 2HDM by performing numerical parameter scans
in three different physical scenarios. Using \code{2HDME}, we have performed
2-loop RG running to study the properties under RG evolution looking for Landau
poles as well as a breakdown of unitarity or stability. Experimental collider
data has been used to restrict the parameter space with the codes
\code{HiggsBounds} and \code{HiggsSignals} and we also checked the oblique
parameters $S$, $T$ and $U$. The amount of \CP~violation was constrained by
calculating the eEDM; which now is an implemented feature of \code{2HDME}.

The physical scenarios we have investigated start from a 2HDM with a softly
broken \Zsym symmetry and quartic couplings $|\lambda_i| \leq 2$, to which we
add additional sources of (hard) \Zsym breaking.

With a softly broken \Zsym symmetry and \CP~violation in the scalar potential,
we found that having $|\lambda_i|\leq 2$ gives an aligned 2HDM with the
alignment parameter
$q_{11}\rightarrow 1$ as the BSM Higgs masses become heavy. We also find that
the amount of \CP~violation
is severely constrained by the eEDM. The limit from ACMEII requires the base
invariant quantities Im$(Z_5^* Z_6^2)$ and Im$(Z_6Z_7^*)$ to be $\sim 0.1$ for
parameter points to be allowed. For the points that are valid up to the Planck
scale, these quantities are even further constrained, $\sim 0.01$.

We also investigated the complex 2HDM with a small hard breaking of the \Zsym
symmetry in the scalar potential by having non-zero $\lambda_{6,7}$ at the EW
scale. While finding similar findings as the softly broken symmetry case, one
also gets the effect of inducing a \Zsym symmetry breaking in the Yukawa sector
during the RG running. Although, we found that there are no sizeable FCNCs being
produced, the \CP~violation in the scalar sector can spread to the Yukawa sector
by a non-trivial amount.

Lastly, we investigated three scenarios of aligned Yukawa sector based on type I, II
and X, but with complex coefficients. 
For the type I based scenario, we find that it is severely constrained by the 
eEDM, which requires Im$(a_La_U^*)\lesssim 10^{-2}$.
This is most easily satisfied if $\tan\beta$ is large.
For the type II and X based scenarios, the constraint on Im$(a_La_U^*)$ tends
to be weaker due to cancellations between different contributions to the eEDM.
At the same time, these scenarios are in general much worse than that of type I.
During RG running, the symmetry breaking spreads to the scalar sector and
induces complex $\lambda_{6,7}$ as well as FCNCs. The FCNCs are, however, not
very large for parameter points that have an allowed eEDM.

%%%%%%%%%%%%%%%%%%%%%%%%%%%%%%%%%%%%%%%%%%%%%%%%%%%%%%%%%%%%%%%%%%%%%%%%%%%%%%%%
\newpage
\begin{acknowledgments}

The authors would like to thank Nils Hermansson Truedsson for useful
    discussions.

This work is supported in part by the Swedish Research Council grants contract
numbers 621-2013-4287 and 2016-05996 and by the European Research Council (ERC)
under the European Union's Horizon 2020 research and innovation programme (grant
agreement No 668679).

\end{acknowledgments}
%%%%%%%%%%%%%%%%%%%%%%%%%%%%%%%%%%%%%%%%%%%%%%%%%%%%%%%%%%%%%%%%%%%%%%%%%%%%%%%%

%%%%%%%%%%%%%%%%%%%%%%%%%%%%%%%%%%%%%%%%%%%%%%%%%%%%%%%%%%%%%%%%%%%%%%%%%%%%%%%%
\appendix
%%%%%%%%%%%%%%%%%%%%%%%%%%%%%%%%%%%%%%%%%%%%%%%%%%%%%%%%%%%%%%%%%%%%%%%%%%%%%%%%

%%%%%%%%%%%%%%%%%%%%%%%%%%%%%%%%%%%%%%%%%%%%%%%%%%%%%%%%%%%%%%%%%%%%%%%%%%%%%%%%
\section{Generic basis in scenario I}\label{app:GenBase}
%%%%%%%%%%%%%%%%%%%%%%%%%%%%%%%%%%%%%%%%%%%%%%%%%%%%%%%%%%%%%%%%%%%%%%%%%%%%%%%%

\begin{figure}[h!]
\begin{center}
\textbf{Scenario I}\\
\begin{tabular}{cc}
\includegraphics[trim=0.5cm 1.2cm 5cm 0.5cm, clip, height=0.3\textwidth]
    {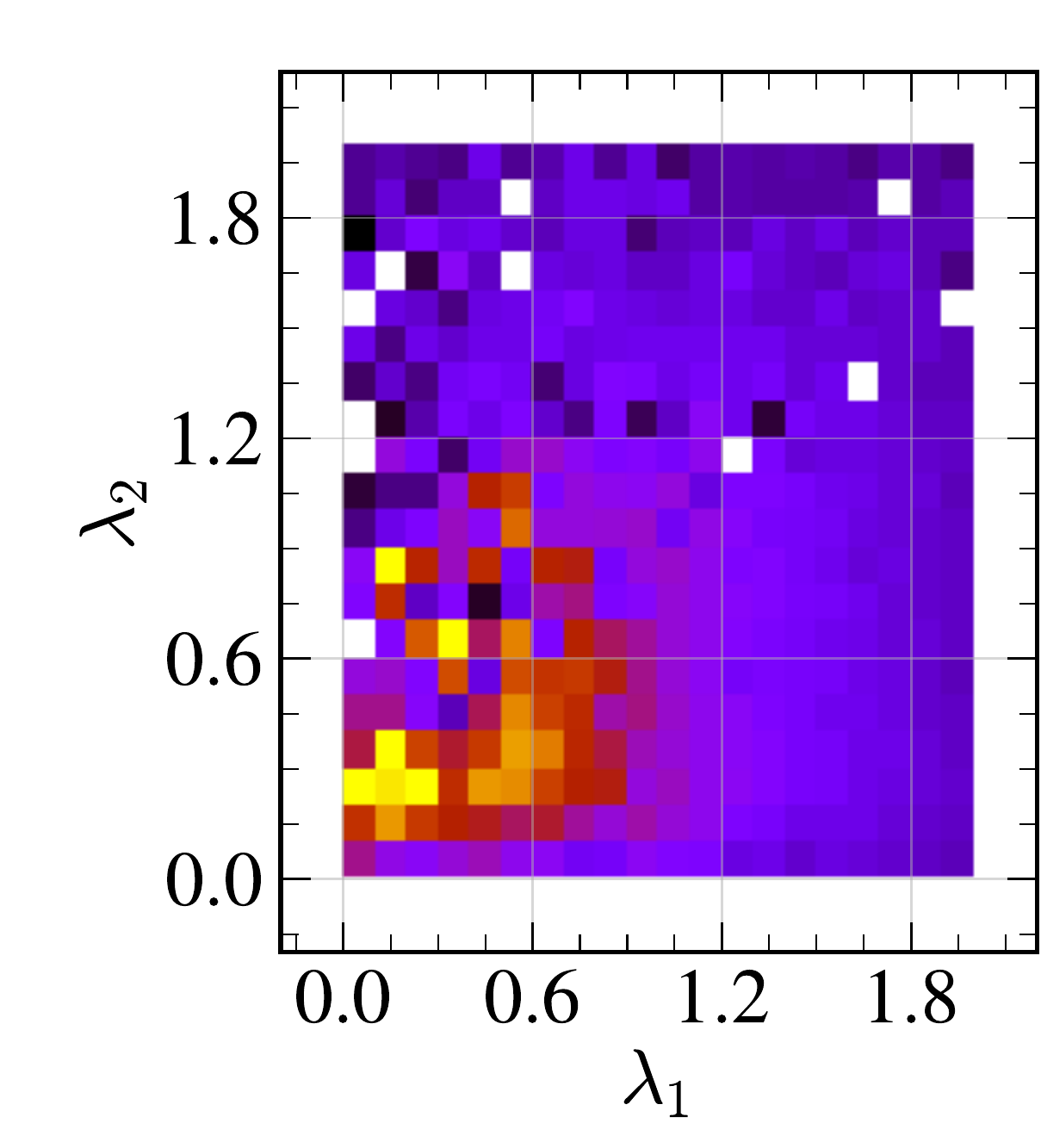} &
\includegraphics[trim=0.5cm 1.2cm 0.5cm 0.5cm,clip,height=0.3\textwidth]
    {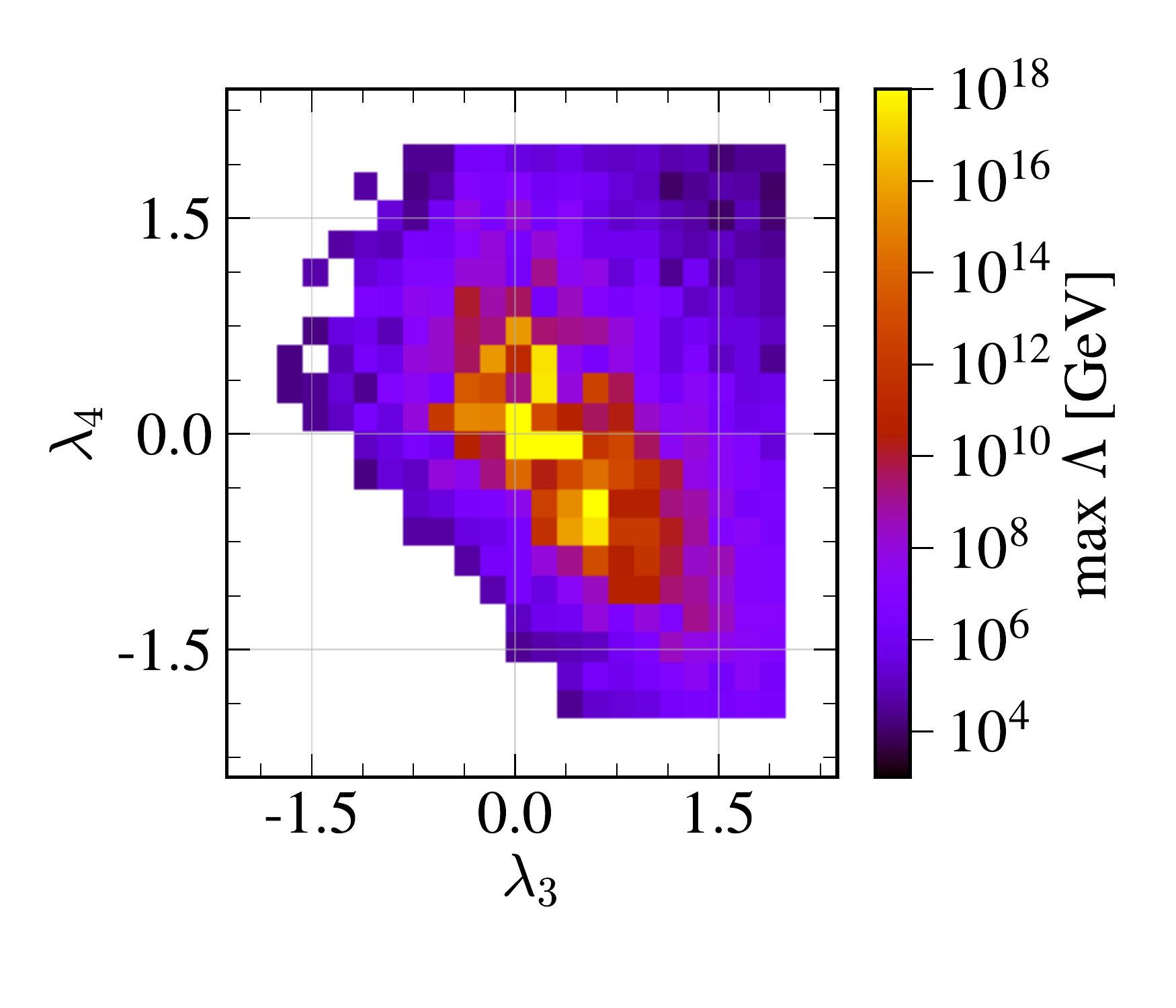} \\
\includegraphics[trim=0.5cm 1.5cm 5cm 0cm, clip, height=0.3\textwidth]
    {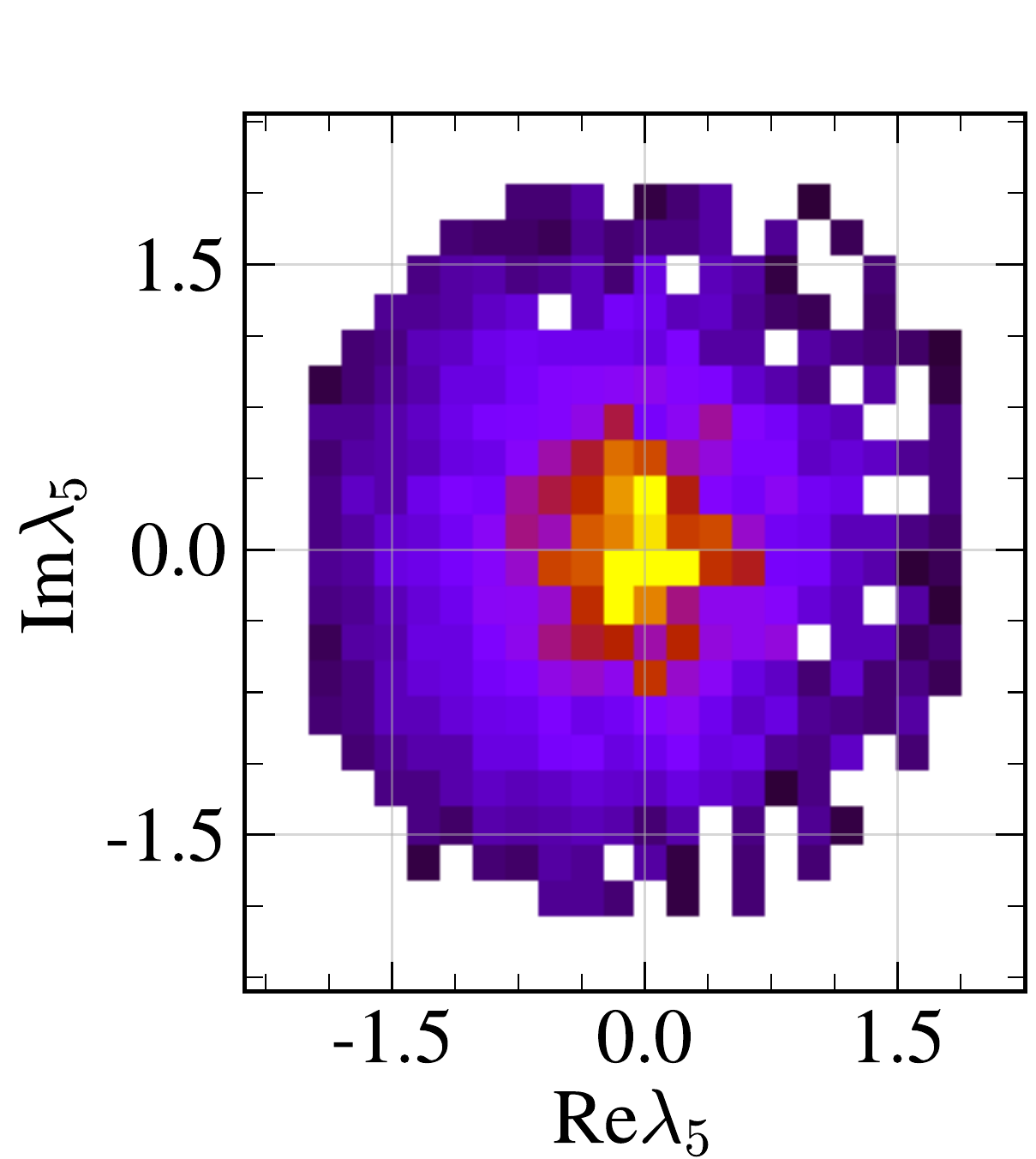} &
\includegraphics[trim=0.5cm 1.5cm 0.5cm 0cm,clip,height=0.3\textwidth]
    {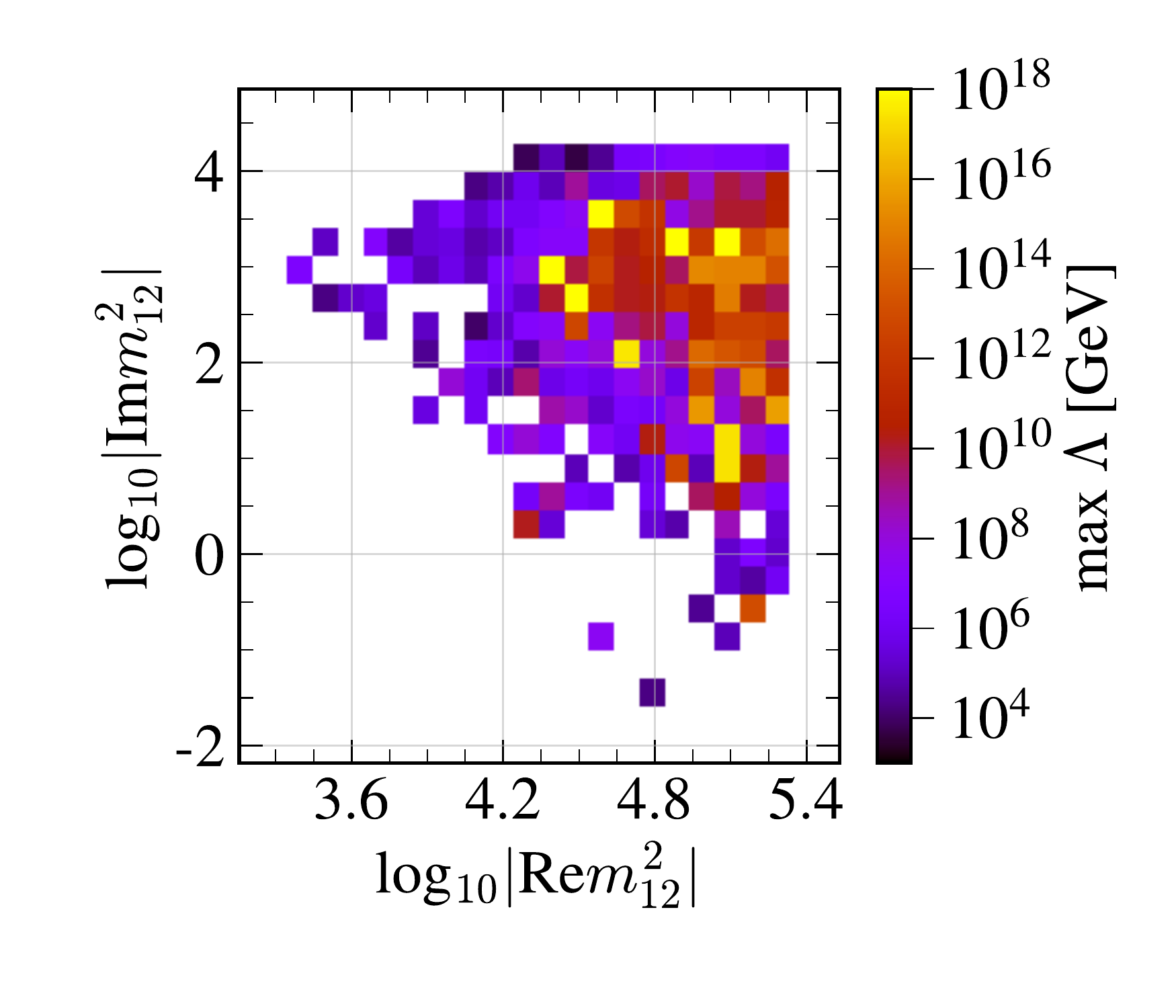}
\end{tabular}
    \caption{The maximum breakdown energy as a function of the parameters in the generic
    basis in scenario I with a type I \Zsym symmetry.}
\label{fig:CPVIgenBase}
\end{center}
\end{figure}

The parameter scans are generating the scalar potential in the generic basis
with flat random distributions. The breakdown energy scale as a function of
the quartic couplings and $m_{12}^2$ is shown in \fig{CPVIgenBase}.

%%%%%%%%%%%%%%%%%%%%%%%%%%%%%%%%%%%%%%%%%%%%%%%%%%%%%%%%%%%%%%%%%%%%%%%%%%%%%%%%
\section{Barr-Zee diagrams for EDM}\label{app:BarrZee}
%%%%%%%%%%%%%%%%%%%%%%%%%%%%%%%%%%%%%%%%%%%%%%%%%%%%%%%%%%%%%%%%%%%%%%%%%%%%%%%%

The largest contributions to light fermions' EDM comes from 2-loop
\textit{Barr-Zee} diagrams \cite{Barr:1990vd}, as shown in \fig{BarrZee}.  The
computation of these diagrams in the context of the 2HDM  was first done in
\mycite{Chang:1990sf}. A general framework to compute them is presented in
\mycite{Nakai:2016atk}.  The results for the 2HDM can be found in various
sources, \textit{e.g.} \mycite{Inoue:2014nva}.  Most of the literature, however,
deal with a softly broken \Zsym symmetry.  We found that these results are
incomplete, \textit{e.g.} when investigating the 2HDM with a complex aligned
Yukawa sector, and therefore compliment the EDM computation with diagrams
involving $W^\pm,H^\pm$ and a fermion loop.  This additional contribution as
well as all other contributions, for completeness, are presented below.

We denote each contribution to the electrons EDM as 
\begin{align}\label{eq:deDef}
	(d_e)^{VS}_l,
\end{align}
where $l$  denotes the particles of the 1-loop blob in \fig{BarrZee}.
Although some formulas below are written for a general loop particle, diagrams 
with light particles participating in the loops are very suppressed. Therefore
we only include the third generation in the total electron EDM; which then 
reduces to
\begin{align}
    d_e = \sum_h\left\{\sum_{f=t,b,\tau} \left[ (d_e)^{\gamma h}_f + (d_e)^{Z h}_f\right]
	 + (d_e)^{\gamma h}_W + 
	(d_e)^{\gamma h}_{H} 
    + (d_e)^{Z h}_{H}+ (d_e)_{hH}^{WH}\right\} + (d_e)^{W H}_{tb} 
	,
\end{align}
where $H$ denotes the charged Higgs.

The tree-level Higgs masses vary with renormalization scale. To circumvent this,
we always use the Higgs mass which satisfies $m_{h}(\mu) = \mu$, which is
independent on renormalization scale.

All the couplings in each contribution  are defined at the mass scale of
the heaviest particle participating in the loop diagram and we use full 2-loop
RGEs to run between the energy scales. The theoretical uncertainty in the
calculation is rather high since the running of couplings change some quantities
rather dramatically and the choice of renormalization scale for each diagram is
somewhat arbitrary. To get an estimate, we varied the renormalization scale for
each diagram to be twice or half the highest participating mass in the loop,
which can change $d_e$ by a factor of 2. This is, however, good enough for any
conclusions we make in this work.

\subsection*{Fermion loops}

For each fermion $f$, the contribution to the electrons EDM is
\begin{align}
	(d_e)^{\gamma h}_f =~& \frac{N_c Q_f^2 e^3}{32\pi^4m_f} 
\times \sum_{k=1}^3\left[f(z_f^k)(c_k^F)_{ff}(\tilde{c}_{k}^L)_{ee} 
 						+ g(z_f^k)(\tilde{c}_{k}^F)_{ff}(c_k^L)_{ee}\right],
\end{align}
where the loop functions are listed at the end of this section. We define the
ratio of masses as $z_x^k \equiv m_x^2/m_{h_k}^2$.

The similar diagram with a $Z$ boson instead of the internal $\gamma$ is 
\begin{align}
	(d_e)^{Z h}_f =~& \frac{N_ce g_{Zee}^Vg_{Zff}^V}{32\pi^4m_f}
\times \sum_{k=1}^3\left[\tilde{f}(z_f^k, m_f^2/m_Z^2)(c_k^F)_{ff}(\tilde{c}_{k}^L)_{ee} 
 						+ \tilde{g}(z_f^k,
                        m_f^2/m_Z^2)(\tilde{c}_{k}^F)_{ff}(c_k^L)_{ee}\right],
\end{align}
where $g_{Zff}^V= g(T_3^f - 2 Q_f \sin^2\theta_W)/(2\cos\theta_W)$.

For a general Yukawa sector, there is also an important contribution coming from a
$W^\pm-H^\pm$ diagram with a fermion loop. This single contribution is
investigated in \mycite{BowserChao:1997bb}, where they only keep the term
proportional $\rho^U$.  We have computed this again with the framework in
\mycite{Nakai:2016atk} and kept also the $\rho^D$ term. The result for the third
generation fermions is
\begin{align}\label{eq:deWH}
	(d_e)^{W H}_{tb} =~& 
    \frac{N_c e^3 |(V_{CKM})_{tb}|^2}{512\pi^4\sin^2\theta_W(m_{H^\pm}^2-m_W^2)}
        \int_0^1 \df x \left[Q_t x + Q_b(1-x)\right]\nn\\
    &\times \left\{\text{Im}\left[(\rho^L)_{ee}(\rho^{U})^*_{tt}\right]m_t x(1+x) +
                    \text{Im}\left[(\rho^L)_{ee}(\rho^{D})^*_{bb}\right]m_b x(1-x)\right\}\nn\\
    &\times \left[ G\left(\frac{m_t^2}{m_{H^\pm}^2}, \frac{m_b^2}{m_{H^\pm}^2}\right)
                - G\left(\frac{m_t^2}{m_{W}^2},
                \frac{m_b^2}{m_{W}^2}\right)\right].
\end{align}
This is the same formula as one gets for the case of \textit{magnetic dipole
moment},
derived in  \mycite{Ilisie:2015tra}; except that it is the imaginary part of the
couplings instead of the real part.

\subsection*{$W$ loops}

With a $W$ internal gauge boson, one gets the contributions
\begin{align}
	(d_e)^{\gamma h}_W =~& -\frac{e^3}{128 \pi^4 v} 
					\sum_{k=1}^3 \left[\left(6+\frac{1}{z_W^k}\right)f(z_W^k) 
									+ \left(10 -
                                    \frac{1}{z_W^k}\right)g(z_W^k)\right]g_{kVV}
                                    (\tilde{c}_k^L)_{ee}.
\end{align}
and
\begin{align}
    (d_e)^{Z h}_W =~& \frac{e g_{ZWW}g_{Zee}^V}{128 \pi^4 v} 
    \sum_{k=1}^3 \left[\left(6-\text{sec}^2\theta_W +
    \frac{2-\text{sec}^2\theta_W}{2z_W^k}\right)
    \tilde{f}(z_W^k,\cos^2\theta_W)\right.
    \nn\\
    &\left.+ \left(10 - 3 \text{sec}^2\theta_W -
    \frac{2-\text{sec}^2\theta_W}{2z_W^k}\right)\tilde{g}(z_W^k,
    \cos^2\theta_W)\right] g_{kVV} (\tilde{c}_{k}^L)_{ee},
\end{align}
where $g_{ZWW} \equiv e \cot\theta_W$.

\subsection*{Scalar loops}

The charged scalar contribution is
\begin{align}
	(d_e)^{\gamma h}_{H} =~& - \frac{e^3 v}{128 \pi^4m_{H^\pm}^2} \sum_{k=1}^3
 \left[f(z_H^k) - g(z_H^k)\right]\lambda_{kH^\pm} (\tilde{c}_{k}^L)_{ee}
\end{align}
and
\begin{align}
	(d_e)^{Z h}_{H} =~& - \frac{e v g_{Zee}^V g_{ZH^\pm}}{128 \pi^4m_{H^\pm}^2}
 \sum_{k=1}^3 \left[\tilde{f}(z_H^k,m_{H^\pm}^2/m_Z^2) 
- \tilde{g}(z_H^k, m_{H^\pm}^2/m_Z^2)\right]\lambda_{kH^\pm} (\tilde{c}_{k}^L)_{ee},
\end{align}
where $g_{Z H^\pm}\equiv e \cot\theta_W (1-\tan^2\theta_W)/2$.

The gauge invariant contribution from charged and neutral scalars has been
calculated in \mycite{Abe:2013qla},
\begin{align}
	(d_e)_{hH}^{WH} =~& \frac{e}{256 \pi^4 v} 
\times \sum_{k=1}^3 \left[ \frac{e^2}{2\sin^2\theta_W} 
        \mathcal{I}_4(m_{h_k}^2, m_{H^\pm}^2) g_{kVV}
        -\mathcal{I}_5(m_{h_k}^2, m_{H^\pm}^2)
        \lambda_{kH^\pm}\right](\tilde{c}_k^L)_{ee}.
\end{align}

\subsection*{Miscellaneous functions}

\begin{align}
	f(z) =~& \frac{z}{2}\int_0^1 \df x \frac{1-2x(1-x)}{x(1-x)-z}\log\left(\frac{x(1-x)}{z}\right),\\
	g(z) =~& \frac{z}{2}\int_0^1 \df x
    \frac{1}{x(1-x)-z}\log\left(\frac{x(1-x)}{z}\right),\\
	\tilde{f}(x,y) =~& \frac{yf(x)}{y-x} + \frac{xf(y)}{x-y},\\
	\tilde{g}(x,y) =~& \frac{yg(x)}{y-x} + \frac{xg(y)}{x-y},
\end{align}

\begin{align}
	G(r_1, r_2) = \frac{\log\left(\frac{r_1 x +
    r_2(1-x)}{x(1-x)}\right)}{x(1-x)-r_1 x - r_2 (1-x)},
\end{align}

\begin{align}
	\mathcal{I}_{4,5}(m_1^2,m_2^2) =~& \frac{m_W^2}{m_{H^\pm}^2-m_W^2}
\left[I_{4,5}(m_W^2,m_1^2) - I_{4,5}(m_2^2,m_1^2)\right],\\
	I_4(m_1^2,m_2^2) =~& \int_0^1 \df z (1-z)^2 
	\left(z-4+z \frac{m_{H^\pm}^2-m_2^2}{m_W^2}\right)\nn\\
	&\times\frac{m_1^2}{m_W^2(1-z) + m_2^2 z - m_1^2 z(1-z)}
	\log\left(\frac{m_W^2 (1-z) + m_2^2z}{m_1^2z(1-z)}\right),\\
	I_5(m_1^2,m_2^2) =~& \int_0^1 \df z 
	\frac{m_1^2z(1-z)^2}{m_W^2(1-z) + m_2^2 z - m_1^2 z(1-z)}
	\log\left(\frac{m_W^2 (1-z) + m_2^2z}{m_1^2z(1-z)}\right).
\end{align}

%% Bibliography
\bibliographystyle{jhep}
\bibliography{2HDMbib}

\end{document}